\documentclass[a4paper,11pt]{article}
\pdfoutput=1
\usepackage{jcappub}
\usepackage[T2A]{fontenc}
\usepackage{graphicx}
\usepackage{amsthm,amssymb,amsmath}
\usepackage{amsfonts}
\usepackage{revsymb}
\graphicspath{{Figures/}}
\DeclareGraphicsExtensions{.pdf,.eps}

\title{\boldmath Gravitational radiation and dynamical degrees of freedom in DGP gravity}

\author[a,b]{M. Khlopunov,}
\author[a]{D.V. Gal'tsov}

\affiliation[a]{Faculty of Physics, Lomonosov Moscow State University,\\Moscow, 119899, Russia}
\affiliation[b]{Institute of Theoretical and Mathematical Physics, Lomonosov Moscow State University,\\Moscow, 119991, Russia}

\emailAdd{khlopunov.mi14@physics.msu.ru}
\emailAdd{galtsov@phys.msu.ru}

\abstract{The paper is devoted to the study of gravitational radiation in the DGP model of gravity. It includes a detailed analysis of dynamical degrees of freedom of tensionless DGP and their comparison with other related theories. We develop a procedure for constructing the effective energy-momentum tensor of the gravitational field on the brane from the non-local effective action of the DGP model based on the Noether method used to determine the conservation laws corresponding to the symmetries of the action. Using it we construct the effective energy-momentum tensor of the dynamical degrees of freedom of DGP gravity on the brane and obtain the effective four-dimensional gravitational radiation energy flux. We then obtain an analog of the quadrupole formula for the effective gravitational radiation power of an arbitrary non-relativistic source on the brane. We also discuss the possibility of detecting the effect of leakage of gravitational waves into the extra dimension resulting from the metastable character of the effective graviton on the brane. In particular, in accordance with the infrared transparency of the bulk of DGP model, we find that the frequency of gravitational-wave signal controls the effective crossover radius, beyond which the leakage of gravitational waves manifests itself.}

\begin{document}

\maketitle
\flushbottom

\section{Introduction}
Over the past two decades, many modifications of the general theory of relativity (GR) have been developed \cite{Clifton:2011jh,Li:2020uaz}. Among them, an important class are theories of gravity with extra space-time dimensions, motivated mainly by the development of string theory, which is currently the main model of quantum gravity and requires the existence of extra dimensions for its self-consistency \cite{Green:1987sp}.   
In particular, braneworld models have been widely developed, aimed at solving certain problems of high-energy physics and cosmology \cite{Rubakov:2001kp,Barvinsky:2005ak,Cheng:2010pt,Maartens:2010ar}. Namely, the Arkani–Hamed–Dimopoulos–Dvali (ADD) \cite{Arkani-Hamed:1998jmv} and Randall–Sundrum (RS) \cite{Randall:1999ee,Randall:1999vf} models provide solutions to the hierarchy problem, while the Dvali–Gabadze–Porrati (DGP) \cite{Dvali:2000hr,Dvali:2000xg} model has been considered as a possible solution to the cosmological constant problem \cite{Deffayet:2000uy,Deffayet:2001pu,Deffayet:2002sp}. In contrast to the Kaluza–Klein \cite{Duff:1986hr} models, an important feature of the braneworld models is that they allow large volumes of extra dimensions, up to infinite volume, due to the localization of the Standard Model fields on the brane. As a result, the production of light Kaluza-Klein gravitons can significantly change astrophysical processes \cite{Arkani-Hamed:1998sfv}, opening up the possibility of experimental study of extra dimensions using astrophysical and cosmological observations.

Recently, the actively developing gravitational-wave astronomy has become one of the most promising tools for the experimental search and study of extra dimensions \cite{Ezquiaga:2018btd,Yu:2019jlb}. Extra spacetime dimensions may manifest themselves through additional polarizations of gravitational waves \cite{Andriot2017,Chu:2021uea,Khlopunov:2022ubp}, differences in the propagation of gravitational-wave and counterpart electromagnetic signals from binary neutron star mergers \cite{Yu:2016tar,Visinelli:2017bny,Lin:2020wnp,Lin:2022hus}, additional contributions to the gravitational field source in Einstein's equations \cite{Shiromizu:1999wj,Maeda:2003ar}, non-local tail signals in gravitational radiation \cite{Barvinsky:2003jf,Chu:2021uea,Khlopunov:2023hnl}, as well as modifications of quasinormal modes \cite{Chakraborty:2017qve,Mishra:2021waw} and tidal deformations of black holes and neutron stars \cite{Chakravarti:2018vlt,Cardoso:2019vof,Chakravarti:2019aup}.
 
In this paper, we consider the process of gravitational radiation and the effect of gravitational wave leakage into an extra dimension in the framework of the DGP gravity model with one infinite extra spacelike dimension. The general DGP action may contain a Nambu-Goto term for the brane, which is equivalent to the four-dimensional cosmological constant term leading to the de Sitter background solution of the theory. In principle, the DGP model contains two branches of solutions that differ in the choice of boundary conditions in the bulk \cite{Gabadadze:2004dq,Charmousis:2006pn,Brown:2016gwv,Khlopunov:2022jaw}: the normal branch corresponds to the radiation boundary conditions and admits a metastable effective graviton quasi-localized on the brane, while the self-accelerated branch corresponds to the opposite boundary condition and admits the de Sitter cosmological solution for the vacuum brane \cite{Deffayet:2000uy}. However, in the latter case the graviton contains a ghost degree of freedom \cite{Nicolis:2004qq,Gorbunov:2005zk,Charmousis:2006pn,Koyama:2007za,Gregory:2008bf}, which makes it impossible to solve the cosmological constant problem within the DGP model. Nevertheless, the normal branch of DGP model remains an interesting diffeomorphism-invariant model of metastable graviton with a soft mass \cite{Gabadadze:2003ck}, admitting the modification of laws of gravitation at cosmological scales. In what follows, we study the gravitational-wave effects in this branch of DGP model. For this purpose we can use the simpler model without Nambu-Goto term.

Since the DGP graviton effective action on the brane \cite{Luty:2003vm,Nicolis:2004qq} lacks gauge symmetry, unlike in general relativity where it is commonly used to extract the dynamical degrees of freedom of the gravitational field transporting radiation to infinity, we also investigate the dynamical degrees of freedom of DGP gravity. Based on an analogy with massive Fierz-Pauli gravity, we construct a non-local effective action of the dynamical degrees of freedom of the DGP graviton on the brane and the corresponding brane-localized effective stress-energy tensor. Since we consider the DGP model with a tensionless brane, in this paper we study the spectrum of gravitational perturbations on the background of a flat brane embedded in a five-dimensional Minkowski bulk, while most of the literature has considered the case of a de Sitter vacuum brane (see, e.g., \cite{Koyama:2005tx,Charmousis:2006pn,Koyama:2007za}).

The metastable character of effective DGP-graviton on the brane should lead to the leakage of gravitational waves into the extra dimension at cosmological scales \cite{Deffayet:2007kf}, which manifests in a faster, compared to GR, attenuation of the amplitude of gravitational waves with distance from the source. In particular, based on the behavior of the Newtonian potential in the DGP model \cite{Dvali:2000hr}, in Ref. \cite{Deffayet:2007kf} an empirical formula qualitatively describing the intensity of leakage of gravitational waves into the extra dimension was proposed
\begin{equation}
\label{eq:DGP_Deff_men_form}
h_{\times/+} \propto \frac{1}{d_L ( 1 + \left( d_L/r_c \right)^{n/2} )^{1/n}},
\end{equation}
where $\times$ and $+$ denote the polarizations of gravitational waves. Here $d_L$ is the luminosity distance to the source of gravitational waves, $r_c$ is the crossover radius determined by the parameters of the DGP model \cite{Dvali:2000hr}, and $n$ is an empirical parameter characterizing the steepness of transition of gravitational field from four-dimensional to five-dimensional behavior at large distances from the source
\begin{equation}
h_{\times/+} \xrightarrow{d_L \ll r_c} d_L^{-1}, \quad h_{\times/+} \xrightarrow{d_L \gg r_c} d_L^{-3/2}.
\end{equation}
This effect opens up the way to experimentally test the DGP model by use of joint observations of gravitational-wave and counterpart electromagnetic signals from the mergers of binary neutron stars \cite{Deffayet:2007kf} as at cosmological scales the distance to the binary star inferred from the gravitational-wave signal should appear to be larger than the one inferred from the electromagnetic signal due to the leakage of gravitational waves into the extra dimension. The first constraints on the number of spacetime dimensions and the parameters of the DGP model have already been obtained based on the Eq. \eqref{eq:DGP_Deff_men_form} from the observation of gravitational-wave signal from the merger of binary neutron star GW170817 and the counterpart gamma-ray burst GRB170817A \cite{Pardo:2018ipy,Corman:2021avn}. Also, the possibility of testing the DGP model by observations at the space gravitational-wave observatory LISA depending on the values of crossover radius $r_c$ and parameter $n$ in Eq. \eqref{eq:DGP_Deff_men_form} has been discussed \cite{Corman:2020pyr}.

Note that the process of gravitational radiation in the DGP model was recently studied in Ref. \cite{Poddar:2021yjd} by use of the effective field theory approach to the problems of radiation \cite{Cardoso:2008gn,Porto:2016pyg,Birnholtz:2013ffa}. However, being based on the momentum space calculations, this approach does not capture information about the structure of gravitational field in the wave zone and, as a consequence, the leakage of gravitational waves into the extra dimension. Moreover, this approach also obscures the study of non-local tail signals in gravitational radiation \cite{Barvinsky:2003jf,Chu:2021uea,Khlopunov:2023hnl} resulting from the Huygens principle violation in the five-dimensional bulk of DGP model \cite{hadamard2014lectures,courant2008methods,Ivanenko_book} (e.g., see also \cite{Cardoso:2002pa,Galtsov:2001iv,Galtsov:2020hhn}). Solutions to these problems were proposed in Ref. \cite{Khlopunov:2022jaw} within the framework of the scalar-field analogue of DGP model by using the Rohrlich-Teitelboim approach to radiation \cite{Rohrlich1961,Teitelboim1970,Kosyakov:1992qx}. Based on the results of this work, here we present a study of the process of gravitational radiation and the effect of leakage of gravitational waves into the extra dimension in the DGP model using classical field theory methods.

We obtain an analog of the quadrupole formula for the effective four-dimensional gra\-vi\-ta\-tio\-nal radiation power of an arbitrary non-relativistic source on the brane. To construct the effective energy-momentum tensor of the gravitational field on the brane from the non-local effective action of DGP model \cite{Luty:2003vm,Hinterbichler:2011tt,deRham:2014zqa} we give a generalization of the Noether method to the case of such non-local theories. In particular, we demonstrate that the effective energy-momentum tensor of gravitational field on the brane can be constructed using the standard formula for the canonical energy-momentum tensor by neglecting the non-local mass terms. We additionally test this result in the framework of scalar-field analog of DGP model by computing the brane localised part of the Hamiltonian of the full five-dimensional model. Based on the obtained quadrupole formula, we also calculate the parameters of the empirical formula \eqref{eq:DGP_Deff_men_form}. We find that the parameter $n$ characterizing the steepness of transition of gravitational field from four-dimensional to five-dimensional behavior is equal to unity, and the crossover radius $r_c$ should be replaced with an effective crossover radius depending on the frequency of gravitational wave. In accordance with the infrared transparency of the bulk of DGP model \cite{Dvali:2000xg,Brown:2016gwv,Khlopunov:2022jaw}, for gravitational waves with frequencies in the sensitivity ranges of current and future gravitational-wave observatories \cite{Moore:2014lga} the effective crossover radius is found to take values many orders in magnitude larger than the crossover radius $r_c$, making it difficult to observe the leakage of gravitational waves and potentially explaining the results of Refs. \cite{Pardo:2018ipy,Corman:2021avn}.

This paper is organised as follows. In section \ref{II} we briefly recall the geometric formulation of the full nonlinear DGP model of gravity, as well as the derivation of the linear equation of motion of the effective metastable graviton on the brane. Section \ref{III} is devoted to the generalization of Noether method to the case of constructing the effective energy-momentum tensor of the gravitational field from the non-local effective action of DGP graviton on the brane. As the effective action of DGP graviton on the brane does not have a gauge symmetry, in contrast with the graviton action in GR, in section \ref{IV} we construct the effective action and the energy-momentum tensor of the dynamical degrees of freedom of the DGP graviton on the brane, which carry the gravitational radiation energy flux. Finally, section \ref{V} is devoted to the calculation of quadrupole formula for the effective gravitational radiation power of an arbitrary non-relativistic source on the brane and the estimation of parameters of the empirical formula \eqref{eq:DGP_Deff_men_form}. In section \ref{VI} we discuss the obtained results and their relation to other works.

\section{DGP model of gravity}\label{II}

The DGP model of gravity \cite{Dvali:2000hr,Dvali:2000xg} is a braneworld model with the dynamical 3-brane $\Sigma$ embedded into the five-dimensional bulk $\cal M$ with one infinite extra spacelike dimension. All matter fields and particles are assumed to be localised on $\Sigma$ and only gravity propagates through the extra dimension. The dynamics of gravitational field is determined by the five-dimensional Einstein-Hilbert action in the bulk $\cal M$, as well as by the additional four-di\-men\-sio\-nal Einstein-Hilbert term induced on the brane $\Sigma$ due to the interaction of bulk gravitons with the quantum loops of matter particles on the brane \cite{Dvali:2000hr}.

\subsection{Non-linear theory}

Let us start with briefly recalling the geometrical formulation of full non-linear DGP gravity model (see, e.g., \cite{Hinterbichler:2011tt,deRham:2014zqa}).

\begin{figure}[t]
\center{\includegraphics[width=0.7\linewidth]{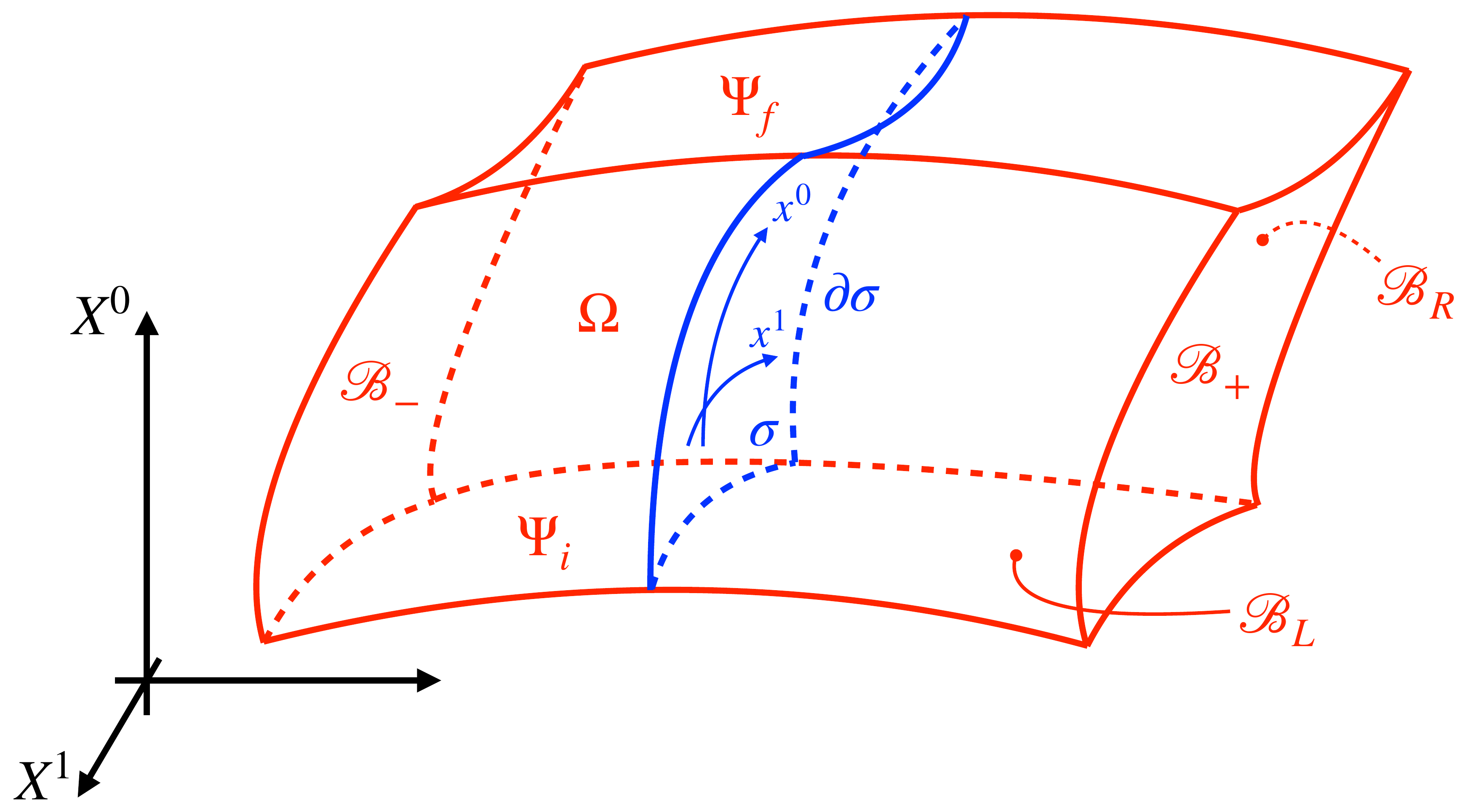}}
\caption{Domain for formulating the variational principle in the DGP model.}
\label{fig:4.1}
\end{figure}

We cover $\cal M$ with coordinates $X^A$, $A=\overline{0,4}$ and denote the bulk metric as
\begin{equation}
ds^2 = {\cal G}_{AB} dX^A dX^B, \quad {\cal G}_{AB} \propto (-++++).
\end{equation}
The brane $\Sigma$ is a timelike hypersurface defined both by the bulk embedding equation and by parameterizing its worldvolume coordinates $Z^A$
\begin{equation}
\label{eq:brane_embedding}
y(X)=0 \quad \Longleftrightarrow \quad Z^A = Z^A(x),
\end{equation}
where $x^\alpha$, $\alpha=\overline{0,3}$ are coordinates on the brane (see Fig. \eqref{fig:4.1}). We assume that $y(X)$ increases in the direction of increasing coordinate along the extra dimension. Accordingly, the induced metric on the brane is defined as
\begin{equation}
g_{\mu\nu} = e_\mu^M e_\nu^N {\cal G}_{MN}, \quad e^M_\mu = \partial_\mu Z^M.
\end{equation}
Further, we denote all the coordinate transformations Jacobians as $e^A_\alpha$, where the upper and lower indices determine the specific coordinate transformation.

The unit normal vector $n^A$ to the brane $\Sigma$ pointing in the direction of increasing $y(X)$ is given by (see, e.g., \cite{Poisson:2009pwt})
\begin{equation}
\label{eq:brane_normal_vect}
n^A = \frac{\varepsilon {\cal G}^{AB} \partial_B y}{|{\cal G}^{MN} \partial_M y \partial_N y|^{1/2}}, \quad \varepsilon = n_A^2 = {\rm sgn}((\partial_M y)^2) = 1.
\end{equation}
Correspondingly, the brane extrinsic curvature is written as \cite{Poisson:2009pwt}
\begin{equation}
\label{eq:br_ext_curv}
K_{\alpha\beta} = e_\alpha^A e_\beta^B \nabla_B n_A.
\end{equation}
By analogy with the coordinate transformations Jacobians, indices of covariant derivatives determine by which coordinates and metric they are calculated.

In DGP model, the variational principle is defined in a bounded region $\Omega$ of the bulk $\cal M$ with a boundary $\partial \Omega$ (see Fig. \eqref{fig:4.1}). We choose the region $\Omega$ so that it encloses the brane $\Sigma$ on both sides. The boundary $\partial \Omega$ consists of two spacelike hypersurfaces $\Psi_{i/f}$ and a timelike hypersurface $\cal B$ closing $\partial \Omega$ (see Fig. \eqref{fig:4.1})
\begin{equation}
\partial \Omega = \Psi_i \cup {\cal B} \cup \Psi_f, \quad {\cal B} = {\cal B}_- \cup {\cal B}_L \cup {\cal B}_R \cup {\cal B}_+.
\end{equation}
On the boundary $\partial \Omega$, we introduce the coordinate system ${\cal Y}^a$ and the unit normal vector ${\cal N}^A$ directed outward from $\Omega$
\begin{equation}
\label{eq:dOmega_norm_vec}
{\cal N}_A e_a^A = 0, \quad {\cal N}_A^2 = {\cal E} = \pm 1.
\end{equation}
Here, the coordinate transformation Jacobian $e_a^A$ forms a set of five-vectors tangent to the boundary $\partial \Omega$ \cite{Poisson:2009pwt}. We also define the induced metric ${\cal H}_{ab}$ and the extrinsic curvature ${\cal K}_{ab}$ on $\partial \Omega$ as
\begin{equation}
\label{eq:dOmega_ind_metr_&_ext_curv}
{\cal H}_{ab} = e_a^A e_b^B {\cal G}_{AB}, \quad {\cal K}_{ab} = e_a^A e_b^B \nabla_B {\cal N}_A.
\end{equation}

Finally, the boundary $\partial \Omega$ cuts out from the brane $\Sigma$ a region $\sigma$ with a boundary $\partial \sigma$
\begin{equation}
\sigma = \Omega \cap \Sigma, \quad \partial \sigma = \partial \Omega \cap \Sigma.
\end{equation}
We cover $\partial \sigma$ with coordinates $\theta^i$ and introduce the unit 4-vector $r^\mu$ normal to it, as well as the set of 4-vectors $e^\alpha_i$ tangent to $\partial \sigma$
\begin{equation}
\partial \sigma: \ x^\mu = x^\mu(\theta), \quad r_\mu^2 = \epsilon = \pm 1, \quad r_\alpha e^\alpha_i = 0.
\end{equation}
Thus, we construct the induced metric $\mathbb{H}_{ij}$ and the extrinsic curvature $\mathbb{K}_{ij}$ on $\partial \sigma$ as
\begin{align}
& \mathbb{H}_{ij} = e^\mu_i e^\nu_j g_{\mu\nu} = e^M_i e^N_j {\cal G}_{MN}, \\
& \mathbb{K}_{ij} = e^\alpha_i e^\beta_j \nabla_\beta r_\alpha = e^\alpha_i e^\beta_j \mathbb{K}_{\alpha\beta}, \quad \mathbb{K} = \mathbb{H}^{ij} \mathbb{K}_{ij}.
\end{align}

As a result, inside $\Omega$ the action of DGP gravity model with one extra dimension is written as \cite{Dvali:2000hr}
\begin{align}
\label{eq:5D_DGP_act_gen}
S & = \frac{1}{2} M_5^3 \int_{\Omega} d^5X \sqrt{-{\cal G}} {\cal R} + M_5^3 \oint_{\partial \Omega} d^4{\cal Y} \sqrt{|{\cal H}|} {\cal E} {\cal K} + \frac{1}{2} M_4^2 \int_\sigma d^4x \sqrt{-g} R \nonumber \\ & + M_4^2 \oint_{\partial \sigma} d^3 \theta \sqrt{|\mathbb{H}|} \epsilon \mathbb{K} - \Lambda \int_{\sigma} d^4x \sqrt{-g} + \int_{\sigma} d^4 x \sqrt{-g} {\cal L}_{\rm m}.
\end{align}
Here, $\cal R$ and $R$ are the Ricci scalars constructed from the bulk metric ${\cal G}_{MN}$ and the induced metric on the brane $g_{\mu\nu}$, correspondingly, ${\cal L}_{\rm m}$ schematically denotes the brane localised matter's Lagrangian, and we have introduced the Nambu-Goto action for the brane with tension $\Lambda$. We have also added to the action the Gibbons-Hawking boundary terms on $\partial \Omega$ and $\partial \sigma$ for the well-posedness of the variational principle \cite{Gibbons:1976ue,Hawking:1995fd,Poisson:2009pwt}.

For the convenience of calculations, we rewrite the action \eqref{eq:5D_DGP_act_gen} in the Gaussian coordinate system of the brane $\Sigma$ -- its rest frame
\begin{equation}
\label{eq:Gauss_coord_system}
Z^4=0, \quad Z^\mu = x^\mu,
\end{equation}
where the brane is a fixed hypersurface in $\cal M$. By analogy with the Hamiltonian formulation of General Relativity \cite{Arnowitt:1962hi,Wald:1984rg,Poisson:2009pwt}, this coordinate system is constructed from the $(4+1)$-foliation of the bulk by non-intersecting timelike hypersurfaces (see, e.g., \cite{Hinterbichler:2011tt}). In particular, parameterizing the coordinate along the extra dimension by the values of the brane embedding function $y(X)$ we rewrite the bulk metric in $(4+1)$-foliated coordinates as
\begin{equation}
\label{eq:5D_metric_foliation}
ds^2 = {\cal G}_{AB} dX^A dX^B = g_{\alpha\beta} dx^\alpha dx^\beta + 2 N_\alpha dx^\alpha dy + \left( N_\alpha^2 + N^2 \right) dy^2,
\end{equation}
where $N(x,y)$ is a lapse function, and $N^\alpha(x,y)$ is a shift 4-vector. Therefore, in the Gaussian coordinate system of the brane the action of DGP gravity model takes the following form \cite{Luty:2003vm,Charmousis:2006pn,Hinterbichler:2011tt} (for technical details see, also, \cite{Poisson:2009pwt})
\begin{align}
S & = \frac{1}{2} M_5^3 \int_{y_-}^{y_+} dy \int_{\sigma_y} d^4x \sqrt{-g} N \! \left( R + K^2 - K^{\alpha\beta} K_{\alpha\beta} \right) + M_5^3 \int_{y_-}^{y_+} dy \oint_{\partial \sigma_y} d^3\theta \sqrt{|\mathbb{H}|} \epsilon \mathbb{K} N \nonumber \\
& + \frac{1}{2} M_4^2 \int_{y_-}^{y_+} dy \int_{\sigma_y} d^4x \sqrt{-g} R \, \delta(y) + M_4^2 \int_{y_-}^{y_+} dy \oint_{\partial \sigma_y} d^3\theta \sqrt{|\mathbb{H}|} \epsilon \mathbb{K} \, \delta(y) \nonumber \\
\label{eq:5D_DGP_4+1_act}
& - \Lambda \int_{y_-}^{y_+} dy \int_{\sigma_y} d^4x \sqrt{-g} \, \delta(y) + \int_{y_-}^{y_+} dy \int_{\sigma_y} d^4x \sqrt{-g} {\cal L}_{\rm mat} \, \delta(y).
\end{align}
Here, $\sigma_y$ denotes a timelike hypersurface parallel to the brane bounded by $\partial \Omega$ and corresponding to a certain value of the embedding function: $y(X) \vert_{\sigma_y} = {\rm Const}$. In particular, in these notations the brane is a hypersurface $\sigma_0$ and the values $y_{\pm}$ correspond to the ${\cal B}_{\pm}$ parts of the boundary $\partial \Omega$. Finally, we also take into account that the extrinsic curvature of the hypersurfaces $\sigma_y$ depends on the induced metric on them and the shift vector
\begin{equation}
\label{eq:brane_ext_curv_interm_ind_metr}
\pounds_y g_{\mu\nu} = 2 N K_{\mu\nu} + \nabla_\mu N_\nu + \nabla_\nu N_\mu = \partial_y g_{\mu\nu} \quad \Longrightarrow \quad K_{\mu\nu} = \frac{1}{2N} \left( \partial_y g_{\mu\nu} - \nabla_\mu N_\nu - \nabla_\nu N_\mu \right).
\end{equation}

The gravitational field equations of motion resulting from the Eq. \eqref{eq:5D_DGP_4+1_act} are written as
\begin{align}
\label{eq:DGP_constr_eq_1}
& R - K^2 + K^{\alpha\beta} K_{\alpha\beta} = 0, \\
\label{eq:DGP_constr_eq_2}
& \nabla_\alpha \left( K^{\mu\alpha} - g^{\mu\alpha} K \right) = 0, \\
\label{eq:DGP_dynam_eq}
& \frac{1}{2} M_5^3 \sqrt{-g} N \left( R_{\alpha\beta} - \frac{1}{2} g_{\alpha\beta} R - \frac{1}{2} g_{\alpha\beta} K^2 + \frac{1}{2} g_{\alpha\beta} K^{\mu\nu} K_{\mu\nu} \right) \nonumber \\ & \quad + \frac{1}{2} M_5^3 \sqrt{-g} \, \big( g_{\alpha\beta} \nabla_\sigma \nabla^\sigma N - \nabla_\alpha \nabla_\beta N \big) + \frac{1}{2} M_5^3 g_{\alpha\beta} \, \partial_y \! \left( \sqrt{-g} K \right) - \frac{1}{2} M_5^3 g_{\alpha\mu} g_{\beta\nu} \, \partial_y \! \left( \sqrt{-g} K^{\mu\nu} \right) \nonumber \\ & \quad - M_5^3 \sqrt{-g} N g^{\mu\nu} K_{\alpha\mu} K_{\beta\nu} - M_5^3 \sqrt{-g} K_{\nu(\alpha} \nabla^\nu N_{\beta)} + \frac{1}{2} M_5^3 \sqrt{-g} \, \nabla_\rho \left( \left( K_{\alpha\beta} - g_{\alpha\beta} K \right) N^\rho \right) \nonumber \\ & \quad + \frac{1}{2} M_4^2 \sqrt{-g} \left( R_{\alpha\beta} - \frac{1}{2} g_{\alpha\beta} R \right) \delta(y) = \frac{1}{2} \sqrt{-g} \, {T_{\rm m}}_{\alpha\beta} \, \delta(y) - \frac{1}{2} \sqrt{-g} \, \Lambda g_{\alpha\beta} \, \delta(y).
\end{align}
The energy-momentum tensor of brane localised matter ${T_{\rm m}}_{\alpha\beta}$ is defined in a standard manner as
\begin{equation}
\label{eq:brane_mat_EMT_def}
T_{\alpha\beta} = - \frac{2}{\sqrt{-g}} \frac{\delta S_{\rm m}}{\delta g^{\alpha\beta}}.
\end{equation}
We also obtain the Israel junction conditions on the brane \cite{Israel:1966rt,Poisson:2009pwt,Charmousis:2006pn,deRham:2014zqa} from the Eq. \eqref{eq:DGP_dynam_eq}
\begin{align}
& \left. g_{\alpha\beta} \right \vert_{y=-0} = \left. g_{\alpha\beta} \right \vert_{y=+0}, \\
\label{eq:DGP_Israel_junc_cond_2}
& - M_5^3 \left \lbrack K_{\alpha\beta} - g_{\alpha\beta} K \right \rbrack_{y=0} + M_4^2 \left. \left( R_{\alpha\beta} - \frac{1}{2} g_{\alpha\beta} R \right) \right \vert_{y=0} = {T_{\rm m}}_{\alpha\beta} - \Lambda \left. g_{\alpha\beta} \right \vert_{y=0}.
\end{align}
Here, $[ \ldots ]_{y=0}$ denotes the jump of the corresponding quantity on the brane -- its value on the right from the brane $y=+0$ minus its value on the left $y=-0$.

In what follows we consider tensionless brane setting $\Lambda = 0$, so that the ground state of the theory is the flat brane embedded into the five-dimensional Minkowski space. It can be checked that this is indeed a solution of the vacuum equations of motion (\ref{eq:DGP_constr_eq_1}--\ref{eq:DGP_dynam_eq})
\begin{equation}
{T_{\rm m}}_{\alpha\beta} = 0, \quad \Lambda = 0 \quad \Longrightarrow \quad g_{\mu\nu} = \eta_{\mu\nu}, \quad N=1, \quad N^\alpha=0, \quad K_{\mu\nu}=0.
\end{equation}
We then  linearize the equations of motion of the gravitational field on the flat background and study dynamical degrees of freedom of the resulting theory. Note that the  
perturbations spectrum in the tensional DGP model with de Sitter backgrouns state \cite{Charmousis:2006pn} generically is different.

\subsection{Linearized theory}

Let us consider small perturbations of the gravitational field on the flat background
\begin{equation}
\label{eq:DGP_grav_field_pert}
g_{\mu\nu} = \eta_{\mu\nu} + h_{\mu\nu}, \quad N = 1 + n, \quad N^\mu = n^\mu, \quad |h_{\mu\nu}|, \, |n^\mu|, \, |n| \ll 1.
\end{equation}
Further, our derivation of linear equations of motion of DGP gravity follows the Ref. \cite{deRham:2014zqa}.

In the linear approximation, the equations of motion (\ref{eq:DGP_constr_eq_1}--\ref{eq:DGP_dynam_eq}) take the form
\begin{align}
\label{eq:DGP_lin_EoM_1}
& \partial^\alpha \partial^\beta h_{\alpha\beta} - \partial^\alpha \partial_\alpha h = 0, \\
\label{eq:DGP_lin_EoM_2}
& \partial_y \partial_\alpha h^{\alpha\mu} - \partial_y \partial^\mu h + \partial^\mu \partial_\alpha n^\alpha - \partial^\alpha \partial_\alpha n^\mu = 0, \\
& \frac{1}{2} M_5^3 \big( \partial^\sigma \partial_\alpha h_{\beta\sigma} + \partial^\sigma \partial_\beta h_{\alpha\sigma} - \partial^\sigma \partial_\sigma h_{\alpha\beta} - \partial_\alpha \partial_\beta h - \eta_{\alpha\beta} \partial^\mu \partial^\nu h_{\mu\nu} + \eta_{\alpha\beta} \partial^\sigma \partial_\sigma h \big) \nonumber \\ & \quad + M_5^3 \big( \eta_{\alpha\beta} \partial^\sigma \partial_\sigma n - \partial_\alpha \partial_\beta n \big) + \frac{1}{2} M_5^3 \big( \eta_{\alpha\beta} \partial_y^2 h - \partial_y^2 h_{\alpha\beta} - 2 \eta_{\alpha\beta} \partial_y \partial_\mu n^\mu + \partial_y \partial_\alpha n_\beta + \partial_y \partial_\beta n_\alpha \big) \nonumber \\ & \quad + \frac{1}{2} M_4^2 \big( \partial^\sigma \partial_\alpha h_{\beta\sigma} + \partial^\sigma \partial_\beta h_{\alpha\sigma} - \partial^\sigma \partial_\sigma h_{\alpha\beta} - \partial_\alpha \partial_\beta h - \eta_{\alpha\beta} \partial^\mu \partial^\nu h_{\mu\nu} + \eta_{\alpha\beta} \partial^\mu \partial_\mu h \big) \delta(y) \nonumber \\ \label{eq:DGP_lin_EoM_3} & \quad = {T_{\rm m}}_{\alpha\beta} \, \delta(y). 
\end{align}
Analogously, in the linear approximation Israel junction condition \eqref{eq:DGP_Israel_junc_cond_2} is written as
\begin{align}
\label{eq:DGP_lin_Israel_junc_cond_2}
& \frac{1}{2} M_5^3 \big \lbrack \eta_{\alpha\beta} \partial_y h - \partial_y h_{\alpha\beta} - 2 \eta_{\alpha\beta} \partial_\sigma n^\sigma + \partial_\alpha n_\beta + \partial_\beta n_\alpha \big \rbrack_{y=0} + \frac{1}{2} M_4^2 \big( \partial^\sigma \partial_\alpha h_{\beta\sigma} + \partial^\sigma \partial_\beta h_{\alpha\sigma} \nonumber \\ & \quad - \left. \partial^\sigma \partial_\sigma h_{\alpha\beta} - \partial_\alpha \partial_\beta h - \eta_{\alpha\beta} \partial^\mu \partial^\nu h_{\mu\nu} + \eta_{\alpha\beta} \partial^\mu \partial_\mu h \big) \right \vert_{y=0} = {T_{\rm m}}_{\alpha\beta}.
\end{align}
The obtained equations of motion have a gauge symmetry \cite{Hinterbichler:2011tt}
\begin{equation}
\label{eq:DGP_gauge_sym}
h_{\alpha\beta} \to h_{\alpha\beta} + \partial_\alpha \xi_\beta + \partial_\beta \xi_\alpha, \quad n^\mu \to n^\mu + \partial_y \xi^\mu, \quad n \to n,
\end{equation}
corresponding to the residual bulk diffeomorphisms that do not affect the Gaussian coordinate system of the brane \eqref{eq:Gauss_coord_system} \cite{Hinterbichler:2011tt}. To fix this gauge symmetry, we impose a five-dimensional Lorentz gauge condition in the bulk
\begin{equation}
\label{eq:DGP_Lor_gauge_cond}
\partial^A H_{AB} - \frac{1}{2} \partial_B H = 0, \quad H = \eta^{AB} H_{AB},
\end{equation}
where $H_{AB}$ denotes the perturbations of five-dimensional bulk metric ${\cal G}_{AB} = \eta_{AB} + H_{AB}$. As follows from the Eq. \eqref{eq:5D_metric_foliation}, they are related to the perturbations of $(4+1)$-foliated metric as
\begin{equation}
\label{eq:5D_GR_bulk-foliation_perturb_rel}
H_{\mu\nu} = h_{\mu\nu}, \quad H_{\mu y} = n_\mu, \quad H_{yy} = 2n.
\end{equation}
Thus, the linear equations of motion of DGP gravity take the form \cite{deRham:2014zqa}
\begin{align}
\label{eq:DGP_lin_bulk_EoM_Lor_gauge}
& {_5}\square \big( h_{\alpha\beta} - \eta_{\alpha\beta} h \big) = 0, \quad y \neq 0, \\
\label{eq:DGP_lin_Israel_junc_cond_2_Lor_gauge}
& M_4^2 \! \left. \big( {_4}\square h_{\alpha\beta} - \partial_\alpha \partial_\beta h \big) \right \vert_{y=0} + M_5^3 \big \lbrack \partial_y h_{\alpha\beta} - \eta_{\alpha\beta} \partial_y h \big \rbrack_{y=0} = - 2 {T_{\rm m}}_{\alpha\beta}, \\
\label{eq:DGP_lin_eq_3}
& \partial^\alpha h_{\alpha\beta} = \partial_\beta h, \quad n = \frac{1}{2} h, \quad n^\mu = 0.
\end{align}
Here, ${_{5/4}} \square$ denotes the five/four-dimensional d'Alembert operators in Minkowski spacetime, correspondingly. We split Eq. \eqref{eq:DGP_lin_EoM_3} into the bulk equation of motion \eqref{eq:DGP_lin_bulk_EoM_Lor_gauge} and the junction condition on the brane \eqref{eq:DGP_lin_Israel_junc_cond_2_Lor_gauge}, and the last equation in \eqref{eq:DGP_lin_eq_3} fixes the residual gauge symmetry \eqref{eq:DGP_gauge_sym} with transformation parameters satisfying the condition ${_5}\square \xi_\mu = 0$.  Also, the energy-momentum tensor of the brane localised matter have to be conserved $\partial^\alpha {T_{\rm m}}_{\alpha\beta} = 0$ for the self-consistency of the obtained equations.

Imposing the radiation boundary condition on the gravitational field in the bulk (see, e.g., \cite{Khlopunov:2022ubp})
\begin{equation}
h_{\mu\nu} \propto \int d\omega \, e^{-i \omega t \pm i k(\omega) y} \tilde{h}_{\mu\nu}(\omega), \quad y \gtrless 0,
\end{equation}
we find from the Eq. \eqref{eq:DGP_lin_bulk_EoM_Lor_gauge} the following expression for the jump of the induced metric derivative on the brane
\begin{equation}
\big \lbrack \partial_y h_{\alpha\beta} - \eta_{\alpha\beta} \partial_y h \big \rbrack_{y=0} = 2 \sqrt{-{_4}\square} \left. \big( h_{\alpha\beta} - \eta_{\alpha\beta} h \big) \right \vert_{y=0}.
\end{equation}
As a result, the junction condition \eqref{eq:DGP_lin_Israel_junc_cond_2_Lor_gauge}, being the effective four-dimensional equation of motion of the gravitational field on the brane, takes the form
\begin{equation}
\label{eq:DGP_eff_EoM}
\left. \big( {_4}\square h_{\alpha\beta} - \partial_\alpha \partial_\beta h \big) \right \vert_{y=0} + M_c \sqrt{-{_4}\square} \left. \big( h_{\alpha\beta} - \eta_{\alpha\beta} h \big) \right \vert_{y=0} = - \frac{2}{M_4^2} {T_{\rm m}}_{\alpha\beta}, \quad M_c = \frac{2M_5^3}{M_4^2}.
\end{equation}

Note that the presence of a strong coupling regime in the DGP model \cite{Luty:2003vm,Rubakov:2003zb,Nicolis:2004qq} imposes constraints on the validity range of the linear approximation. Namely, the further calculations are valid only for the processes of emission of gravitational waves by systems of gravitating objects separated from each other by a distance $r \gg r_{\rm sc} \sim (M_c^2 M_4)^{-1/3}$ and moving with characteristic frequencies $\omega \ll m_{\rm sc} \sim (M_c^2 M_4)^{1/3}$.

\section{Effective theory on the brane}\label{III}

To calculate the effective four-dimensional gravitational radiation power of matter on the brane, one needs to construct the effective energy-momentum tensor (EMT) of the gra\-vi\-ta\-tio\-nal field on the brane. One way to construct it is to project and localise on the brane the full five-dimensional EMT of the gravitational field in the bulk, by analogy with the effective EMT of the scalar DGP model obtained in Ref. \cite{Khlopunov:2022jaw}. However, in this section we develop a procedure for constructing the effective EMT of the gravitational field from the non-local effective action of the DGP model on the brane based on the Noether method.

\subsection{Effective energy-momentum tensor of DGP scalar}

First, we illustrate our method for constructing the effective EMT of the field on the brane within the framework of scalar field analog of DGP model. Due to its simplicity, it is easy to additionally verify the proposed method by calculating the brane localized part of the full five-dimensional Hamiltonian of the scalar field in the bulk.

\subsubsection{Effective action on the brane}

For the reader's convenience, we first briefly recall the method for constructing the effective action of scalar field DGP model on the brane \cite{Luty:2003vm}.

Let us rewrite the action of the scalar field DGP model \cite{Dvali:2000hr} as
\begin{equation}
\label{eq:DGP_sc_act_aux}
S = M_5^3 \int d^4x \, dy \left( ( \partial_\alpha \varphi )^2 - ( \partial_y \varphi )^2 \right) + M_4^2 \int d^4x \, dy \, \delta(y) \, ( \partial_\alpha \varphi )^2, \quad \eta_{\mu\nu} \propto (+--\,-).
\end{equation}
Fourier transforming the scalar field over the coordinates on the brane, we obtain the solution to the corresponding equation of motion in the bulk
\begin{equation}
{_5}\square \varphi(x;y) = 0, \quad y \neq 0,
\end{equation}
satisfying the radiation boundary condition in the bulk as \cite{Luty:2003vm,Khlopunov:2022jaw}
\begin{equation}
\label{eq:DGP_sc_in_bulk_gen}
\varphi(x;y) = \int \frac{d^4p}{(2\pi)^4} \, \tilde{\varphi}(p) e^{-ipx} e^{\sqrt{-p^2}|y|} = e^{\sqrt{{_4}\square}|y|} \psi(x), \quad \psi(x) = \varphi(x;0).
\end{equation}
To construct the effective action of the scalar field $\psi(x)$ on the brane, we substitute Eq. \eqref{eq:DGP_sc_in_bulk_gen} into the five-dimensional action \eqref{eq:DGP_sc_act_aux} and compute the integral over the extra dimension.

Using the Eq. \eqref{eq:DGP_sc_in_bulk_gen} and the Riemann-Lebesgue lemma, one finds that the parts of the bulk action \eqref{eq:DGP_sc_act_aux} to the left and right of the brane have the same form
\begin{equation}
M_5^3 \int d^4x \int_{\mathbb{R}^\pm} dy \left( \partial^\alpha \varphi \partial_\alpha \varphi - \partial_y \varphi \partial_y \varphi \right) = M_5^3 \int \frac{d^4p}{(2\pi)^4} \sqrt{-p^2} \tilde{\varphi}(p) \tilde{\varphi}(-p).
\end{equation}
Combining them and substituting the $\delta$-function into the momentum integral,  we rewrite the bulk part of the action \eqref{eq:DGP_sc_act_aux} in the coordinate space as
\begin{align}
M_5^3 \int d^4x \, dy \left( (\partial_\alpha \varphi)^2 - (\partial_y \varphi)^2 \right) & = 2 M_5^3 \int \frac{d^4p}{(2\pi)^4} \int d^4k \, \delta^{(4)}(k+p) \sqrt{-p^2} \tilde{\varphi}(p) \tilde{\varphi}(k) \nonumber \\ & = 2 M_5^3 \int d^4x \, \psi \sqrt{{_4}\square} \, \psi.
\end{align}
Note that, due to the definition of the square root of d'Alembertian based on the field's Fourier transform, one can <<integrate by parts>> the integrands containing it without changing the sign
\begin{align}
\label{eq:sq_dAlemb_by_parts}
\int d^4x \, A(x) \sqrt{{_4}\square} \, B(x) = \int d^4x \, B(x) \sqrt{{_4}\square} \, A(x).
\end{align}

Thus, the effective action of the DGP scalar field on the brane takes the form \cite{Luty:2003vm}
\begin{equation}
\label{eq:DGP_sc_eff_act}
S_{\rm eff} = M_4^2 \int d^4x \left( \partial^\alpha \psi \partial_\alpha \psi + M_c \psi \sqrt{{_4}\square} \, \psi \right).
\end{equation}
Correspondingly, the effective equation of motion of the field on the brane is found as \cite{Luty:2003vm,Khlopunov:2022jaw}
\begin{equation}
\label{eq:DGP_sc_eff_EoM}
\frac{\delta S_{\rm eff}}{\delta \psi} = 0 \quad \Longrightarrow \quad {_4}\square \psi - M_c \sqrt{{_4}\square} \, \psi = 0.
\end{equation}
Here, when varying the action we integrated by parts the square root of d'Alembertian without changing the sign \eqref{eq:sq_dAlemb_by_parts}.

\subsubsection{Effective energy-momentum tensor on the brane}

The effective EMT of the DGP scalar field on the brane cannot be constructed from the action \eqref{eq:DGP_sc_eff_act} using the standard formula for the canonical EMT (see, e.g., \cite{Maggiore:2005qv,Stepanyanz_book}), since the effective Lagrangian contains not only the first derivatives of the field but also a non-local mass term
\begin{equation}
\label{eq:DGP_sc_eff_Lagrangian}
{\cal L}_{\rm eff} = M_4^2 \partial^\alpha \psi \partial_\alpha \psi + 2 M_5^3 \psi \sqrt{{_4}\square} \, \psi.
\end{equation}
We construct it by the Noether method used to obtain the conservation laws corresponding to the symmetries of the action \cite{Maggiore:2005qv,Stepanyanz_book}. In particular, in the standard local theory the conservation of EMT corresponds to the invariance of the action with respect to the global shifts of coordinates.

We infinitesimally shift coordinates in the action \eqref{eq:DGP_sc_eff_act}, temporarily assuming the trans\-for\-ma\-tion parameters to depend on initial coordinates
\begin{equation}
x^\mu \to x^{\prime \mu} = x^\mu + \xi^\mu(x), \quad \psi(x) \to \psi'(x') = \psi(x), \quad |\xi^\mu| \ll 1.
\end{equation}
The field derivatives and the transformation Jacobian are expanded in $\xi$ up to the first order contributions as
\begin{align}
\label{eq:der_trans_1}
& \partial^\prime_\alpha = \partial_\alpha - \partial_\alpha \xi^\mu \partial_\mu + {\cal O}(\xi^2), \quad \det \left( \frac{\partial x^{\prime \alpha}}{\partial x^\beta} \right) = 1 + \partial_\sigma \xi^\sigma + {\cal O}(\xi^2), \\
\label{eq:der_trans_2}
& \sqrt{{_4}\square'} = \sqrt{{_4}\square} - \partial^\alpha \xi^\beta \frac{1}{\sqrt{{_4}\square}} \partial_\alpha \partial_\beta - \frac{1}{2} {_4}\square \xi^\alpha \frac{1}{\sqrt{{_4}\square}} \partial_\alpha + {\cal O}(\xi^2).
\end{align}
As a result, the action \eqref{eq:DGP_sc_eff_act} receives an infinitesimal increment, which up to the boundary terms is written as
\begin{align}
S_{\rm eff}' \lbrack \psi' \rbrack & = S_{\rm eff} \lbrack \psi \rbrack + \delta S_{\rm eff} \lbrack \psi \rbrack + {\cal O}(\xi^2), \\
\delta S_{\rm eff} \lbrack \psi \rbrack & = \int d^4x \bigg( - 2 M_4^2 \partial^\alpha \xi^\beta \partial_\alpha \psi \partial_\beta \psi + M_4^2 \partial_\sigma \xi^\sigma \partial^\alpha \psi \partial_\alpha \psi + 2 M_5^3 \partial_\sigma \xi^\sigma \psi \sqrt{{_4}\square} \, \psi \nonumber \\ & - M_5^3 \partial^\alpha \xi^\beta \psi \frac{1}{\sqrt{{_4}\square}} \partial_\alpha \partial_\beta \psi + M_5^3 \partial^\alpha \xi^\beta \partial_\alpha \psi \frac{1}{\sqrt{{_4}\square}} \partial_\beta \psi \bigg).
\end{align}
We remove the derivatives from $\xi^\alpha$ by integrating by parts and after that assume the trans\-for\-ma\-tion parameters to be constants
\begin{align}
\label{eq:DGP_sc_act_inc_unmas}
\delta S_{\rm eff} & \overset{\to}{=} \int d^4x \, \xi^\beta \partial^\alpha \bigg( 2 M_4^2 \partial_\alpha \psi \partial_\beta \psi - \eta_{\alpha\beta} M_4^2 \partial^\sigma \psi \partial_\sigma \psi - \eta_{\alpha\beta} 2 M_5^3 \psi \sqrt{{_4}\square} \, \psi \nonumber \\ & + M_5^3 \psi \frac{1}{\sqrt{{_4}\square}} \partial_\alpha \partial_\beta \psi - M_5^3 \partial_\alpha \psi \frac{1}{\sqrt{{_4}\square}} \partial_\beta \psi \bigg), \quad \xi^\alpha = {\rm Const}.
\end{align}
Here $\overset{\to}{=}$ denotes the result of integration by parts up to the boundary terms discarded in the action. In the limit $M_5 \to 0$, we obtain in Eq. \eqref{eq:DGP_sc_act_inc_unmas} under the 4-divergence operator the canonical EMT of the massless scalar field, being conserved on shell.

In the case of DGP model $M_5 \neq 0$, we can eliminate some terms under the divergence operator in the Eq. \eqref{eq:DGP_sc_act_inc_unmas}. Integrating by parts, we find that the divergence of the non-local part of the action increment is zero off shell
\begin{align}
\delta S_{\rm eff}^{\rm nl} & = \int d^4x \, \xi^\beta \partial^\alpha \bigg( - \eta_{\alpha\beta} 2 M_5^3 \psi \sqrt{{_4}\square} \, \psi + M_5^3 \psi \frac{1}{\sqrt{{_4}\square}} \partial_\alpha \partial_\beta \psi - M_5^3 \partial_\alpha \psi \frac{1}{\sqrt{{_4}\square}} \partial_\beta \psi \bigg) \nonumber \\ & \overset{\to}{=} \int d^4x \, \xi^\beta \left( - 2 M_5^3 \partial_\beta \psi \sqrt{{_4}\square} \, \psi - 2 M_5^3 \psi \sqrt{{_4}\square} \, \partial_\beta \psi \right) \overset{\to}{=} 0.
\end{align}
Thus, the increment of the effective action of DGP scalar takes the form
\begin{equation}
\delta S_{\rm eff} = \int d^4x \, \xi^\beta \partial^\alpha \bigg( 2 M_4^2 \partial_\alpha \psi \partial_\beta \psi - \eta_{\alpha\beta} M_4^2 \partial^\sigma \psi \partial_\sigma \psi \bigg).
\end{equation}
The tensor under the divergence operator, which does not vanish off shell when the divergence acts on it, is the effective EMT of the DGP scalar on the brane
\begin{equation}
\label{eq:DGP_sc_eff_EMT}
T_{\alpha\beta} = 2 M_4^2 \left( \partial_\alpha \psi \partial_\beta \psi - \frac{1}{2} \eta_{\alpha\beta} \partial^\sigma \psi \partial_\sigma \psi \right).
\end{equation}
The EMT \eqref{eq:DGP_sc_eff_EMT} coincides with the effective EMT obtained in Ref. \cite{Khlopunov:2022jaw} by projecting the canonical EMT of the DGP scalar in the bulk on the brane and extracting from it the part localized on the brane. Unlike the local theory of a massless scalar field $M_5 = 0$, the obtained effective EMT \eqref{eq:DGP_sc_eff_EMT} is not conserved on the equation of motion \eqref{eq:DGP_sc_eff_EoM}
\begin{equation}
\partial^\alpha T_{\alpha\beta} = 2 M_4^2 {_4}\square \, \psi \partial_\beta \psi \neq 0
\end{equation}
in accordance with the metastable character of the effective graviton on the brane \cite{Gabadadze:2004dq}.

Note that the effective EMT on the brane \eqref{eq:DGP_sc_eff_EMT} could be obtained from the effective Lagrangian \eqref{eq:DGP_sc_eff_Lagrangian} using the standard formula for the canonical EMT \cite{Maggiore:2005qv,Stepanyanz_book} by neglecting the non-local mass term. Further, an analogous result will also be obtained for the effective EMT of DGP gravity.

\subsubsection{Brane localized part of the five-dimensional Hamiltonian}

To verify the obtained expression for the effective EMT of the field on the brane \eqref{eq:DGP_sc_eff_EMT}, we calculate the Hamiltonian of the five-dimensional theory \eqref{eq:DGP_sc_act_aux} and extract from it the part localized on the brane. It should coincide with the $T^{00}$ component of the effective EMT, since they both give the energy density of the DGP scalar localized on the brane.

From the action \eqref{eq:DGP_sc_act_aux} we find the conjugate momentum of the field in the bulk
\begin{equation}
\pi = \frac{\partial {\cal L}}{\partial (\partial_0 \varphi)} = 2 M_5^3 \partial_0 \varphi + 2 M_4^2 \delta(y) \partial_0 \varphi.
\end{equation}
Accordingly, the DGP scalar field's Hamiltonian in the bulk is written as
\begin{equation}
{\cal H} = M_5^3 \left( (\partial_0 \varphi)^2 + (\mathbf{\nabla} \varphi)^2 + (\partial_y \varphi)^2 \right) + M_4^2 \delta(y) \left( (\partial_0 \varphi)^2 + (\mathbf{\nabla} \varphi)^2 \right).
\end{equation}
Integrating it over a small interval around the brane $y \in (-\epsilon, \epsilon)$, $\epsilon \to +0$ by use of Eq. \eqref{eq:DGP_sc_in_bulk_gen}, we find the brane localized part of the Hamiltonian
\begin{equation}
\label{eq:DGP_sc_Hamilt_local}
{\cal H}_{\rm br} = \int_{-\epsilon}^{\epsilon} dy \, {\cal H} = M_4^2 \left( (\partial_0 \psi)^2 + (\mathbf{\nabla} \psi)^2 \right).
\end{equation}
Obviously, the $T^{00}$ component of the effective EMT of the DGP scalar \eqref{eq:DGP_sc_eff_EMT} coincides with the obtained brane localised part of five-dimensional Hamiltonian. Thus, Eq. \eqref{eq:DGP_sc_eff_EMT} provides the correct expression for the effective EMT of the DGP scalar field on the brane.

\subsection{Effective energy-momentum tensor of DGP gravity}

Now we turn to calculating the effective EMT of the DGP gravity on the brane from the quadratic effective action of DGP graviton \cite{Luty:2003vm,Hinterbichler:2011tt,deRham:2014zqa}
\begin{align}
\label{eq:DGP_grav_eff_act}
S_{\rm eff} & = \frac{1}{8} M_4^2 \int d^4x \Big( - \partial^\sigma h^{\mu\nu} \partial_\sigma h_{\mu\nu} + 2 \partial_\alpha h^{\mu\alpha} \partial^\sigma h_{\mu\sigma} - 2 \partial^\alpha h \partial^\beta h_{\alpha\beta} + \partial^\sigma h \partial_\sigma h \nonumber \\ & - M_c \big( h \sqrt{-{_4}\square} h - h^{\mu\nu} \sqrt{-{_4}\square} h_{\mu\nu} \big) \Big).
\end{align}
One can easily show that the equations of motion resulting from this action coincide with Eq. \eqref{eq:DGP_eff_EoM}, obtained by linearising the equations of motion of gravitational field in the bulk. Note also that the action \eqref{eq:DGP_grav_eff_act} is the Fierz-Pauli massive gravity action \cite{Fierz:1939ix,Hinterbichler:2011tt,deRham:2014zqa} up to replacing the non-local mass with the standard one
\begin{equation}
\label{eq:DGP-mas_grav_relation}
- M_c \sqrt{-{_4}\square} \leftrightarrow m^2.
\end{equation}
In particular, due to the lack of gauge symmetry, the effective DGP graviton on the brane has five dynamical degrees of freedom, by analogy with massive graviton.

We obtain the effective EMT of DGP graviton on the brane from the effective action \eqref{eq:DGP_grav_eff_act} by use of the Noether method. For this, we infinitesimally shift the coordinates and accordingly transform the gravitational field as
\begin{equation}
\label{eq:loc_coor_transl}
x^\mu \to x^{\prime \mu} = x^\mu + \xi^\mu(x), \quad h_{\mu\nu}(x) \to h_{\mu\nu}^{\prime}(x') = h_{\mu\nu}(x), \quad |\xi^\mu| \ll 1.
\end{equation}
The transformations of field derivatives and non-local mass terms are given by Eqs. \eqref{eq:der_trans_1} and \eqref{eq:der_trans_2}. Let us consider separately the increments received by the local and non-local parts of the effective action under this transformations.

We start with the non-local part of the effective action
\begin{equation}
S_{\rm eff}^{\rm nl} = \frac{1}{4} M_5^3 \int d^4x \left( h^{\mu\nu} \sqrt{-{_4}\square} \, h_{\mu\nu} - h \sqrt{-{_4}\square} \, h \right),
\end{equation}
as the increment of local part is expected to yield the canonical EMT of a massless graviton. Up to the first order contributions, we find the increment of the non-local part of effective action \eqref{eq:DGP_grav_eff_act} as
\begin{align}
& {S_{\rm eff}^{\rm nl}}'[h_{\mu\nu}^\prime] = S_{\rm eff}^{\rm nl}[h_{\mu\nu}] + \delta S_{\rm eff}^{\rm nl}[h_{\mu\nu}] + {\cal O}(\xi^2), \\
& \delta S_{\rm eff}^{\rm nl} = \frac{1}{4} M_5^3 \int d^4x \left( h^{\mu\nu} \partial^\alpha \xi^\beta \frac{1}{\sqrt{-{_4}\square}} \partial_\alpha \partial_\beta h_{\mu\nu} - h \partial^\alpha \xi^\beta \frac{1}{\sqrt{-{_4}\square}} \partial_\alpha \partial_\beta h + \partial_\sigma \xi^\sigma h^{\mu\nu} \sqrt{-{_4}\square} h_{\mu\nu} \right. \nonumber \\ & \quad - \left. \partial_\sigma \xi^\sigma h \sqrt{-{_4}\square} h + \frac{1}{2} h^{\mu\nu} \partial_\beta \partial^\beta \xi^\alpha \frac{1}{\sqrt{-{_4}\square}} \partial_\alpha h_{\mu\nu} - \frac{1}{2} h \partial_\beta \partial^\beta \xi^\alpha \frac{1}{\sqrt{-{_4}\square}} \partial_\alpha h \right).
\end{align}
Integrating by parts we remove the derivatives from the local shifts of coordinates $\xi^\alpha$ and then assume them to be global
\begin{align}
\label{eq:eff_act_incr_nl_part_aux_1}
\delta S_{\rm eff}^{\rm nl} & \overset{\to}{=} \frac{1}{4} M_5^3 \int d^4x \, \xi^\beta \partial^\alpha \left( - \frac{1}{2} h^{\mu\nu} \frac{1}{\sqrt{-{_4}\square}} \partial_\alpha \partial_\beta h_{\mu\nu} + \frac{1}{2} h \frac{1}{\sqrt{-{_4}\square}} \partial_\alpha \partial_\beta h + \frac{1}{2} \partial_\alpha h^{\mu\nu} \frac{1}{\sqrt{-{_4}\square}} \partial_\beta h_{\mu\nu} \right. \nonumber \\ & \left. - \frac{1}{2} \partial_\alpha h \frac{1}{\sqrt{-{_4}\square}} \partial_\beta h - \eta_{\alpha\beta} h^{\mu\nu} \sqrt{-{_4}\square} h_{\mu\nu} + \eta_{\alpha\beta} h \sqrt{-{_4}\square} h \right), \quad \xi^\alpha = {\rm Const}.
\end{align}
Due to the $\xi^\alpha$ being constant, we can transform the first two terms in Eq. \eqref{eq:eff_act_incr_nl_part_aux_1} to the bilinear form in field derivatives and combine them with the next two terms
\begin{align}
\delta S_{\rm eff}^{\rm nl} & \overset{\to}{=} \frac{1}{4} M_5^3 \int d^4x \, \xi^\beta \partial^\alpha \bigg( \partial_\alpha h^{\mu\nu} \frac{1}{\sqrt{-{_4}\square}} \partial_\beta h_{\mu\nu} - \partial_\alpha h \frac{1}{\sqrt{-{_4}\square}} \partial_\beta h - \eta_{\alpha\beta} h^{\mu\nu} \sqrt{-{_4}\square} h_{\mu\nu} \nonumber \\ & + \eta_{\alpha\beta} h \sqrt{-{_4}\square} h \bigg).
\end{align}
Finally, integrating by parts we find that the 4-divergence participating in the increment of the non-local part of effective action vanishes off shell
\begin{align}
\delta S_{\rm eff}^{\rm nl} & \overset{\to}{=} \frac{1}{4} M_5^3 \int d^4x \, \xi^\beta \bigg( {_4}\square h^{\mu\nu} \frac{1}{\sqrt{-{_4}\square}} \partial_\beta h_{\mu\nu} - {_4}\square h^{\mu\nu} \frac{1}{\sqrt{-{_4}\square}} \partial_\beta h_{\mu\nu} - {_4}\square h \frac{1}{\sqrt{-{_4}\square}} \partial_\beta h \nonumber \\ & + {_4}\square h \frac{1}{\sqrt{-{_4}\square}} \partial_\beta h - \partial_\beta h^{\mu\nu} \sqrt{-{_4}\square} h_{\mu\nu} + \partial_\beta h^{\mu\nu} \sqrt{-{_4}\square} h_{\mu\nu} + \partial_\beta h \sqrt{-{_4}\square} h \nonumber \\ & - \partial_\beta h \sqrt{-{_4}\square} h \bigg) = 0.
\end{align}
Thus, the non-local part of effective action \eqref{eq:DGP_grav_eff_act} does not contribute to the effective EMT of DGP graviton on the brane, by analogy with the effective EMT of the DGP scalar \eqref{eq:DGP_sc_eff_EMT}.

Now we consider the local part of effective action \eqref{eq:DGP_grav_eff_act}
\begin{align}
S_{\rm eff}^{\rm loc} = \frac{1}{8} M_4^2 \int d^4x \left( - \partial^\sigma h^{\mu\nu} \partial_\sigma h_{\mu\nu} + 2 \partial_\alpha h^{\mu\alpha} \partial^\sigma h_{\mu\sigma} - 2 \partial^\alpha h \partial^\beta h_{\alpha\beta} + \partial^\sigma h \partial_\sigma h \right).
\end{align}
After the local shift of coordinate \eqref{eq:loc_coor_transl} it is rewritten as
\begin{align}
& {S_{\rm eff}^{\rm loc}}'[h_{\mu\nu}^\prime] = S_{\rm eff}^{\rm loc}[h_{\mu\nu}] + \delta S_{\rm eff}^{\rm loc}[h_{\mu\nu}] + {\cal O}(\xi^2), \\
& \delta S_{\rm eff}^{\rm loc} = \frac{1}{2} M_4^2 \int d^4x \, \partial^\alpha \xi^\beta \left( \frac{1}{2} \partial_\alpha h^{\mu\nu} \partial_\beta h_{\mu\nu} - \partial_\beta h_{\mu\alpha} \partial_\nu h^{\mu\nu} + \frac{1}{2} \partial^\mu h \partial_\beta h_{\mu\alpha} + \frac{1}{2} \partial_\beta h \partial^\mu h_{\mu\alpha} \right. \nonumber \\ & \quad - \frac{1}{2} \partial_\alpha h \partial_\beta h - \frac{1}{4} \eta_{\alpha\beta} \partial^\sigma h^{\mu\nu} \partial_\sigma h_{\mu\nu} + \frac{1}{2} \eta_{\alpha\beta} \partial_\nu h^{\mu\nu} \partial^\sigma h_{\mu\sigma} - \frac{1}{2} \eta_{\alpha\beta} \partial^\mu h \partial^\nu h_{\mu\nu} \nonumber \\ & \quad \left. + \frac{1}{4} \eta_{\alpha\beta} \partial^\sigma h \partial_\sigma h \right).
\end{align}
Integrating by parts we remove derivatives from $\xi^\alpha$, in what follows assuming them constant, and combine the third and fourth terms in the increment of the action
\begin{align}
\delta S_{\rm eff}^{\rm loc} & \overset{\to}{=} \frac{1}{2} M_4^2 \int d^4x \, \xi^\beta \partial^\alpha \left( - \frac{1}{2} \partial_\alpha h^{\mu\nu} \partial_\beta h_{\mu\nu} + \partial_\beta h_{\mu\alpha} \partial_\nu h^{\mu\nu} - \partial_\beta h \partial^\mu h_{\mu\alpha} + \frac{1}{2} \partial_\alpha h \partial_\beta h \right. \nonumber \\ & \left. + \frac{1}{4} \eta_{\alpha\beta} \partial^\sigma h^{\mu\nu} \partial_\sigma h_{\mu\nu} - \frac{1}{2} \eta_{\alpha\beta} \partial_\nu h^{\mu\nu} \partial^\sigma h_{\mu\sigma} + \frac{1}{2} \eta_{\alpha\beta} \partial^\mu h \partial^\nu h_{\mu\nu} - \frac{1}{4} \eta_{\alpha\beta} \partial^\sigma h \partial_\sigma h \right).
\end{align}
Given the mostly positive signature of Minkowski metric, the tensor under the 4-divergence is equal to the effective EMT of DGP graviton on the brane with the opposite sign
\begin{align}
T_{\alpha\beta} & = \frac{1}{2} M_4^2 \left( \frac{1}{2} \partial_\alpha h^{\mu\nu} \partial_\beta h_{\mu\nu} - \partial_\beta h_{\mu\alpha} \partial_\nu h^{\mu\nu} + \partial_\beta h \partial^\mu h_{\mu\alpha} - \frac{1}{2} \partial_\alpha h \partial_\beta h \right. \nonumber \\ & \left. - \frac{1}{4} \eta_{\alpha\beta} \partial^\sigma h^{\mu\nu} \partial_\sigma h_{\mu\nu} + \frac{1}{2} \eta_{\alpha\beta} \partial_\nu h^{\mu\nu} \partial^\sigma h_{\mu\sigma} - \frac{1}{2} \eta_{\alpha\beta} \partial^\mu h \partial^\nu h_{\mu\nu} + \frac{1}{4} \eta_{\alpha\beta} \partial^\sigma h \partial_\sigma h \right).
\end{align}
By analogy with the DGP scalar, the obtained effective EMT coincides with the one calculated from the effective Lagrangian \eqref{eq:DGP_grav_eff_act} by use of the standard formula for the canonical EMT neglecting the non-local terms.

Note that, in accordance with the equivalence principle, only the energy density of gravitational field averaged over the region of spacetime with a size $\cal S$ of order of several gravitational field wavelengths centered on the observation point $O_{\cal S}(x)$ has a physical meaning \cite{Isaacson:1968hbi,Isaacson:1968zza,maggiore2008}
\begin{equation}
\label{eq:Isaac_averaging}
\langle T_{\alpha\beta}(x) \rangle = \frac{1}{{\cal S}^4} \int_{O_{\cal S}(x)} d^4x' \, T_{\alpha\beta}(x'), \quad {\cal S} > \lambdabar.
\end{equation}
In this case, one can integrate by parts under the averaging sign and neglect the boundary terms of order $\lambdabar / {\cal S}$ choosing a sufficiently large averaging region. As a result, given the Eqs. \eqref{eq:DGP_eff_EoM}, the averaged effective EMT of DGP gravity on the brane can be rewritten as
\begin{equation}
\label{eq:DGP_grav_EMT_eff}
\left \langle T_{\alpha\beta} \right \rangle \overset{\to}{=} \frac{1}{4} M_4^2 \left \langle \partial_\alpha h^{\mu\nu} \partial_\beta h_{\mu\nu} - \partial_\alpha h \partial_\beta h - \frac{1}{2} \eta_{\alpha\beta} \partial^\sigma h^{\mu\nu} \partial_\sigma h_{\mu\nu} + \frac{1}{2} \eta_{\alpha\beta} \partial^\sigma h \partial_\sigma h \right \rangle.
\end{equation}

\section{Dynamical degrees of freedom of tensionless DGP gravity}\label{IV}

As mentioned above, the effective action of DGP graviton on the brane \eqref{eq:DGP_grav_eff_act} does not have a gauge symmetry, by analogy with the Fierz-Pauli massive gravity theory \cite{Fierz:1939ix} (see, also, \cite{Hinterbichler:2011tt,deRham:2014zqa}). Therefore, the effective DGP graviton on the brane has more degrees of freedom compared to the massless graviton. Some of them are non-dynamical and, thus, do not carry the gravitational radiation from the field source to infinity. However, their contribution into the effective EMT \eqref{eq:DGP_grav_EMT_eff} can not be eliminated by an appropriate gauge choice, in contrast with massless gravity \cite{Flanagan:2005yc,maggiore2008}, due to the lack of gauge symmetry in the effective action \eqref{eq:DGP_grav_eff_act}. Moreover, taking into account their contribution to the gravitational radiation leads to the non-physical results. As a result, one has to separately construct the effective EMT of dynamical degrees of freedom of DGP graviton to calculate the gravitational radiation in DGP model.

We construct the effective EMT of the dynamical degrees of freedom of DGP graviton from the dynamical part of effective action \eqref{eq:DGP_grav_eff_act}. For this, one has to decouple the contributions of dynamical (dDOF) and non-dynamical (ndDOF) degrees of freedom into the effective action, which is a non-trivial task in the case of DGP gravity.

Further, we define the dDOF as the degrees of freedom whose equations of motion contain the second time derivatives and for which there are no constraint equations rewriting their first time derivatives in terms of other degrees of freedom.

\subsection{Dynamical DOFs of electrodynamics}

Let us illustrate the main steps of our approach within a simple example of electrodynamics. Here, the contribution of ndDOF into the EMT is automatically eliminated, in contrast with gravity. Therefore, one can calculate the electromagnetic radiation without fixing a unitary gauge.

\subsubsection{Counting DOFs}

Consider the free electromagnetic field with the equation of motion \cite{Landau:1975pou,Jackson:1998nia}
\begin{equation}
\label{eq:ED_free_EoM}
\partial_\mu F^{\mu\nu} = 0, \quad F_{\mu\nu} = \partial_\mu A_\nu - \partial_\nu A_\mu, \quad \eta_{\mu\nu} \propto (+--\,-).
\end{equation}
To determine the dDOF of electromagnetic field, we perform its scalar-vector (SV) de\-com\-po\-si\-tion \cite{Flanagan:2005yc}
\begin{equation}
\label{eq:4D_ED_sc-vec_decomp}
A_\mu = \left \lbrace A_0, \partial_i \varphi + A_i^{\rm t} \right \rbrace, \quad \partial_i A_i^{\rm t} = 0, \quad i,j = \overline{1,3},
\end{equation}
where the four degrees of freedom of electromagnetic field are divided into two scalar, with respect to the spatial coordinates transformations, degrees of freedom $A_0$ and $\varphi$ and two vector degrees of freedom $A_i^{\rm t}$. Correspondingly, the Eq. \eqref{eq:ED_free_EoM} is rewritten as
\begin{equation}
\label{eq:4D_ED_gen_EoM}
- \Delta A_0 + \partial_0 \Delta \varphi = 0, \quad - \partial_0^2 \partial_i \varphi - \partial_0^2 A_i^{\rm t} + \Delta A_i^{\rm t} + \partial_i \partial_0 A_0 = 0, \quad \Delta \equiv \partial_i \partial_i.
\end{equation}
Here, only $\varphi$ and $A_i^{\rm t}$ have second time derivatives. However, due to the gauge symmetry of electromagnetic field
\begin{equation}
\label{eq:4D_ED_gauge_sym_sc-vec_form}
A_\mu \to A_\mu + \partial_\mu \alpha \quad \Longrightarrow \quad A_0 \to A_0 + \partial_0 \alpha, \quad \varphi \to \varphi + \alpha, \quad A_i^{\rm t} \to A_i^{\rm t}
\end{equation}
the field $\varphi$ is the Stuckelberg field and is not a dDOF. In particular, in the unitary gauge $\varphi = 0$ the equations of motion take the form
\begin{equation}
\Delta A_0 = 0, \quad - \partial_0^2 A_i^{\rm t} + \Delta A_i^{\rm t} + \partial_i \partial_0 A_0 = 0.
\end{equation}
Thus, the electromagnetic field has only two dDOF $A_i^{\rm t}$ \cite{Flanagan:2005yc}.

One can also demonstrate it without fixing the unitary gauge. For this, we solve the first equation in \eqref{eq:4D_ED_gen_EoM} taking into account the boundary condition $A_0, \varphi \xrightarrow{r \to \infty} 0$
\begin{equation}
\Delta \left( \partial_0 \varphi - A_0 \right) = 0 \quad \Longrightarrow \quad A_0 = \partial_0 \varphi.
\end{equation}
Substituting the solution into the second equation in \eqref{eq:4D_ED_gen_EoM}, we obtain the following system of equations of motion
\begin{equation}
\label{eq:4D_ED_resoleved_EoM}
\partial_0^2 A_i^{\rm t} - \Delta A_i^{\rm t} = 0, \quad A_0 = \partial_0 \varphi.
\end{equation}
Analogously, here only $A_i^{\rm t}$ has the second time derivative and, thus, is dDOF, while $A_0$ and $\varphi$ are ndDOF.

Note that the gauge transformations \eqref{eq:4D_ED_gauge_sym_sc-vec_form} include only the ndDOF $A_0$ and $\varphi$, while $A_i^{\rm t}$ remains unaffected.

\subsubsection{Dynamical DOFs contribution to the EMT}

Let us demonstrate that the on shell ndDOF of electromagnetic field do not contribute into the symmetrized gauge-invariant EMT \cite{Landau:1975pou,Jackson:1998nia,Stepanyanz_book}
\begin{equation}
T_{\alpha\beta} = F_{\alpha\mu} F^\mu_{\,\cdot\,\beta} + \frac{1}{4} \eta_{\alpha\beta} F_{\mu\nu}^2.
\end{equation}
Substituting into it the SV decomposition of electromagnetic field \eqref{eq:4D_ED_sc-vec_decomp} and rewriting it by components using the Eqs. \eqref{eq:4D_ED_resoleved_EoM}, we obtain
\begin{align}
T_{00} & = \frac{1}{2} \partial_0 A_i^{\rm t} \partial_0 A_i^{\rm t} + \frac{1}{4} \left( \partial_i A_j^{\rm t} - \partial_j A_i^{\rm t} \right) \left( \partial_i A_j^{\rm t} - \partial_j A_i^{\rm t} \right), \quad T_{0k} = \partial_0 A_i^{\rm t} \left( \partial_k A_i^{\rm t} - \partial_i A_k^{\rm t} \right), \\
T_{kl} & = - \partial_0 A_k^{\rm t} \partial_0 A_l^{\rm t} + \left( \partial_k A_i^{\rm t} - \partial_i A_k^{\rm t} \right) \left( \partial_l A_i^{\rm t} - \partial_i A_l^{\rm t} \right) + \frac{1}{2} \delta_{kl} \partial_0 A_i^{\rm t} \partial_0 A_i^{\rm t} \nonumber \\ & - \frac{1}{4} \delta_{kl} \left( \partial_i A_j^{\rm t} - \partial_j A_i^{\rm t} \right) \left( \partial_i A_j^{\rm t} - \partial_j A_i^{\rm t} \right).
\end{align}
Thus, all the components of symmetrized EMT depend only on the dDOF $A_i^{\rm t}$. Given the gauge invariance of symmetrized EMT, it makes it possible to compute the electromagnetic radiation without fixing the unitary gauge $\varphi = 0$ and/or separately extracting the EMT of dDOF of electromagnetic field.

Note that we can discard some of the terms in the EMT components, as they contain the derivatives of transverse vector along itself $A_j^{\rm t} \partial_j A_i^{\rm t}$, which have to vanish. This can be easily understood using the example of a plane monochromatic wave
\begin{align}
& A_i^{\rm t} = \int \frac{d^4k}{(2\pi)^4} \, e^{-ikx} \tilde{A}_i^{\rm t}(k), \quad \tilde{A}_i^{\rm t}(k) = \tilde{A}_i^{\rm t}(k_0) \delta^{(4)}(k - k_0), \\
& \partial_i A_i^{\rm t} = 0 \quad \Longrightarrow \quad \tilde{A}_i^{\rm t}(k_0) k_{0i} = 0.
\end{align}
In this case, for the derivative of transverse vector along itself we obtain
\begin{equation}
A_j^{\rm t} \partial_j A_i^{\rm t} = \int \frac{d^4k d^4k'}{(2\pi)^8} \, e^{-ix(k+k')} \tilde{A}_j^{\rm t}(k') i k_j \tilde{A}_i^{\rm t}(k) = \frac{e^{-2ik_0x}}{(2\pi)^8} i k_{0j} \tilde{A}_j^{\rm t}(k_0) \tilde{A}_i^{\rm t}(k_0) = 0.
\end{equation}
Thus, the symmetrized EMT of electromagnetic field is written as
\begin{align}
T_{00} & = \frac{1}{2} \partial_0 A_i^{\rm t} \partial_0 A_i^{\rm t} + \frac{1}{2} \partial_i A_j^{\rm t} \partial_i A_j^{\rm t}, \\
T_{0k} & = \partial_0 A_i^{\rm t} \partial_k A_i^{\rm t}, \\
T_{kl} & = \partial_k A_i^{\rm t} \partial_l A_i^{\rm t} + \frac{1}{2} \delta_{kl} \partial_0 A_i^{\rm t} \partial_0 A_i^{\rm t} - \frac{1}{2} \delta_{kl} \partial_i A_j^{\rm t} \partial_i A_j^{\rm t}.
\end{align}
Here, some terms in the $(kl)$-components of EMT have been discarded, due to the Eqs. \eqref{eq:4D_ED_resoleved_EoM} and the corresponding dispersion relation
\begin{equation}
{_4}\square A_i^{\rm t} = 0 \quad \Longrightarrow \quad k_{0\mu}^2 = 0.
\end{equation}
From this one easily finds that the symmetrized EMT of electromagnetic field can be rewritten in terms of dDOF $A_i^{\rm t}$ as
\begin{equation}
\label{eq:4D_ED_sym_EMT_ddofs}
T_{\mu\nu} = \partial_\mu A_i^{\rm t} \partial_\nu A_i^{\rm t} - \frac{1}{2} \eta_{\mu\nu} \partial^\alpha A_i^{\rm t} \partial_\alpha A_i^{\rm t}.
\end{equation}
It has the structure of the canonical EMT of a set of massless <<scalar fields>> $\left \lbrace A_i^{\rm t} \right \rbrace$. Further, we demonstrate that the EMT of both massless gravity and DGP gravity also has an analogous structure.

\subsubsection{Action and EMT of dynamical DOFs}

Finally, we extract from the action of free electromagnetic field \cite{Landau:1975pou,Jackson:1998nia,Stepanyanz_book}
\begin{equation}
S = - \frac{1}{4} \int d^4x \, F_{\mu\nu}^2
\end{equation}
the part corresponding to the dDOF and obtain from it the canonical EMT of dDOF of electromagnetic field coinciding with the Eq. \eqref{eq:4D_ED_sym_EMT_ddofs}.

Integrating by parts and taking into account the transversality of the dDOF $A_i^{\rm t}$, we rewrite the action of electromagnetic field as
\begin{equation}
\label{eq:4D_ED_act_sc-vec_decomp}
S \overset{\to}{=} \int d^4x \left( \frac{1}{2} \partial_0 \partial_i \varphi \partial_0 \partial_i \varphi + \frac{1}{2} \partial_i A_0 \partial_i A_0 - \partial_0 \partial_i \varphi \partial_i A_0 + \frac{1}{2} \partial_0 A_i^{\rm t} \partial_0 A_i^{\rm t} - \frac{1}{2} \partial_j A_i^{\rm t} \partial_j A_i^{\rm t} \right).
\end{equation}
Here, the contributions of dynamical and non-dynamical degrees of freedom are decoupled and we can consider the actions for them separately
\begin{align}
\label{eq:ED_dyn_non-dyn_act_split_1}
& S = S_{\rm d} \lbrack A_i^{\rm t} \rbrack + S_{\rm nd} \lbrack \varphi, A_0 \rbrack, \\
\label{eq:ED_dyn_non-dyn_act_split_2}
& S_{\rm d} = \frac{1}{2} \int d^4x \, \partial^\mu A_i^{\rm t} \partial_\mu A_i^{\rm t}, \\
\label{eq:ED_dyn_non-dyn_act_split_3}
& S_{\rm nd} = \frac{1}{2} \int d^4x \, \big( \partial_0 \partial_i \varphi \partial_0 \partial_i \varphi + \partial_i A_0 \partial_i A_0 - 2 \partial_0 \partial_i \varphi \partial_i A_0 \big).
\end{align}
One can demonstrate that the non-dynamical part of the action is gauge-invariant \eqref{eq:4D_ED_gauge_sym_sc-vec_form}. Also, it is easy to show that the obtained actions yield the equations of motion \eqref{eq:4D_ED_resoleved_EoM}.

We obtain the canonical EMT of dDOF of electromagnetic field from the action \eqref{eq:ED_dyn_non-dyn_act_split_2} using the standard formula \cite{Landau:1975pou,Jackson:1998nia,Stepanyanz_book}
\begin{equation}
\label{eq:4D_ED_ddof_canon_EMT_def}
{T_{\rm d}}^\mu_{\,\cdot\,\nu} = \frac{\partial {\cal L}_{\rm d}}{\partial(\partial_\mu A_i^{\rm t})} \partial_\nu A_i^{\rm t} - \delta^\mu_\nu {\cal L}_{\rm d}, \quad {\cal L}_{\rm d} = \frac{1}{2} \partial^\mu A_i^{\rm t} \partial_\mu A_i^{\rm t}.
\end{equation}
As a result, we find the EMT of dDOF of electromagnetic field $A_i^{\rm t}$ coinciding with the symmetrized EMT \eqref{eq:4D_ED_sym_EMT_ddofs}
\begin{equation}
\label{eq:4D_ED_ddof_canon_EMT}
{T_{\rm d}}_{\mu\nu} = \partial_\mu A_i^{\rm t} \partial_\nu A_i^{\rm t} - \frac{1}{2} \eta_{\mu\nu} \partial^\alpha A_i^{\rm t} \partial_\alpha A_i^{\rm t}.
\end{equation}

Therefore, in electrodynamics there are two equivalent ways of computing the electromagnetic radiation:
\begin{itemize}
\item
\textit{By splitting the degrees of freedom.} One has to extract the action of dDOF of electromagnetic field and obtain the corresponding equations of motion and the canonical EMT. Next, one has to solve the equations of motion for the dDOF and calculate the radiation energy flux carried by them.
\item
\textit{Without splitting the degrees of freedom.} One has to construct the symmetrized EMT of electromagnetic field and solve the equations of motion without splitting the dynamical and non-dynamical degrees of freedom. After this, one computes the corresponding radiation energy flux, in which the contributions of ndDOF mutually cancel out.
\end{itemize}
Further, we demonstrate that in the case of gravity the second method for calculating the gravitational radiation is not available (at least when using the Isaacson EMT).

\subsection{Dynamical DOFs of gravity}

Let us demonstrate that in the case of massless gravity Isaacson EMT, being gauge invariant, contains contributions from the ndDOF of gravitational field, in contrast with electrodynamics. Therefore, to correctly compute the gravitational radiation one has either to fix the unitary gauge to use the Isaacson EMT or to construct the canonical EMT of dDOF of gravitational field.

\subsubsection{Isaacson energy-momentum tensor of gravitational field}

In accordance with the Isaacson approach \cite{Isaacson:1968hbi,Isaacson:1968zza} (see, also, \cite{maggiore2008}), in GR the EMT of gravitational field is constructed from the quadratic part of expansion of Einstein tensor in metric perturbations
\begin{align}
& t_{\mu\nu} = - \frac{1}{8 \pi G} \left \langle {_{(2)}}G_{\mu\nu} \right \rangle, \quad G_{\mu\nu} = R_{\mu\nu} - \frac{1}{2} g_{\mu\nu} R, \quad g_{\mu\nu} = \bar{g}_{\mu\nu} + h_{\mu\nu}, \quad |h_{\mu\nu}| \ll 1,
\end{align}
where $\bar{g}_{\mu\nu}$ is some background metric, and ${_{(2)}}G_{\mu\nu}$ is the quadratic part of Einstein tensor expansion in $h_{\mu\nu}$. Here, the averaging \eqref{eq:Isaac_averaging} guarantees the gauge-invariance of EMT.

Therefore, on the Minkowski space background $\bar{g}_{\mu\nu} = \eta_{\mu\nu} \propto (-++\,+)$ the Isaacson EMT of gravitational field is written as
\begin{equation}
t_{\mu\nu} = - \frac{1}{8 \pi G} \left \langle {_{(2)}}R_{\mu\nu} - \frac{1}{2} \eta_{\mu\nu} {_{(2)}}R - \frac{1}{2} h_{\mu\nu} {_{(1)}}R \right \rangle,
\end{equation}
where ${_{(1)}}R$, ${_{(2)}}R$, and ${_{(2)}}R_{\mu\nu}$ denote the Ricci tensor and scalar expansions up to the first and second order contributions in $h_{\mu\nu}$, correspondingly. Integrating by parts under the averaging operator, we obtain the Isaacson EMT in the following form \cite{Isaacson:1968hbi,Isaacson:1968zza,maggiore2008}
\begin{align}
\label{eq:GR_Isaac_EMT_gen}
t_{\mu\nu} & \overset{\to}{=} \frac{1}{8 \pi G} \left \langle \frac{1}{4} \partial_\mu h^{\alpha\beta} \partial_\nu h_{\alpha\beta} - \frac{1}{4} \partial_\mu h \partial^\beta h_{\beta\nu} - \frac{1}{4} \partial_\nu h \partial^\beta h_{\mu\beta} + \frac{3}{4} \partial^\beta h \partial_\beta h_{\mu\nu} - \frac{1}{2} \partial^\beta h_{\mu\nu} \partial^\alpha h_{\alpha\beta} \right. \nonumber \\ & \left. + \frac{1}{2} \partial^\beta h_{\beta\nu} \partial^\sigma h_{\mu\sigma} - \frac{1}{2} \partial_\sigma h_{\beta\nu} \partial^\sigma h_\mu^\beta - \frac{1}{8} \eta_{\mu\nu} \partial^\sigma h \partial_\sigma h + \frac{1}{4} \eta_{\mu\nu} \partial_\alpha h^{\alpha\beta} \partial^\sigma h_{\sigma\beta} - \frac{1}{8} \eta_{\mu\nu} \partial_\sigma h_{\alpha\beta} \partial^\sigma h^{\alpha\beta} \right \rangle.
\end{align}
Note that until this point we assumed that the metric perturbations satisfy the general linear equation of motion (see, e.g., \cite{maggiore2008})
\begin{equation}
\label{eq:4D_Einst_eq_lin_gen}
\partial^\sigma \partial_\mu h_{\sigma\nu} + \partial^\sigma \partial_\nu h_{\sigma\mu} - {_4}\square h_{\mu\nu} - \partial_\mu \partial_\nu h - \eta_{\mu\nu} \partial^\alpha \partial^\beta h_{\alpha\beta} + \eta_{\mu\nu} {_4}\square h = 0.
\end{equation}
We simplify the equation of motion by fixing the Lorentz gauge
\begin{equation}
\label{eq:4D_EoM_in_Lor_gauge}
\partial^\mu h_{\mu\nu} = \frac{1}{2} \partial_\nu h \quad \Longrightarrow \quad {_4}\square \left( h_{\mu\nu} - \frac{1}{2} \eta_{\mu\nu} h \right) = 0.
\end{equation}
Correspondingly, the EMT \eqref{eq:GR_Isaac_EMT_gen} is also simplified significantly given the Eq. \eqref{eq:4D_EoM_in_Lor_gauge}
\begin{equation}
\label{eq:4D_GR_Isaac_EMT}
t_{\mu\nu} \overset{\to}{=} \frac{1}{32 \pi G} \left \langle \partial_\mu h^{\alpha\beta} \partial_\nu h_{\alpha\beta} - \frac{1}{2} \partial_\mu h \partial_\nu h \right \rangle.
\end{equation}
The obtained Isaacson EMT is invariant under the residual gauge transformations \cite{maggiore2008}
\begin{equation}
h_{\mu\nu} \to h_{\mu\nu} + \partial_\mu \xi_\nu + \partial_\nu \xi_\mu, \quad {_4}\square \xi_\mu = 0
\end{equation}
up to the 4-divergence, which can be neglected under the averaging operator
\begin{align}
t_{\mu\nu} & \to \frac{1}{32 \pi G} \bigg \langle \partial_\mu h^{\alpha\beta} \partial_\nu h_{\alpha\beta} - \frac{1}{2} \partial_\mu h \partial_\nu h + 2 \partial_\mu h^{\alpha\beta} \partial_\nu \partial_\alpha \xi_\beta + 2 \partial_\nu h^{\alpha\beta} \partial_\mu \partial_\alpha \xi_\beta - \partial_\mu h \partial_\nu \partial_\sigma \xi^\sigma \nonumber \\ & - \partial_\nu h \partial_\mu \partial_\sigma \xi^\sigma + 2 \partial_\mu \partial^\alpha \xi^\beta \partial_\nu \partial_\alpha \xi_\beta + 2 \partial_\mu \partial^\alpha \xi^\beta \partial_\nu \partial_\beta \xi_\alpha - 2 \partial_\mu \partial_\sigma \xi^\sigma \partial_\nu \partial_\gamma \xi^\gamma \bigg \rangle \nonumber \\ & \overset{\to}{=} \frac{1}{32 \pi G} \bigg \langle \partial_\mu h^{\alpha\beta} \partial_\nu h_{\alpha\beta} - \frac{1}{2} \partial_\mu h \partial_\nu h + \partial_\mu h \partial_\nu \partial_\alpha \xi^\alpha + \partial_\nu h \partial_\mu \partial_\alpha \xi^\alpha - \partial_\mu h \partial_\nu \partial_\sigma \xi^\sigma \nonumber \\ & - \partial_\nu h \partial_\mu \partial_\sigma \xi^\sigma - 2 \partial_\mu \xi^\beta \partial_\nu {_4}\square \xi_\beta + 2 \partial_\mu \partial_\beta \xi^\beta \partial_\nu \partial^\alpha \xi_\alpha - 2 \partial_\mu \partial_\sigma \xi^\sigma \partial_\nu \partial_\gamma \xi^\gamma \bigg \rangle = t_{\mu\nu}.
\end{align}

Thus, one can expect that the gravitational radiation energy flux calculated using the Isaacson EMT both in the Lorentz gauge \eqref{eq:4D_GR_Isaac_EMT} preserving the part of ndDOF's contributions and in the transverse-traceless (unitary) gauge eliminating all the ndDOF's contributions should be the same. However, the EMT \eqref{eq:4D_GR_Isaac_EMT} does not lead to the correct quadrupole formula for the gravitational radiation power of a non-relativistic source without fixing the transverse-traceless gauge. In particular, in Appendix \ref{A} we demonstrate that this EMT yields a non-physical negative result for the gravitational radiation power of a non-relativistic harmonically oscillating point mass. Below, we demonstrate that this is due to the remaining contributions of ndDOF of gravitational field into the Isaacson EMT \eqref{eq:4D_GR_Isaac_EMT}.

\subsubsection{Counting DOFs}

To introduce our notation, in this section we determine the dDOF of massless gravitational field. For this, we perform its scalar-vector-tensor (SVT) decomposition \cite{Lifshitz:1945du,Bardeen:1980kt,Stewart:1990fm,Flanagan:2005yc}
\begin{align}
\label{eq:sc-vec-ten_decomp_1}
& h_{00} = - 2 \Phi, \quad h_{0i} = \partial_i B + S_i, \quad h_{ij} = - 2 \delta_{ij} \Psi + 2 \partial_i \partial_j E + \partial_i F_j + \partial_j F_i + h_{ij}^{\rm tt}, \\
\label{eq:sc-vec-ten_decomp_2}
& \partial_i S_i = 0, \quad \partial_i F_i = 0, \quad  \partial_i h_{ij}^{\rm tt} = 0, \quad h_{ii}^{\rm tt} = 0.
\end{align}
As a result, the linearized Einstein equations \eqref{eq:4D_Einst_eq_lin_gen} are rewritten as
\begin{align}
\label{eq:4D_Einst_eq_sc-vec-ten_decomp_1}
& 2 \Delta \Psi = 0, \quad 4 \partial_0 \partial_i \Psi + \Delta \left( \partial_0 F_i - S_i \right) = 0, \\
& \partial_0^2 h_{ij}^{\rm tt} - \Delta h_{ij}^{\rm tt} + 2 \delta_{ij} \left( \Delta \left( \Phi - \Psi \right) + 2 \partial_0^2 \Psi - \partial_0^2 \Delta E + \partial_0 \Delta B \right) - 2 \partial_i \partial_j \left( \Phi - \Psi \right) - 2 \partial_0 \partial_i \partial_j B \nonumber \\ \label{eq:4D_Einst_eq_sc-vec-ten_decomp_2} & \quad - \partial_0 \left( \partial_i S_j + \partial_j S_i \right) + 2 \partial_0^2 \partial_i \partial_j E + \partial_0^2 \left( \partial_i F_j + \partial_j F_i \right) = 0.
\end{align}
We partially solve these equations to obtain the constraints equations for the gravitational field components. From the first equation in \eqref{eq:4D_Einst_eq_sc-vec-ten_decomp_1}, given the boundary condition $h_{\mu\nu} \xrightarrow{r \to \infty} 0$, we find
\begin{equation}
\label{eq:4D_Einst_eq_sc-vec-ten_decomp_sol_1}
\Delta \Psi = 0 \quad \Longrightarrow \quad \Psi = 0.
\end{equation}
Substituting this solution into the second equation in \eqref{eq:4D_Einst_eq_sc-vec-ten_decomp_1}, we arrive at
\begin{equation}
\label{eq:4D_Einst_eq_sc-vec-ten_decomp_sol_2}
\Delta \left( \partial_0 F_i - S_i \right) = 0 \quad \Longrightarrow \quad \partial_0 F_i = S_i.
\end{equation}
Using Eqs. \eqref{eq:4D_Einst_eq_sc-vec-ten_decomp_sol_1} and \eqref{eq:4D_Einst_eq_sc-vec-ten_decomp_sol_2} we find the trace of Eq. \eqref{eq:4D_Einst_eq_sc-vec-ten_decomp_2} in the form
\begin{equation}
\label{eq:4D_Einst_eq_sc-vec-ten_decomp_sol_3}
\Delta \left( \Phi - \partial_0^2 E + \partial_0 B \right) = 0 \quad \Longrightarrow \quad \Phi + \partial_0 B = \partial_0^2 E.
\end{equation}
Finally, substituting Eqs. \eqref{eq:4D_Einst_eq_sc-vec-ten_decomp_sol_1}, \eqref{eq:4D_Einst_eq_sc-vec-ten_decomp_sol_2}, and \eqref{eq:4D_Einst_eq_sc-vec-ten_decomp_sol_3} into the Eq. \eqref{eq:4D_Einst_eq_sc-vec-ten_decomp_2}, we rewrite the system of gravitational field equations of motion as
\begin{equation}
\label{eq:4D_Einst_eq_sc-vec-ten_decomp_decoupled}
\partial_0^2 h_{ij}^{\rm tt} - \Delta h_{ij}^{\rm tt} = 0, \quad \Psi = 0, \quad \partial_0 F_i = S_i, \quad \Phi + \partial_0 B = \partial_0^2 E.
\end{equation}
Here, only the fields $h_{ij}^{\rm tt}$ and $E$ contain the second time derivatives. However, $E$ is not a dDOF, due to the gauge symmetry of gravitational field \cite{maggiore2008}
\begin{equation}
h_{\mu\nu} \to h_{\mu\nu} + \partial_\mu \xi_\nu + \partial_\nu \xi_\mu, \quad \xi_\mu = \left \lbrace \xi_0, \partial_i \xi + \xi_i^{\rm t} \right \rbrace, \quad \partial_i \xi_i^{\rm t} = 0,
\end{equation}
which in SVT decomposition \eqref{eq:sc-vec-ten_decomp_1} is written as
\begin{align}
\label{eq:4D_GR_lin_gauge_sym_sc-vec-ten_decomp_1}
& \Phi \to \Phi - \partial_0 \xi_0, \quad B \to B + \xi_0 + \partial_0 \xi, \quad S_i \to S_i + \partial_0 \xi_i^{\rm t}, \\
\label{eq:4D_GR_lin_gauge_sym_sc-vec-ten_decomp_2}
& \Psi \to \Psi, \quad E \to E + \xi, \quad F_i \to F_i + \xi_i^{\rm t}, \quad h_{ij}^{\rm tt} \to h_{ij}^{\rm tt}.
\end{align}
Thus, $E$ is the Stuckelberg field and is not a dDOF. In particular, in the unitary gauge $B = E = F_i = 0$ all the fields except $h_{ij}^{\rm tt}$ vanish on the equations of motion \eqref{eq:4D_Einst_eq_sc-vec-ten_decomp_decoupled}, while the latter satisfies the free wave equation. Therefore, the massless gravitational field has two dDOF $h_{ij}^{\rm tt}$, as expected \cite{Flanagan:2005yc}.

\subsubsection{Non-dynamical DOFs contribution into the Isaacson EMT}

Substituting the SVT expansion of gravitational field \eqref{eq:sc-vec-ten_decomp_1} into the Eq. \eqref{eq:4D_GR_Isaac_EMT}, integrating by parts, and using the equations of motion \eqref{eq:4D_Einst_eq_sc-vec-ten_decomp_decoupled}, we arrive at the following expression for the Isaacson EMT of gravitational field
\begin{align}
t_{\mu\nu} & \overset{\to}{=} \frac{1}{32 \pi G} \left \langle \partial_\mu h_{ij}^{\rm tt} \partial_\nu h_{ij}^{\rm tt} + 2 \partial_\mu \Phi \partial_\nu \Phi - 2 \partial_\mu \partial_i B \partial_\nu \partial_i B - 2 \partial_\mu S_i \partial_\nu S_i + 2 \partial_\mu \partial_i F_j \partial_\nu \partial_i F_j \right. \nonumber \\ & \left. - 2 \partial_\mu \Phi \partial_\nu \Delta E - 2 \partial_\mu \Delta E \partial_\nu \Phi + 2 \partial_\mu \Delta E \partial_\nu \Delta E \right \rangle.
\end{align}
Here, we have used the Isaacson EMT in the Lorentz gauge \eqref{eq:4D_EoM_in_Lor_gauge}, which in the SVT decomposition \eqref{eq:sc-vec-ten_decomp_1} is written as
\begin{equation}
\partial_0 \Phi + \Delta B + 3 \partial_0 \Psi - \partial_0 \Delta E = 0, \quad - \partial_0 S_i + \Delta F_i = 0, \quad - \partial_0 B + \Psi + \Delta E - \Phi = 0.
\end{equation}
Thus, outside the unitary gauge the contributions of ndDOF are preserved in the Isaacson EMT \eqref{eq:4D_GR_Isaac_EMT}, vanishing only in the transverse-traceless (unitary) gauge on the equations of motion \eqref{eq:4D_Einst_eq_sc-vec-ten_decomp_decoupled}
\begin{equation}
\label{eq:Isaac_EMT_unit_gauge}
B = E = F_i = 0 \quad \Longrightarrow \quad t_{\mu\nu} = \frac{1}{32 \pi G} \left \langle \partial_\mu h_{ij}^{\rm tt} \partial_\nu h_{ij}^{\rm tt} \right \rangle.
\end{equation}

As a result, in contrast with the electrodynamics, Isaacson gauge-invariant EMT generally preserves the contributions of ndDOF of gravitational field, leading to the need to either calculate gravitational radiation in the unitary gauge or to separately extract the EMT of dDOF.

\subsubsection{Action and EMT of dynamical DOFs}

Let us extract the dynamical part of the action of massless gravitational field. We start with the quadratic approximation of the Einstein-Hilbert action \cite{maggiore2008}
\begin{equation}
S = - \frac{1}{64 \pi G} \int d^4x \left( \partial^\mu h^{\alpha\beta} \partial_\mu h_{\alpha\beta} + 2 \partial_\mu h^{\mu\nu} \partial_\nu h - 2 \partial_\mu h^{\mu\nu} \partial^\sigma h_{\sigma\nu} - \partial^\mu h \partial_\mu h \right).
\end{equation}
Substituting the Eqs. \eqref{eq:sc-vec-ten_decomp_1} and \eqref{eq:sc-vec-ten_decomp_2} into it, we obtain
\begin{align}
\label{eq:4D_GR_lin_act_ddof-nddof_split}
& S = S_{\rm d} \lbrack h_{ij}^{\rm tt} \rbrack + S_{\rm nd} \lbrack \Phi, B, S_i, \Psi, E, F_i \rbrack, \\
\label{eq:4D_grav_dyn_act}
& S_{\rm d} = - \frac{1}{64 \pi G} \int d^4x \, \partial^\alpha h_{ij}^{\rm tt} \partial_\alpha h_{ij}^{\rm tt}, \\
& S_{\rm nd} = \frac{1}{64 \pi G} \int d^4x \, \big( 16 \partial_0 \Psi \partial_0 \Delta E - 24 \partial_0 \Psi \partial_0 \Psi + 2 \partial_0 \partial_i F_j \partial_0 \partial_i F_j + 2 \partial_k S_i \partial_k S_i \nonumber \\ \label{eq:4D_grav_non-dyn_act} & \quad - 16 \Delta B \partial_0 \Psi - 4 \partial_0 S_i \Delta F_i + 8 \partial_i \Psi \partial_i \Psi - 16 \partial_i \Phi \partial_i \Psi \big).
\end{align}
Therefore, in massless gravity the contributions of dynamical and non-dynamical degrees of freedom of gravitational field into the action are automatically decoupled, by analogy with electrodynamics (\ref{eq:ED_dyn_non-dyn_act_split_1}--\ref{eq:ED_dyn_non-dyn_act_split_3}) and, thus, one can extract the dynamical part of the action. Note also that the gauge symmetry (\ref{eq:4D_GR_lin_gauge_sym_sc-vec-ten_decomp_1}--\ref{eq:4D_GR_lin_gauge_sym_sc-vec-ten_decomp_2}) is entirely contained in the non-dynamical part of the action, while the dynamical part is not affected by the gauge transformations, by analogy with electrodynamics. One easily finds that the action \eqref{eq:4D_GR_lin_act_ddof-nddof_split} yields the equations of motion
\begin{equation}
\label{eq:GR_lin_EoM_dyn_non-dyn_split}
{_4}\square h_{ij}^{\rm tt} = 0, \quad \Delta \Psi = 0, \quad \partial_0 B + \Phi = \partial_0^2 E, \quad S_i = \partial_0 F_i,
\end{equation}
coinciding with the Eq. \eqref{eq:4D_Einst_eq_sc-vec-ten_decomp_decoupled} obtained by linearizing the Einstein equations.

Finally, we construct the canonical EMT of the dDOF of massless gravitational field \cite{Landau:1975pou,Jackson:1998nia,Stepanyanz_book}
\begin{equation}
{T_{\rm d}}^\mu_{\,\cdot\,\nu} = - \frac{\partial {\cal L}_{\rm d}}{\partial (\partial_\mu h_{ij}^{\rm tt})} \partial_\nu h_{ij}^{\rm tt} + \delta^\mu_\nu {\cal L}_{\rm d}, \quad {\cal L}_{\rm d} = -\frac{1}{64 \pi G} \partial^\alpha h_{ij}^{\rm tt} \partial_\alpha h_{ij}^{\rm tt}.
\end{equation}
Here, the canonical EMT is defined with the opposite sign, compared to the Eq. \eqref{eq:4D_ED_ddof_canon_EMT_def}, due to the use of Minkowski metric with the mostly positive signature. As a result, the canonical EMT of the dDOF of massless gravitational field is written as
\begin{equation}
\label{eq:4D_GR_lin_dyn_EMT}
{T_{\rm d}}_{\mu\nu} = \frac{1}{32 \pi G} \left( \partial_\mu h_{ij}^{\rm tt} \partial_\nu h_{ij}^{\rm tt} - \frac{1}{2} \eta_{\mu\nu} \partial^\alpha h_{ij}^{\rm tt} \partial_\alpha h_{ij}^{\rm tt} \right).
\end{equation}
By analogy with the EMT of electromagnetic field \eqref{eq:4D_ED_ddof_canon_EMT}, it also has the structure of the EMT of a set of <<scalar fields>> $\big \lbrace h_{ij}^{\rm tt} \big \rbrace$. After averaging \eqref{eq:Isaac_averaging}, the obtained EMT coincides on the equations of motion \eqref{eq:GR_lin_EoM_dyn_non-dyn_split} with the Isaacson EMT in the unitary gauge \eqref{eq:Isaac_EMT_unit_gauge}
\begin{equation}
\label{eq:4D_GR_ddof_canon_EMT}
{_4}\square h_{ij}^{\rm tt} = 0 \quad \Longrightarrow \quad \left \langle {T_{\rm d}}_{\mu\nu} \right \rangle \overset{\to}{=} \frac{1}{32 \pi G} \left \langle \partial_\mu h_{ij}^{\rm tt} \partial_\nu h_{ij}^{\rm tt} \right \rangle.
\end{equation}

Thus, in massless gravity, besides using the Isaacson EMT in the unitary gauge, the gravitational radiation can be equivalently computed as the energy flux through a distant surface enclosing the radiation source determined by the canonical EMT of dDOF of gra\-vi\-ta\-tio\-nal field \eqref{eq:4D_GR_ddof_canon_EMT}.

\subsection{Dynamical DOFs of massive electrodynamics}

In contrast with electrodynamics and massless gravity, the effective action of DGP gravity \eqref{eq:DGP_grav_eff_act} does not have a gauge symmetry, obscuring the extraction of dynamical part of the action. Therefore, before proceeding to the construction of the effective EMT of dDOF of DGP gravity, we first consider a couple of simple examples of gauge-noninvariant theories. Let us start with the massive electrodynamics.

The action of free massive electrodynamics has the form \cite{Hinterbichler:2011tt,deRham:2014zqa}
\begin{equation}
\label{eq:4D_mas_ED_act}
S = \int d^4x \left( - \frac{1}{4} F_{\mu\nu}^2 + \frac{1}{2} m^2 A_\mu^2 \right), \quad \eta_{\mu\nu} \propto(+--\,-),
\end{equation}
and the corresponding equations of motion are written as
\begin{equation}
\label{eq:PT_EoM_start}
\partial^\alpha F_{\alpha\mu} + m^2 A_\mu = 0.
\end{equation}
Acting on it by the 4-divergence, we obtain the constraint equation $\partial^\mu A_\mu = 0$ for the components of the vector field. Using it, we rewrite the Eq. \eqref{eq:PT_EoM_start} as the system of equations
\begin{equation}
\label{eq:4D_mas_ED_EoM+constr}
{_4}\square A_\mu + m^2 A_\mu = 0, \quad \partial^\mu A_\mu = 0.
\end{equation}

As is demonstrated below, to decouple the dynamical and non-dynamical parts of the action \eqref{eq:4D_mas_ED_act} one has to introduce the gauge symmetry into the action by use of the Stuckelberg change of field variables \cite{Hinterbichler:2011tt,deRham:2014zqa} and then fix it by imposing such a gauge condition that in terms of the new field variables the contributions of dynamical and non-dynamical degrees of freedom into the action are decoupled.

\subsubsection{Counting DOFs}

To determine the dDOF of massive vector field, we perform its SV decomposition in accordance with the Eq. \eqref{eq:4D_ED_sc-vec_decomp}. As a result, the equations of motion \eqref{eq:4D_mas_ED_EoM+constr} are rewritten as
\begin{align}
\label{eq:4D_mas_ED_EoM_sc-vec_decomp_1}
& \partial_0^2 A_0 - \Delta A_0 + m^2 A_0 = 0, \\
\label{eq:4D_mas_ED_EoM_sc-vec_decomp_2}
& \partial_0^2 \partial_i \varphi + \partial_0^2 A_i^{\rm t} - \Delta \partial_i \varphi - \Delta A_i^{\rm tt} + m^2 \partial_i \varphi + m^2 A_i^{\rm tt} = 0, \\
\label{eq:4D_mas_ED_constr_sc-vec_decomp}
& \partial_0 A_0 = \Delta \varphi.
\end{align}
Acting by the 3-divergence on the Eq. \eqref{eq:4D_mas_ED_EoM_sc-vec_decomp_2}, we split it into the equations of motion for $\varphi$ and $A_i^{\rm t}$
\begin{equation}
\Delta \left( {_4}\square \varphi + m^2 \varphi \right) = 0 \quad \Longrightarrow \quad {_4}\square \varphi + m^2 \varphi = 0, \quad {_4}\square A_i^{\rm t} + m^2 A_i^{\rm t} = 0.
\end{equation}
Finally, substituting the constraint equation \eqref{eq:4D_mas_ED_constr_sc-vec_decomp} into the Eq. \eqref{eq:4D_mas_ED_EoM_sc-vec_decomp_1}, we arrive at the following system of equations of motion for the massive vector field components
\begin{align}
& \partial_0 \Delta \varphi - \Delta A_0 + m^2 A_0 = 0, \quad \partial_0 A_0 = \Delta \varphi, \\
& {_4}\square \varphi + m^2 \varphi = 0, \quad {_4}\square A_i^{\rm t} + m^2 A_i^{\rm t} = 0.
\end{align}
Thus, the massive vector field has three dDOF $\varphi$ and $A_i^{\rm t}$, in contrast with massless electro\-dy\-na\-mics, where $\varphi$ is the Stuckelberg field \eqref{eq:4D_ED_gauge_sym_sc-vec_form}.

\subsubsection{Action and EMT of dynamical DOFs}

Decomposing the action of massive electrodynamics \eqref{eq:4D_mas_ED_act} by use of Eq. \eqref{eq:4D_ED_sc-vec_decomp}, we find that the contributions of dynamical and non-dynamical degrees of freedom are not decoupled
\begin{align}
\label{eq:4D_mas_ED_act_sc-vec_decomp}
S & = \int d^4x \left( \frac{1}{2} \partial_0 A_i^{\rm t} \partial_0 A_i^{\rm t} - \frac{1}{2} \partial_j A_i^{\rm t} \partial_j A_i^{\rm t} - \frac{1}{2} m^2 A_i^{\rm t} A_i^{\rm t} + \frac{1}{2} \partial_0 \partial_i \varphi \partial_0 \partial_i \varphi - \frac{1}{2} m^2 \partial_i \varphi \partial_i \varphi \right. \nonumber \\ & \left. + \frac{1}{2} \partial_i A_0 \partial_i A_0 + \frac{1}{2} m^2 A_0 A_0 - \partial_0 \partial_i \varphi \partial_i A_0 \right).
\end{align}
Here, the last term in the action \eqref{eq:4D_mas_ED_act_sc-vec_decomp} contains the interaction of dynamical degree of freedom $\varphi$ and non-dynamical $A_0$. Analogous term is also present in the action of massless electrodynamics \eqref{eq:4D_ED_act_sc-vec_decomp}, however, in this case $\varphi$ is a non-dynamical degree of freedom.

We decouple the contributions of dDOF and ndDOF into the action using the Stuckelberg method \cite{Hinterbichler:2011tt,deRham:2014zqa}, introducing the gauge symmetry into the action by the appropriate change of field variables. Namely, we make the following change of field variables
\begin{equation}
A_\mu = \tilde{A}_\mu + \partial_\mu \phi,
\end{equation}
after which the action \eqref{eq:4D_mas_ED_act} is rewritten as
\begin{equation}
\label{eq:mas_ED_Stuck_act}
S = \int d^4x \left( - \frac{1}{4} \tilde{F}_{\mu\nu}^2 + \frac{1}{2} m^2 \tilde{A}_\mu^2 + m^2 \tilde{A}^\mu \partial_\mu \phi + \frac{1}{2} m^2 \partial^\mu \phi \partial_\mu \phi \right).
\end{equation}
Here, we obtained the kinetic interaction of fields $\tilde{A}_\mu$ and $\phi$, and the action acquired the gauge symmetry
\begin{equation}
\label{eq:4D_mas_ED_Stuck_gauge_sym}
\tilde{A}_\mu \to \tilde{A}_\mu + \partial_\mu \alpha, \quad \phi \to \phi - \alpha.
\end{equation}
The equations of motion corresponding to the obtained action have the form
\begin{equation}
\label{eq:mas_ED_Stuck_EoMs}
\partial^\alpha \tilde{F}_{\alpha\mu} + m^2 \tilde{A}_\mu + m \partial_\mu \varphi = 0, \quad {_4}\square \varphi + m \partial^\mu \tilde{A}_\mu = 0,
\end{equation}
where we have rescaled the scalar field as $\phi = \varphi/m$ (note that here $\varphi$ does not coincide with the longitudinal part of the vector field in Eq. \eqref{eq:4D_ED_sc-vec_decomp}).

We fix the gauge symmetry \eqref{eq:4D_mas_ED_Stuck_gauge_sym} by imposing the gauge condition similar to the Lorentz gauge in massless electrodynamics, which decouples the equations of motion \eqref{eq:mas_ED_Stuck_EoMs} \cite{Hinterbichler:2011tt,deRham:2014zqa}
\begin{equation}
\label{eq:4D_mas_ED_Stuck_trans_Lor_gauge}
\partial^\mu \tilde{A}_\mu = m \varphi \quad \Longrightarrow \quad {_4}\square \tilde{A}_\mu + m^2 \tilde{A}_\mu = 0, \quad {_4}\square \varphi + m^2 \varphi = 0.
\end{equation}
Adding to the action \eqref{eq:mas_ED_Stuck_act} the corresponding gauge-fixing term
\begin{equation}
S_{\rm gf} = - \frac{1}{2} \int d^4x \left( \partial^\mu \tilde{A}_\mu - m \varphi \right)^2
\end{equation}
and integrating by parts, we rewrite the action \eqref{eq:mas_ED_Stuck_act} as
\begin{equation}
S + S_{\rm gf} \overset{\to}{=} \int d^4x \left( - \frac{1}{2} \partial^\mu \tilde{A}^\nu \partial_\mu \tilde{A}_\nu + \frac{1}{2} m^2 \tilde{A}_\mu^2 + \frac{1}{2} \partial^\mu \varphi \partial_\mu \varphi - \frac{1}{2} m^2 \varphi^2 \right).
\end{equation}
As a result, we have decoupled the $\tilde{A}_\mu$ and $\varphi$ contributions into the action, obtaining in both cases the canonical kinetic and mass terms in the action. The obtained action still has the residual gauge symmetry
\begin{equation}
\tilde{A}_\mu \to \tilde{A}_\mu + \partial_\mu \alpha, \quad \varphi \to \varphi - m \alpha, \quad {_4}\square \alpha + m^2 \alpha = 0.
\end{equation}
As the field $\varphi$ also satisfies the free massive wave equation, we completely fix the gauge symmetry by imposing the unitary gauge condition $\varphi = 0$. Note that one can fix this gauge even in the presence of a conserved vector current as the source of massive vector field \cite{Hinterbichler:2011tt,deRham:2014zqa}. Thus, in the unitary gauge the action of massive electrodynamics takes the form
\begin{equation}
\label{eq:4D_mas_ED_act_after_Stuck_trick}
S + S_{\rm gf} = \int d^4x \left( - \frac{1}{2} \partial^\mu \tilde{A}^\nu \partial_\mu \tilde{A}_\nu + \frac{1}{2} m^2 \tilde{A}_\mu^2 \right).
\end{equation}
By fixing the unitary gauge, we have returned to the original field variables $\tilde{A}_\mu = A_\mu$ satisfying, by virtue of Eq. \eqref{eq:4D_mas_ED_Stuck_trans_Lor_gauge}, the system of equations of motion \eqref{eq:4D_mas_ED_EoM+constr}
\begin{equation}
{_4}\square \tilde{A}_\mu + m^2 \tilde{A}_\mu = 0, \quad \partial^\mu \tilde{A}_\mu = 0.
\end{equation}

As a result of this change of field variables, the dDOF and ndDOF contributions into the action \eqref{eq:4D_mas_ED_act_after_Stuck_trick} are decoupled. Performing the SV decomposition of the vector field \eqref{eq:4D_ED_sc-vec_decomp}
\begin{equation}
\label{eq:4D_mas_ED_Stuck_trans_sc-vec_decomp}
\tilde{A}_\mu = \big \lbrace \tilde{A}_0, \partial_i \tilde{A} + \tilde{A}_i^{\rm t} \big \rbrace, \quad \partial_i \tilde{A}_i^{\rm t} = 0,
\end{equation}
we rewrite the action \eqref{eq:4D_mas_ED_act_after_Stuck_trick} as
\begin{align}
S + S_{\rm gf} & \overset{\to}{=} \int d^4x \left( - \frac{1}{2} \partial^\mu \tilde{A}_0 \partial_\mu \tilde{A}_0 + \frac{1}{2} m^2 \tilde{A}_0^2 + \frac{1}{2} \partial^\mu \tilde{A}_i^{\rm l} \partial_\mu \tilde{A}_i^{\rm l} - \frac{1}{2} m^2 \tilde{A}_i^{\rm l} \tilde{A}_i^{\rm l} \right. \nonumber \\ & \left. + \frac{1}{2} \partial^\mu \tilde{A}_i^{\rm t} \partial_\mu \tilde{A}_i^{\rm t} - \frac{1}{2} m^2 \tilde{A}_i^{\rm t} \tilde{A}_i^{\rm t} \right), \quad \tilde{A}_i^{\rm l} = \partial_i \tilde{A}.
\end{align}
From the corresponding equations of motion
\begin{equation}
{_4}\square \tilde{A}_0 + m^2 \tilde{A}_0 = 0, \quad {_4}\square \tilde{A}_i^{\rm l} + m^2 \tilde{A}_i^{\rm l} = 0, \quad {_4}\square \tilde{A}_i^{\rm t} + m^2 \tilde{A}_i^{\rm t} = 0
\end{equation}
and gauge condition $\partial^\mu \tilde{A}_\mu = 0$ we find that the dynamical degrees of freedom of massive electrodynamics are $\tilde{A}_i^{\rm l}$ and $\tilde{A}_i^{\rm t}$. Therefore, the action of massive electrodynamics \eqref{eq:4D_mas_ED_act_after_Stuck_trick} is splitted into the dynamical and non-dynamical parts as
\begin{align}
& S + S_{\rm gf} = S_{\rm d}[\tilde{A}_i^{\rm l}, \tilde{A}_i^{\rm t}] + S_{\rm nd}[\tilde{A}_0], \\
\label{eq:4D_mas_ED_Stuck_trans_dyn-ndyn_split}
& S_{\rm d} = \int d^4x \left( \frac{1}{2} \partial^\mu \tilde{A}_i^{\rm l} \partial_\mu \tilde{A}_i^{\rm l} - \frac{1}{2} m^2 \tilde{A}_i^{\rm l} \tilde{A}_i^{\rm l} + \frac{1}{2} \partial^\mu \tilde{A}_i^{\rm t} \partial_\mu \tilde{A}_i^{\rm t} - \frac{1}{2} m^2 \tilde{A}_i^{\rm t} \tilde{A}_i^{\rm t} \right), \\
& S_{\rm nd} = \int d^4x \left( - \frac{1}{2} \partial^\mu \tilde{A}_0 \partial_\mu \tilde{A}_0 + \frac{1}{2} m^2 \tilde{A}_0^2 \right).
\end{align}
Note that the non-dynamical part enters the action with the opposite sign -- $\tilde{A}_0$ is the ghost degree of freedom. However, it poses no danger here, being non-dynamical and not interacting with other degrees of freedom.

Finally, from the action \eqref{eq:4D_mas_ED_Stuck_trans_dyn-ndyn_split} we find the canonical EMT of dynamical degrees of freedom of massive electrodynamics
\begin{align}
{T_{\rm d}}_{\alpha\beta} & = \partial_\alpha \tilde{A}_i^{\rm l} \partial_\beta \tilde{A}_i^{\rm l} - \frac{1}{2} \eta_{\alpha\beta} \partial^\mu \tilde{A}_i^{\rm l} \partial_\mu \tilde{A}_i^{\rm l} + \frac{1}{2} \eta_{\alpha\beta} m^2 \tilde{A}_i^{\rm l} \tilde{A}_i^{\rm l} + \partial_\alpha \tilde{A}_i^{\rm t} \partial_\beta \tilde{A}_i^{\rm t} - \frac{1}{2} \eta_{\alpha\beta} \partial^\mu \tilde{A}_i^{\rm t} \partial_\mu \tilde{A}_i^{\rm t} \nonumber \\ & + \frac{1}{2} \eta_{\alpha\beta} m^2 \tilde{A}_i^{\rm t} \tilde{A}_i^{\rm t}.
\end{align}
By analogy with massless electrodynamics \eqref{eq:4D_ED_ddof_canon_EMT}, it has the structure of the EMT of a set of massive <<scalar fields>> $\lbrace \tilde{A}_i^{\rm l}, \tilde{A}_i^{\rm t} \rbrace$. In accordance with the presence of the third dynamical degree of freedom $\tilde{A}_i^{\rm l}$ in the massive vector field, in the massless limit it does not transform into the EMT of massless electrodynamics \eqref{eq:4D_ED_ddof_canon_EMT}
\begin{align}
{T_{\rm d}}_{\alpha\beta} & \xrightarrow{m \to 0} \partial_\alpha \tilde{A}_i^{\rm l} \partial_\beta \tilde{A}_i^{\rm l} - \frac{1}{2} \eta_{\alpha\beta} \partial^\mu \tilde{A}_i^{\rm l} \partial_\mu \tilde{A}_i^{\rm l} + \partial_\alpha \tilde{A}_i^{\rm t} \partial_\beta \tilde{A}_i^{\rm t} - \frac{1}{2} \eta_{\alpha\beta} \partial^\mu \tilde{A}_i^{\rm t} \partial_\mu \tilde{A}_i^{\rm t} \nonumber \\
& \neq \partial_\alpha \tilde{A}_i^{\rm t} \partial_\beta \tilde{A}_i^{\rm t} - \frac{1}{2} \eta_{\alpha\beta} \partial^\mu \tilde{A}_i^{\rm t} \partial_\mu \tilde{A}_i^{\rm t}.
\end{align}

Thus, to decouple the contributions of dynamical and non-dynamical degrees of freedom into the action of a theory without gauge symmetry, such as massive electrodynamics and massive gravity, or DGP gravity, one can use the Stuckelberg trick, introducing the gauge symmetry into the action by the appropriate change of field variables and then fixing it by imposing the decoupling gauge condition.

\subsection{Dynamical DOFs of massive gravity}

Let us now construct the canonical EMT of dDOF of massive gravity. As the effective action of DGP gravity \eqref{eq:DGP_grav_eff_act} coincides with the action of massive gravity up to replacing the non-local mass by the rigid one \eqref{eq:DGP-mas_grav_relation}, the EMT of dDOF of DGP gravity will be constructed by analogy with the derivations presented in this section.

The action of free Fierz-Pauli massive gravity is written as \cite{Fierz:1939ix,Hinterbichler:2011tt,deRham:2014zqa}
\begin{align}
S & = \frac{1}{64 \pi G} \int d^4x \, \big( - \partial^\sigma h^{\alpha\beta} \partial_\sigma h_{\alpha\beta} + 2 \partial^\alpha h_{\mu\alpha} \partial_\beta h^{\mu\beta} - 2 \partial_\alpha h^{\alpha\beta} \partial_\beta h \nonumber \\ \label{eq:4D_mas_grav_act} & + \partial^\alpha h \partial_\alpha h + m^2 (h^2 - h^{\alpha\beta} h_{\alpha\beta}) \big), \quad \eta_{\mu\nu} \propto (-++\,+).
\end{align}
We obtain the corresponding equation of motion of the massive graviton in the form
\begin{equation}
\label{eq:mas_grav_EoM_gen}
{_4}\square h_{\alpha\beta} - \partial_\alpha \partial^\sigma h_{\sigma\beta} - \partial_\beta \partial^\sigma h_{\alpha\sigma} + \partial_\alpha \partial_\beta h + \eta_{\alpha\beta} \partial^\mu \partial^\nu h_{\mu\nu} - \eta_{\alpha\beta} {_4}\square h - m^2 \left( h_{\alpha\beta} - \eta_{\alpha\beta} h \right) = 0.
\end{equation}
By analogy with massive electrodynamics, we reduced it to the simpler form, obtaining from it the constraint equations for the components of gravitational field. Acting on the equation \eqref{eq:mas_grav_EoM_gen} by the 4-divergence, we find the first constraint equation
\begin{equation}
\partial^\alpha h_{\alpha\beta} = \partial_\beta h.
\end{equation}
Substituting it into the Eq. \eqref{eq:mas_grav_EoM_gen}, we arrive at the system of equations of motion
\begin{equation}
{_4}\square h_{\alpha\beta} - \partial_\alpha \partial_\beta h - m^2 \left( h_{\alpha\beta} - \eta_{\alpha\beta} h \right) = 0, \quad \partial^\alpha h_{\alpha\beta} = \partial_\beta h.
\end{equation}
Calculating the trace of the obtained equation, we find the second constraint equation
\begin{equation}
h = 0.
\end{equation}
As a result, we rewrite the equation of motion of the massive gravitational field \eqref{eq:mas_grav_EoM_gen} as the free massive wave equation and two constraint equations \cite{Hinterbichler:2011tt,deRham:2014zqa}
\begin{equation}
\label{eq:mas_grav_EoM_constr}
{_4}\square h_{\alpha\beta} - m^2 h_{\alpha\beta} = 0, \quad \partial^\alpha h_{\alpha\beta} = 0, \quad h = 0.
\end{equation}

\subsubsection{Counting DOFs}

Substituting the SVT decomposition (\ref{eq:sc-vec-ten_decomp_1}--\ref{eq:sc-vec-ten_decomp_2}) into the Eq. \eqref{eq:mas_grav_EoM_constr}, we obtain the equations of motion of massive gravity in the form
\begin{align}
\label{eq:mas_grav_EoM_sc-vec-ten_decomp_1}
& {_4}\square \Phi - m^2 \Phi = 0, \\
\label{eq:mas_grav_EoM_sc-vec-ten_decomp_2}
& {_4}\square \partial_i B + {_4}\square S_i - m^2 \partial_i B - m^2 S_i = 0, \\
& - 2 \delta_{ij} {_4}\square \Psi + 2 {_4}\square \partial_i \partial_j E + {_4}\square \partial_i F_j + {_4}\square \partial_j F_i + {_4}\square h_{ij}^{\rm tt} + 2 \delta_{ij} m^2 \Psi - 2 m^2 \partial_i \partial_j E \nonumber \\ \label{eq:mas_grav_EoM_sc-vec-ten_decomp_3} & \quad - m^2 \partial_i F_j - m^2 \partial_j F_i - m^2 h_{ij}^{\rm tt} = 0,
\end{align}
while the constraint equations for the components of gravitational field are written as
\begin{align}
\label{eq:mas_grav_constr_sc-vec-ten_decomp_1}
& 2 \partial_0 \Phi + \Delta B = 0, \\
\label{eq:mas_grav_constr_sc-vec-ten_decomp_2}
& - \partial_0 \partial_i B - \partial_0 S_i - 2 \partial_i \Psi + 2 \Delta \partial_i E + \Delta F_i = 0, \\
\label{eq:mas_grav_constr_sc-vec-ten_decomp_3}
& \Phi - 3 \Psi + \Delta E = 0.
\end{align}
We decouple the equations of motion by use of the constraint equations. Acting on the Eq. \eqref{eq:mas_grav_EoM_sc-vec-ten_decomp_2} with 3-divergence, we decouple the equations of motion for $B$ and $S_i$
\begin{equation}
\Delta \left( {_4}\square B - m^2 B \right) = 0 \quad \Longrightarrow \quad {_4}\square B - m^2 B = 0, \quad {_4}\square S_i - m^2 S_i = 0.
\end{equation}
Calculating the trace of the Eq. \eqref{eq:mas_grav_EoM_sc-vec-ten_decomp_3}, we obtain
\begin{equation}
\frac{1}{3} \Delta \left( {_4}\square E - m^2 E \right) = {_4}\square \Psi - m^2 \Psi.
\end{equation}
Substituting the obtained relation back into the Eq. \eqref{eq:mas_grav_EoM_sc-vec-ten_decomp_3}, we arrive at the equation
\begin{align}
& - \frac{2}{3} \delta_{ij} \Delta \left( {_4}\square E - m^2 E \right) + 2 \partial_i \partial_j \left( {_4}\square E - m^2 E \right) + {_4}\square \partial_i F_j + {_4}\square \partial_j F_i + {_4}\square h_{ij}^{\rm tt} \nonumber \\ & \quad - m^2 \partial_i F_j - m^2 \partial_j F_i - m^2 h_{ij}^{\rm tt} = 0.
\end{align}
Acting on it twice by the 3-divergence, we decouple the equations of motion for $\Psi$ and $E$
\begin{equation}
\Delta^2 \left( {_4}\square E - m^2 E \right) = 0 \quad \Longrightarrow \quad {_4}\square E - m^2 E = 0, \quad {_4}\square \Psi - m^2 \Psi = 0.
\end{equation}
Finally, given the obtained equations of motion, acting on the Eq. \eqref{eq:mas_grav_EoM_sc-vec-ten_decomp_3} by the 3-divergence, we decouple the equations of motion for $F_i$ and $h_{ij}^{\rm tt}$
\begin{equation}
\Delta \left( {_4}\square F_j - m^2 F_j \right) = 0 \quad \Longrightarrow \quad {_4}\square F_j - m^2 F_j = 0, \quad {_4}\square h_{ij}^{\rm tt} - m^2 h_{ij}^{\rm tt} = 0.
\end{equation}
Thus, all the components of massive gravitational field satisfy the free massive wave equations.

Now we determine the dDOF of massive graviton by use of the constraint equations. For this, we split the Eq. \eqref{eq:mas_grav_constr_sc-vec-ten_decomp_2} into two constraint equations, acting on it by the 3-divergence
\begin{equation}
\Delta \left( - \partial_0 B - 2 \Psi + 2 \Delta E \right) = 0 \quad \Longrightarrow \quad - \partial_0 B - 2 \Psi + 2 \Delta E = 0, \quad - \partial_0 S_i + \Delta F_i = 0.
\end{equation}
Given the Eq. \eqref{eq:mas_grav_constr_sc-vec-ten_decomp_1}, we find that $\Phi$, $B$ and $S_i$ are ndDOF. To obtain another constraint equation, we act by the time derivative on the Eq. \eqref{eq:mas_grav_constr_sc-vec-ten_decomp_3}. Then from the system of equations
\begin{equation}
\partial_0 \Phi - 3 \partial_0 \Psi + \partial_0 \Delta E = 0, \quad 2 \partial_0 \Phi + \Delta B = 0, \quad - \partial_0 B - 2 \Psi + 2 \Delta E = 0,
\end{equation}
we find the following expressions for the first time derivatives of the fields $\Psi$ and $E$
\begin{equation}
\partial_0 \Psi = - \frac{1}{4} m^2 B, \quad \partial_0 \Delta E = \frac{1}{2} \partial_0^2 B - \frac{1}{4} m^2 B.
\end{equation}
Therefore, $\Psi$ is also the ndDOF, in contrast with $E$, as substituting the expression for its time derivative into the equation of motion increases its order in time derivatives. As a result, the massive gravitational field has five dDOF $E$, $F_i$ and $h_{ij}^{\rm tt}$, as expected \cite{Hinterbichler:2011tt,deRham:2014zqa}.

\subsubsection{Action and EMT of dynamical DOFs}

By analogy with massive electrodynamics, substituting the SVT decomposition of the massive gravitational field \eqref{eq:sc-vec-ten_decomp_1} into the action \eqref{eq:4D_mas_grav_act}, we do not decouple the contributions of dynamical and non-dynamical degrees of freedom into the action. One can see this from the decomposition of the action of massless gravity \eqref{eq:4D_GR_lin_act_ddof-nddof_split}, which is also contained in the decomposition of the massive gravity action. There, the interaction term $\propto \partial_0 \Psi \partial_0 \Delta E$ is present in the non-dynamical part of the action \eqref{eq:4D_grav_non-dyn_act}. However, in massive gravity the field $E$ is dDOF and does not allow one to directly split the action \eqref{eq:4D_mas_grav_act} into the dynamical and non-dynamical parts.

To split the action \eqref{eq:4D_mas_grav_act} into the dynamical and non-dynamical parts, we use the Stuckelberg method \cite{Hinterbichler:2011tt,deRham:2014zqa}, by analogy with massive electrodynamics. However, in this case, to decouple the contributions of dDOF and ndDOF into the action, we need to make three changes of field variables. The first one is written as
\begin{equation}
\label{eq:mas_grav_Stuck_var_change_1}
h_{\mu\nu} = \tilde{h}_{\mu\nu} + \partial_\mu \tilde{A}_\nu + \partial_\nu \tilde{A}_\mu.
\end{equation}
After this field variables change the action of massive gravity \eqref{eq:4D_mas_grav_act} is rewritten as
\begin{align}
\label{eq:mas_grav_act_after_1_Stuck_change}
S & = \frac{1}{64 \pi G} \int d^4x \, \Big( - \partial^\sigma \tilde{h}^{\alpha\beta} \partial_\sigma \tilde{h}_{\alpha\beta} + 2 \partial^\alpha \tilde{h}_{\mu\alpha} \partial_\beta \tilde{h}^{\mu\beta} - 2 \partial_\alpha \tilde{h}^{\alpha\beta} \partial_\beta \tilde{h} + \partial^\alpha \tilde{h} \partial_\alpha \tilde{h} \nonumber \\ & + m^2 \big( \tilde{h}^2 - \tilde{h}^{\alpha\beta} \tilde{h}_{\alpha\beta} \big) - m^2 \tilde{F}_{\alpha\beta}^2 + 4 m^2 \big( \partial^\alpha \tilde{A}_\alpha \tilde{h} - \tilde{h}^{\alpha\beta} \partial_\alpha \tilde{A}_\beta \big) \Big)
\end{align}
and acquires the gauge symmetry
\begin{equation}
\label{eq:mass_grav_gauge_sym_1}
\tilde{h}_{\alpha\beta} \to \tilde{h}_{\alpha\beta} + \partial_\alpha \xi_\beta + \partial_\beta \xi_\alpha, \quad \tilde{A}_\alpha \to \tilde{A}_\alpha - \xi_\alpha.
\end{equation}
Making the second change of field variables
\begin{equation}
\label{eq:mas_grav_Stuck_var_change_2}
\tilde{A}_\mu = \bar{A}_\mu + \partial_\mu \phi,
\end{equation}
we arrive at the action \eqref{eq:mas_grav_act_after_1_Stuck_change} in the form
\begin{align}
S & = \frac{1}{64 \pi G} \int d^4x \, \Big( - \partial^\sigma \tilde{h}^{\alpha\beta} \partial_\sigma \tilde{h}_{\alpha\beta} + 2 \partial^\alpha \tilde{h}_{\mu\alpha} \partial_\beta \tilde{h}^{\mu\beta} - 2 \partial_\alpha \tilde{h}^{\alpha\beta} \partial_\beta \tilde{h} + \partial^\alpha \tilde{h} \partial_\alpha \tilde{h} \nonumber \\ & + m^2 \big( \tilde{h}^2 - \tilde{h}^{\alpha\beta} \tilde{h}_{\alpha\beta} \big) - m^2 \bar{F}_{\alpha\beta}^2 + 4 m^2 \big( \partial^\alpha \bar{A}_\alpha \tilde{h} - \tilde{h}^{\alpha\beta} \partial_\alpha \bar{A}_\beta \big) \nonumber \\ & + 4 m^2 \big( \partial^\alpha \partial_\alpha \phi \tilde{h} - \tilde{h}^{\alpha\beta} \partial_\alpha \partial_\beta \phi \big) \Big)
\end{align}
introducing into it the second independent gauge symmetry
\begin{equation}
\label{eq:mass_grav_gauge_sym_2}
\bar{A}_\mu \to \bar{A}_\mu + \partial_\mu \alpha, \quad \phi \to \phi - \alpha.
\end{equation}
Finally, to eliminate the interaction between $\tilde{h}_{\mu\nu}$ and $\phi$, we make the third change of variables
\begin{equation}
\label{eq:mas_grav_Stuck_var_change_3}
\tilde{h}_{\mu\nu} = \bar{h}_{\mu\nu} + m^2 \eta_{\mu\nu} \phi,
\end{equation}
after which the action of massive gravity is written as \cite{Hinterbichler:2011tt,deRham:2014zqa}
\begin{align}
\label{eq:mas_grav_act_after_3_Stuck_change}
S & = \frac{1}{64 \pi G} \int d^4x \Big( - \partial^\sigma \bar{h}^{\alpha\beta} \partial_\sigma \bar{h}_{\alpha\beta} + 2 \partial^\alpha \bar{h}_{\mu\alpha} \partial_\beta \bar{h}^{\mu\beta} - 2 \partial_\alpha \bar{h}^{\alpha\beta} \partial_\beta \bar{h} + \partial^\alpha \bar{h} \partial_\alpha \bar{h} \nonumber \\ & + m^2 ( \bar{h}^2 - \bar{h}^{\alpha\beta} \bar{h}_{\alpha\beta} ) - m^2 \bar{F}_{\alpha\beta}^2 + 4 m^2 ( \partial^\alpha \bar{A}_\alpha \bar{h} - \bar{h}^{\alpha\beta} \partial_\alpha \bar{A}_\beta) - 6 m^4 \partial^\alpha \phi \partial_\alpha \phi + 12 m^6 \phi^2 \nonumber \\ & + 12 m^4 \partial^\alpha \bar{A}_\alpha \phi + 6 m^4 \bar{h} \phi \Big).
\end{align}
The obtained action \eqref{eq:mas_grav_act_after_3_Stuck_change} is invariant under two independent gauge transformations, the combined action of which has the form
\begin{equation}
\label{eq:mass_grav_gauge-sym_comb}
\bar{h}_{\mu\nu} \to \bar{h}_{\mu\nu} + \partial_\mu \xi_\nu + \partial_\nu \xi_\mu + m^2 \eta_{\mu\nu} \alpha, \quad \bar{A}_\mu \to \bar{A}_\mu - \xi_\mu + \partial_\mu \alpha, \quad \phi \to \phi - \alpha.
\end{equation}
Further, it is the two independent gauge symmetries that allows one to decouple the contributions of dDOF and ndDOF into the action of massive gravity by imposing the decoupling gauge conditions.

The action \eqref{eq:mas_grav_act_after_3_Stuck_change} leads to the following equations of motion
\begin{align}
& {_4}\square \bar{h}_{\alpha\beta} - \partial_\alpha \partial^\sigma \bar{h}_{\sigma\beta} - \partial_\beta \partial^\sigma \bar{h}_{\alpha\sigma} + \partial_\alpha \partial_\beta \bar{h} + \eta_{\alpha\beta} \partial^\mu \partial^\nu \bar{h}_{\mu\nu} - \eta_{\alpha\beta} {_4}\square \bar{h} + m^2 \left( \eta_{\alpha\beta} \bar{h} - \bar{h}_{\alpha\beta} \right) \nonumber \\ & \quad + 2 m^2 \eta_{\alpha\beta} \partial^\mu \bar{A}_\mu - m^2 \partial_\alpha \bar{A}_\beta - m^2 \partial_\beta \bar{A}_\alpha + 3 m^4 \eta_{\alpha\beta} \phi = 0, \\
& \partial^\alpha \bar{F}_{\alpha\beta} - \partial_\beta \bar{h} + \partial^\alpha \bar{h}_{\alpha\beta} - 3 m^2 \partial_\beta \phi = 0, \\
& 6 m^2 {_4}\square \phi + 12 m^4 \phi + 6 m^2 \partial^\alpha \bar{A}_\alpha + 3 m^2 \bar{h} = 0.
\end{align}
We decouple them by imposing two gauge conditions, in accordance with the two symmetries of the action \eqref{eq:mass_grav_gauge_sym_1} and \eqref{eq:mass_grav_gauge_sym_2} \cite{Hinterbichler:2011tt,deRham:2014zqa}
\begin{equation}
\label{eq:mas_grav_Stuck_gauge_sym_decoup_cond}
\partial^\alpha \bar{h}_{\alpha\beta} = \frac{1}{2} \partial_\beta \bar{h} - m^2 \bar{A}_\beta, \quad \partial^\mu \bar{A}_\mu = - \frac{1}{2} \bar{h} - 3 m^2 \phi.
\end{equation}
As a result, the equations of motion take the simple form
\begin{equation}
\left( {_4}\square - m^2 \right) \left( \bar{h}_{\alpha\beta} - \frac{1}{2} \eta_{\alpha\beta} \bar{h} \right) = 0, \quad \left( {_4}\square - m^2 \right) \bar{A}_\mu = 0, \quad \left( {_4}\square - m^2 \right) \phi = 0.
\end{equation}
Accordingly, adding two gauge-fixing terms to the action \eqref{eq:mas_grav_act_after_3_Stuck_change}
\begin{align}
& S_{\rm gf1} = - \frac{1}{32 \pi G} \int d^4x \left( \partial^\alpha \bar{h}_{\alpha\beta} - \frac{1}{2} \partial_\beta \bar{h} + m^2 \bar{A}_\beta \right)^2, \\
& S_{\rm gf2} = - \frac{1}{32 \pi G} \int d^4x \left( m \partial^\mu \bar{A}_\mu + \frac{1}{2} m \bar{h} + 3 m^3 \phi \right)^2,
\end{align}
we rewrite the action of massive gravity as \cite{Hinterbichler:2011tt,deRham:2014zqa}
\begin{align}
& S + S_{\rm gf1} + S_{\rm gf2} = \frac{1}{64 \pi G} \int d^4x \, \Big( - \partial^\sigma \bar{h}^{\alpha\beta} \partial_\sigma \bar{h}_{\alpha\beta} + \frac{1}{2} \partial^\alpha \bar{h} \partial_\alpha \bar{h} - m^2 \bar{h}^{\alpha\beta} \bar{h}_{\alpha\beta} + \frac{1}{2} m^2 \bar{h}^2 \nonumber \\ & \quad - 2 m^2 \partial^\alpha \bar{A}^\beta \partial_\alpha \bar{A}_\beta - 2 m^4 \bar{A}^\alpha \bar{A}_\alpha - 6 m^4 \partial^\alpha \phi \partial_\alpha \phi - 6 m^6 \phi^2 \Big).
\end{align}
The obtained action still has the residual gauge symmetry \eqref{eq:mass_grav_gauge-sym_comb} with the transformation parameters satisfying the free massive wave equations
\begin{equation}
{_4}\square \xi_\mu - m^2 \xi_\mu = 0, \quad {_4}\square \alpha - m^2 \alpha = 0.
\end{equation}
We use these two symmetries to impose the unitary gauge $\bar{A}_\mu = 0$, which fixes one of the residual symmetries (in the presence of interaction of the massive gravitational field with the conserved matter EMT, this is the only possible choice of unitary gauge). As a result, the action of massive gravity takes the form
\begin{align}
\label{eq:mas_grav_Stuck_trans_act_gauge_fixed}
& S + S_{\rm gf1} + S_{\rm gf2} = \frac{1}{64 \pi G} \int d^4x \, \Big( - \partial^\sigma \bar{h}^{\alpha\beta} \partial_\sigma \bar{h}_{\alpha\beta} + \frac{1}{2} \partial^\alpha \bar{h} \partial_\alpha \bar{h} - m^2 \bar{h}^{\alpha\beta} \bar{h}_{\alpha\beta} + \frac{1}{2} m^2 \bar{h}^2 \nonumber \\ & \quad - 6 m^4 \partial^\alpha \phi \partial_\alpha \phi - 6 m^6 \phi^2 \Big).
\end{align}
However, the action still retains one residual gauge symmetry
\begin{equation}
\label{eq:mas_grav_resid_guage_sym_after_fixing}
\bar{h}_{\mu\nu} \to \bar{h}_{\mu\nu} + 2 \partial_\mu \partial_\nu \alpha + m^2 \eta_{\mu\nu} \alpha, \quad \phi \to \phi - \alpha, \quad {_4}\square \alpha - m^2 \alpha = 0,
\end{equation}
and the previously imposed decoupling gauge conditions \eqref{eq:mas_grav_Stuck_gauge_sym_decoup_cond} are written as
\begin{equation}
\label{eq:mas_grav_Stuck_trans_gauge_cond_reduced}
\partial^\alpha \bar{h}_{\alpha\beta} = \frac{1}{2} \partial_\beta \bar{h}, \quad \bar{h} = - 6 m^2 \phi.
\end{equation}
Here the first gauge condition coincides with the Lorentz gauge in massless gravity.

To determine the dDOF, we perform the SVT decomposition \eqref{eq:sc-vec-ten_decomp_1} of the tensor field
\begin{align}
\label{eq:mas_grav_Stuck_trans_sc-vec-ten_decomp_1}
& \bar{h}_{00} = - 2 \bar{\Phi}, \quad \bar{h}_{0i} = \partial_i \bar{B} + \bar{S}_i, \quad \bar{h}_{ij} = - 2 \delta_{ij} \bar{\Psi} + 2 \partial_i \partial_j \bar{E} + \partial_i \bar{F}_j + \partial_j \bar{F}_i + \bar{h}_{ij}^{\rm tt}, \\
\label{eq:mas_grav_Stuck_trans_sc-vec-ten_decomp_2}
& \partial_i \bar{S}_i = 0, \quad \partial_i \bar{F}_i = 0, \quad \partial_i \bar{h}_{ij}^{\rm tt} = 0, \bar{h}_{ii}^{\rm tt} = 0.
\end{align}
In this form, the residual gauge symmetry \eqref{eq:mas_grav_resid_guage_sym_after_fixing} is written as
\begin{align}
& \bar{\Phi} \to \bar{\Phi} - \partial_0^2 \alpha + \frac{1}{2} m^2 \alpha, \quad \bar{B} \to \bar{B} + 2 \partial_0 \alpha, \quad \bar{S}_i \to \bar{S}_i, \quad \bar{\Psi} \to \bar{\Psi} - \frac{1}{2} m^2 \alpha, \\
& \bar{E} \to \bar{E} + \alpha, \quad \bar{F}_i \to \bar{F}_i, \quad \bar{h}_{ij}^{\rm tt} \to \bar{h}_{ij}^{\rm tt}, \quad \phi \to \phi, \quad {_4}\square \alpha - m^2 \alpha = 0.
\end{align}
Substituting the decomposition (\ref{eq:mas_grav_Stuck_trans_sc-vec-ten_decomp_1}--\ref{eq:mas_grav_Stuck_trans_sc-vec-ten_decomp_2}) into the action \eqref{eq:mas_grav_Stuck_trans_act_gauge_fixed}, we arrive at
\begin{align}
\label{eq:mas_grav_Stuck_trans_act_reduced_sc-vec-ten_decomp}
& S + S_{\rm gf1} + S_{\rm gf2} = \frac{1}{64 \pi G} \int d^4x \, \Big( - 2 \partial^\sigma \bar{\Phi} \partial_\sigma \bar{\Phi} - 2 m^2 \bar{\Phi}^2 + \partial^\sigma \partial_i \bar{B} \partial_\sigma \partial_i \bar{B} + m^2 \partial_i \bar{B} \partial_i \bar{B} \nonumber \\ & \quad + \partial^\sigma \bar{S}_i \partial_\sigma \bar{S}_i + m^2 \bar{S}_i \bar{S}_i + 6 \partial^\sigma \bar{\Psi} \partial_\sigma \bar{\Psi} + 6 m^2 \bar{\Psi}^2 - 2 \partial^\sigma \Delta \bar{E} \partial_\sigma \Delta \bar{E} - 2 m^2 (\Delta E)^2 \nonumber \\ & \quad - 2 \partial^\sigma \partial_i \bar{F}_j \partial_\sigma \partial_i \bar{F}_j - 2 m^2 \partial_i \bar{F}_j \partial_i \bar{F}_j - \partial^\sigma \bar{h}_{ij}^{\rm tt} \partial_\sigma \bar{h}_{ij}^{\rm tt} - m^2 \bar{h}_{ij}^{\rm tt} \bar{h}_{ij}^{\rm tt} - 6 m^4 \partial^\sigma \phi \partial_\sigma \phi - 6 m^6 \phi^2 \nonumber \\ & \quad - 4 m^2 \bar{\Psi} \Delta \bar{E} - 4 \partial^\sigma \bar{\Psi} \partial_\sigma \Delta \bar{E} - 12 \partial^\sigma \bar{\Phi} \partial_\sigma \bar{\Psi} + 4 \partial^\sigma \bar{\Phi} \partial_\sigma \Delta \bar{E} - 12 m^2 \bar{\Phi} \bar{\Psi} + 4 m^2 \bar{\Phi} \Delta \bar{E} \Big).
\end{align}
The corresponding equations of motion have the form
\begin{align}
& {_4}\square_m \bar{\Phi} + 3 {_4}\square_m \bar{\Psi} - {_4}\square_m \Delta \bar{E} = 0, \quad - 3 {_4}\square_m \bar{\Phi} + 3 {_4}\square_m \bar{\Psi} - {_4}\square_m \Delta \bar{E} = 0, \\
& - {_4}\square_m \bar{\Phi} + {_4}\square_m \bar{\Psi} + {_4}\square_m \Delta \bar{E} = 0, \quad {_4}\square_m \bar{B} = 0, \quad {_4}\square_m \bar{S}_i = 0, \\
& {_4}\square_m \bar{F}_i = 0, \quad {_4}\square_m \bar{h}_{ij}^{\rm tt} = 0, \quad {_4}\square_m \phi = 0,
\end{align}
where we have introduced the short notation for the massive wave operator ${_4}\square_m = {_4}\square - m^2$. Solving the system of the first three equations, we find that all the components of massive gravitational field satisfy the free massive wave equation
\begin{equation}
{_4}\square_m \bar{\Phi} = 0, \quad {_4}\square_m \bar{\Psi} = 0, \quad {_4}\square_m \bar{E} = 0.
\end{equation}
To determine the dDOF, we use the gauge conditions \eqref{eq:mas_grav_Stuck_trans_gauge_cond_reduced}, which in the SVT decomposition are written as
\begin{align}
& \partial_0 \bar{\Phi} + 3 \partial_0 \bar{\Psi} - \partial_0 \Delta \bar{E} = - \Delta \bar{B}, \quad - \partial_0 \bar{B} = \bar{\Phi} - \bar{\Psi} - \Delta \bar{E}, \\
& \partial_0 \bar{S}_i = \Delta \bar{F}_i, \quad - 3 m^2 \phi = \bar{\Phi} - 3 \bar{\Psi} + \Delta \bar{E}.
\end{align}
From these one finds that $\bar{B}$ and $\bar{S}_i$ are ndDOF. Acting on the second and fourth gauge conditions with time derivatives and using the equations of motion, we find the expressions for the time derivatives of the fields $\bar{\Phi}$, $\bar{\Psi}$ and $\bar{E}$
\begin{align}
& \partial_0 \bar{\Psi} = - \frac{1}{4} m^2 \bar{B}, \quad \partial_0 \Delta \bar{E} = - \frac{3}{2} m^2 \partial_0 \phi - \frac{3}{4} m^2 \bar{B} + \frac{1}{2} \Delta \bar{B}, \\
& \partial_0 \bar{\Phi} = - \frac{3}{2} m^2 \partial_0 \phi - \frac{1}{2} \Delta \bar{B}.
\end{align}
Thus, the fields $\phi$, $\bar{F}_i$ and $\bar{h}_{ij}^{\rm tt}$ are the only dDOF of massive gravitational field. Moreover, their contributions into the action \eqref{eq:mas_grav_Stuck_trans_act_reduced_sc-vec-ten_decomp} are automatically decoupled from the ndDOF.

Therefore, the dynamical part of the action of massive gravity is found as
\begin{align}
\label{eq:mas_grav_dyn_act}
& S_{\rm d} = \frac{1}{64 \pi G} \int d^4x \, \Big( - \partial^\sigma \bar{h}_{ij}^{\rm lt} \partial_\sigma \bar{h}_{ij}^{\rm lt} - m^2 \bar{h}_{ij}^{\rm lt} \bar{h}_{ij}^{\rm lt} - \partial^\sigma \bar{h}_{ij}^{\rm tt} \partial_\sigma \bar{h}_{ij}^{\rm tt} - m^2 \bar{h}_{ij}^{\rm tt} \bar{h}_{ij}^{\rm tt} - 6 m^4 \partial^\sigma \phi \partial_\sigma \phi \nonumber \\ & \quad - 6 m^6 \phi^2 \Big), \quad \bar{h}_{ij}^{\rm lt} = \partial_i \bar{F}_j + \partial_j \bar{F}_i.
\end{align}
Correspondingly, we obtain the canonical EMT of dynamical degrees of freedom of massive gravitational field in the form
\begin{align}
{T_{\rm d}}_{\mu\nu} & = \frac{3 m^4}{16 \pi G} \left( \partial_\mu \phi \partial_\nu \phi - \frac{1}{2} \eta_{\mu\nu} \partial^\alpha \phi \partial_\alpha \phi - \frac{1}{2} m^2 \eta_{\mu\nu} \phi^2 \right) \nonumber \\ & + \frac{1}{32 \pi G} \left( \partial_\mu \bar{h}_{ij}^{\rm lt} \partial_\nu \bar{h}_{ij}^{\rm lt} - \frac{1}{2} \eta_{\mu\nu} \partial^\alpha \bar{h}_{ij}^{\rm lt} \partial_\alpha \bar{h}_{ij}^{\rm lt} - \frac{1}{2} m^2 \eta_{\mu\nu} \bar{h}_{ij}^{\rm lt} \bar{h}_{ij}^{\rm lt} \right) \nonumber \\ & + \frac{1}{32 \pi G} \left( \partial_\mu \bar{h}_{ij}^{\rm tt} \partial_\nu \bar{h}_{ij}^{\rm tt} - \frac{1}{2} \eta_{\mu\nu} \partial^\alpha \bar{h}_{ij}^{\rm tt} \partial_\alpha \bar{h}_{ij}^{\rm tt} - \frac{1}{2} m^2 \eta_{\mu\nu} \bar{h}_{ij}^{\rm tt} \bar{h}_{ij}^{\rm tt} \right).
\end{align}
By analogy with massive electrodynamics, it has the structure of the EMT of a set of massive <<scalar fields>> $\lbrace \phi, \bar{h}_{ij}^{\rm lt}, \bar{h}_{ij}^{\rm tt} \rbrace$. Note that in the massless limit $m \to 0$, $m^2 \phi \neq 0$ the obtained EMT preserves the contributions of the additional, compared to massless gravity, dDOF
\begin{align}
{T_{\rm d}}_{\mu\nu} & \xrightarrow{m \to 0} \frac{3 m^4}{16 \pi G} \left( \partial_\mu \phi \partial_\nu \phi - \frac{1}{2} \eta_{\mu\nu} \partial^\alpha \phi \partial_\alpha \phi \right) + \frac{1}{32 \pi G} \left( \partial_\mu \bar{h}_{ij}^{\rm lt} \partial_\nu \bar{h}_{ij}^{\rm lt} - \frac{1}{2} \eta_{\mu\nu} \partial^\alpha \bar{h}_{ij}^{\rm lt} \partial_\alpha \bar{h}_{ij}^{\rm lt} \right) \nonumber \\ & + \frac{1}{32 \pi G} \left( \partial_\mu \bar{h}_{ij}^{\rm tt} \partial_\nu \bar{h}_{ij}^{\rm tt} - \frac{1}{2} \eta_{\mu\nu} \partial^\alpha \bar{h}_{ij}^{\rm tt} \partial_\alpha \bar{h}_{ij}^{\rm tt} \right)
\end{align}
and does not transform into the EMT of dDOF of massless gravity \eqref{eq:4D_GR_lin_dyn_EMT}, in accordance with the vDVZ discontinuity in massive gravity \cite{vanDam:1970vg,Zakharov:1970cc,Hinterbichler:2011tt,deRham:2014zqa}. Under the averaging operator \eqref{eq:Isaac_averaging}, on the equations of motion the dynamical EMT of massive gravity takes the form
\begin{equation}
\left \langle {T_{\rm d}}_{\mu\nu} \right \rangle = \frac{3 m^4}{16 \pi G} \langle \partial_\mu \phi \partial_\nu \phi \rangle + \frac{1}{32 \pi G} \langle \partial_\mu \bar{h}_{ij}^{\rm lt} \partial_\nu \bar{h}_{ij}^{\rm lt} \rangle + \frac{1}{32 \pi G} \langle \partial_\mu \bar{h}_{ij}^{\rm tt} \partial_\nu \bar{h}_{ij}^{\rm tt} \rangle.
\end{equation}

Note that, in the case of interaction of the massive graviton with the conserved matter EMT, the residual gauge symmetry \eqref{eq:mas_grav_resid_guage_sym_after_fixing} can be completely fixed by imposing the condition
\begin{equation}
\phi = 0 \quad \Longleftrightarrow \quad T_{\rm m} = 0,
\end{equation}
as the field $\phi$ would be sourced by the trace of the matter energy-momentum tensor ${T_{\rm m}}_{\alpha\beta}$.

\subsection{Dynamical DOFs of DGP gravity}

Finally, let us construct the effective EMT of dDOF of DGP gravity. As the effective action of DGP gravity \eqref{eq:DGP_grav_eff_act} coincides with the action of massive gravity \eqref{eq:4D_mas_grav_act} up to the replacing the mass \eqref{eq:DGP-mas_grav_relation}, we follow the derivations presented in the previous section and only briefly describe the main steps of calculations.

Making three Stuckelberg changes of field variables analogous to the Eqs. \eqref{eq:mas_grav_Stuck_var_change_1}, \eqref{eq:mas_grav_Stuck_var_change_2}, and \eqref{eq:mas_grav_Stuck_var_change_3} (with the only difference in replacing the rigid mass by a momentum-dependent mass \eqref{eq:DGP-mas_grav_relation}) and fixing the resulting gauge symmetries by imposing the gauge conditions similar to the Eq. \eqref{eq:mas_grav_Stuck_gauge_sym_decoup_cond}, we arrive at the effective action of DGP gravity on the brane in the unitary gauge in the form
\begin{align}
& S_{\rm eff} + S_{\rm gf1} + S_{\rm gf2} = \frac{1}{8} M_4^2 \int d^4x \, \Big( - \partial^\sigma \bar{h}^{\alpha\beta} \partial_\sigma \bar{h}_{\alpha\beta} + \frac{1}{2} \partial^\alpha \bar{h} \partial_\alpha \bar{h} + M_c \bar{h}^{\alpha\beta} \sqrt{-{_4}\square} \, \bar{h}_{\alpha\beta} \nonumber \\ \label{eq:DGP_eff_act_after_Stuck_changes} & \quad - \frac{1}{2} M_c \bar{h} \sqrt{-{_4}\square} \, \bar{h} - 6 M_c^2 \partial^\alpha \partial^\beta \phi \partial_\alpha \partial_\beta \phi + 6 M_c^3 \partial^\alpha \phi \sqrt{-{_4}\square} \,\partial_\alpha \phi \Big),
\end{align}
where the initial induced metric perturbations $h_{\mu\nu}$ are rewritten in terms of the new field variables as
\begin{equation}
h_{\mu\nu} = \bar{h}_{\mu\nu} - M_c \eta_{\mu\nu} \sqrt{-{_4}\square} \, \phi + 2 \partial_\mu \partial_\nu \phi.
\end{equation}
The corresponding equations of motion take the form
\begin{equation}
\left( {_4}\square + M_c \sqrt{-{_4}\square} \right) \left( \bar{h}_{\mu\nu} - \frac{1}{2} \eta_{\mu\nu} \bar{h} \right) = 0, \quad \left( {_4}\square + M_c \sqrt{-{_4}\square} \right) \phi = 0.
\end{equation}
Note that the scalar field enters the action \eqref{eq:DGP_eff_act_after_Stuck_changes} with higher derivatives. However, they are eliminated by the following field rescaling
\begin{equation}
\phi = \frac{\varphi}{M_c \sqrt{-{_4}\square}}.
\end{equation}
As a result, the effective action of DGP gravity is written as
\begin{align}
& S_{\rm eff} + S_{\rm gf1} + S_{\rm gf2} = \frac{1}{8} M_4^2 \int d^4x \, \Big( - \partial^\sigma \bar{h}^{\alpha\beta} \partial_\sigma \bar{h}_{\alpha\beta} + \frac{1}{2} \partial^\alpha \bar{h} \partial_\alpha \bar{h} + M_c \bar{h}^{\alpha\beta} \sqrt{-{_4}\square} \, \bar{h}_{\alpha\beta} \nonumber \\ \label{eq:DGP_eff_act_covar_form} & \quad - \frac{1}{2} M_c \bar{h} \sqrt{-{_4}\square} \, \bar{h} - 6 \partial^\alpha \varphi \partial_\alpha \varphi + 6 M_c \varphi \sqrt{-{_4}\square} \, \varphi \Big).
\end{align}

Performing the SVT decomposition of the tensor field \eqref{eq:mas_grav_Stuck_trans_sc-vec-ten_decomp_1}, we arrive at the effective action of DGP gravity in the form
\begin{align}
& S_{\rm eff} + S_{\rm gf1} + S_{\rm gf2} = \frac{1}{8} M_4^2 \int d^4x \, \Big( - 2 \partial^\sigma \bar{\Phi} \partial_\sigma \bar{\Phi} + 2 M_c \bar{\Phi} \sqrt{-{_4}\square} \, \bar{\Phi} + \partial^\sigma \partial_i \bar{B} \partial_\sigma \partial_i \bar{B} \nonumber \\ & \quad - M_c \partial_i \bar{B} \sqrt{-{_4}\square} \, \partial_i \bar{B} + \partial^\sigma \bar{S}_i \partial_\sigma \bar{S}_i - M_c \bar{S}_i \sqrt{-{_4}\square} \, \bar{S}_i + 6 \partial^\sigma \bar{\Psi} \partial_\sigma \bar{\Psi} - 6 M_c \bar{\Psi} \sqrt{-{_4}\square} \, \bar{\Psi} \nonumber \\ & \quad - 2 \partial^\sigma \Delta \bar{E} \partial_\sigma \Delta \bar{E} + 2 M_c \Delta \bar{E} \sqrt{-{_4}\square} \, \Delta \bar{E} - 2 \partial^\sigma \partial_i \bar{F}_j \partial_\sigma \partial_i \bar{F}_j + 2 M_c \partial_i \bar{F}_j \sqrt{-{_4}\square} \, \partial_i \bar{F}_j \nonumber \\ & \quad - \partial^\sigma \bar{h}_{ij}^{\rm tt} \partial_\sigma \bar{h}_{ij}^{\rm tt} + M_c \bar{h}_{ij}^{\rm tt} \sqrt{-{_4}\square} \, \bar{h}_{ij}^{\rm tt} - 6 \partial^\sigma \varphi \partial_\sigma \varphi + 6 M_c \varphi \sqrt{-{_4}\square} \, \varphi + 4 M_c \bar{\Psi} \sqrt{-{_4}\square} \, \Delta \bar{E} \nonumber \\ & \quad - 4 \partial^\sigma \bar{\Psi} \partial_\sigma \Delta \bar{E} - 12 \partial^\sigma \bar{\Phi} \partial_\sigma \bar{\Psi} + 4 \partial^\sigma \bar{\Phi} \partial_\sigma \Delta \bar{E} + 12 M_c \bar{\Phi} \sqrt{-{_4}\square} \, \bar{\Psi} - 4 M_c \bar{\Phi} \sqrt{-{_4}\square} \, \Delta \bar{E} \Big).
\end{align}
By analogy with massive gravity, from the corresponding equations of motion and gauge conditions we find that the dDOF of DGP gravity on the brane are $\varphi$, $\bar{F}_i$ and $\bar{h}_{ij}^{\rm tt}$. Thus, the dynamical part of effective action of DGP gravity is written as
\begin{align}
& S_{\rm d} = \frac{1}{8} M_4^2 \int d^4x \, \Big( - \partial^\sigma \bar{h}_{ij}^{\rm lt} \partial_\sigma \bar{h}_{ij}^{\rm lt} + M_c \bar{h}_{ij}^{\rm lt} \sqrt{-{_4}\square} \, \bar{h}_{ij}^{\rm lt} - \partial^\sigma \bar{h}_{ij}^{\rm tt} \partial_\sigma \bar{h}_{ij}^{\rm tt} + M_c \bar{h}_{ij}^{\rm tt} \sqrt{-{_4}\square} \, \bar{h}_{ij}^{\rm tt} \nonumber \\ & \quad - 6 \partial^\sigma \varphi \partial_\sigma \varphi + 6 M_c \varphi \sqrt{-{_4}\square} \, \varphi \Big), \quad \bar{h}_{ij}^{\rm lt} = \partial_i \bar{F}_j + \partial_j \bar{F}_i.
\end{align}
As was shown in Section \ref{III}, the EMT of such a non-local theory can be constructed by the standard formula for the canonical EMT neglecting the non-local mass terms in the effective Lagrangian. Therefore, the effective EMT of dDOF of DGP gravity on the brane is given by
\begin{align}
\label{eq:DGP_grav_eff_EMT_dyn}
{T_{\rm d}}_{\mu\nu} & = \frac{3}{2} M_4^2 \left( \partial_\mu \varphi \partial_\nu \varphi - \frac{1}{2} \eta_{\mu\nu} \partial^\alpha \varphi \partial_\alpha \varphi \right) + \frac{1}{4} M_4^2 \left( \partial_\mu \bar{h}_{ij}^{\rm lt} \partial_\nu \bar{h}_{ij}^{\rm lt} - \frac{1}{2} \eta_{\mu\nu} \partial^\alpha \bar{h}_{ij}^{\rm lt} \partial_\alpha \bar{h}_{ij}^{\rm lt} \right) \nonumber \\ & + \frac{1}{4} M_4^2 \left( \partial_\mu \bar{h}_{ij}^{\rm tt} \partial_\nu \bar{h}_{ij}^{\rm tt} - \frac{1}{2} \eta_{\mu\nu} \partial^\alpha \bar{h}_{ij}^{\rm tt} \partial_\alpha \bar{h}_{ij}^{\rm tt} \right).
\end{align}
By analogy with massive gravity, the obtained EMT does not transform into the EMT of massless gravity \eqref{eq:4D_GR_lin_dyn_EMT} in the four-dimensional limit $M_5 \to 0$, $\varphi \neq 0$, due to the effective graviton on the brane acquiring the additional dynamical degrees of freedom.

\section{Gravitational radiation in DGP model}\label{V}

Now we turn to computing the analog of quadrupole formula for the effective gravitational radiation power of an arbitrary non-relativistic source on the brane in DGP gravity. For this, we extract the emitted parts of the gravitational field dDOF $\varphi$, $\bar{h}_{ij}^{\rm lt}$ and $\bar{h}_{ij}^{\rm tt}$ and construct from them, in accordance with the Eq. \eqref{eq:DGP_grav_eff_EMT_dyn}, the effective gravitational radiation energy flux density on the brane.

The interaction of the gravitational field with the conserved energy-momentum tensor of brane localised matter ${T_{\rm m}}_{\alpha\beta}$ is introduced into the action \eqref{eq:DGP_eff_act_covar_form} as \cite{Hinterbichler:2011tt,deRham:2014zqa}
\begin{equation}
S_{\rm int} = \int d^4x \left( \frac{1}{2} \bar{h}^{\alpha\beta} {T_{\rm m}}_{\alpha\beta} - \frac{1}{2} \varphi T_{\rm m} \right), \quad \partial^\alpha{T_{\rm m}}_{\alpha\beta} = 0.
\end{equation}
Accordingly, the dynamical degrees of freedom of gravitational field interact with the corresponding parts of the matter energy-momentum tensor
\begin{align}
& S_{\rm int} = \int d^4x \left( \frac{1}{2} \bar{h}_{ij}^{\rm lt} {T_{\rm m}}_{ij}^{\rm lt} + \frac{1}{2} \bar{h}_{ij}^{\rm tt} {T_{\rm m}}_{ij}^{\rm tt} - \frac{1}{2} \varphi T_{\rm m} \right), \\
& \partial_i {T_{\rm m}}_{ij}^{\rm lt} \neq 0, \quad \partial_i {T_{\rm m}}_{ij}^{\rm tt} = 0, \quad {T_{\rm m}}_{ii}^{\rm lt} = {T_{\rm m}}_{ii}^{\rm tt} = 0.
\end{align}
As a result, the effective equations of motion of dDOF of DGP gravity in the presence of brane localised matter are written as
\begin{align}
\label{eq:DGP_phi_EoM}
& {_4}\square \varphi + M_c \sqrt{-{_4}\square} \, \varphi = \frac{1}{3M_4^2} T_{\rm m}, \\
\label{eq:DGP_h_lt/tt_EoM}
& {_4}\square \bar{h}_{ij}^{\rm lt/tt} + M_c \sqrt{-{_4}\square} \, \bar{h}_{ij}^{\rm lt/tt} = - \frac{2}{M_4^2} {T_{\rm m}}_{ij}^{\rm lt/tt}.
\end{align}

\subsection{Tensor field in the wave zone}

The solutions of the equations of motion \eqref{eq:DGP_h_lt/tt_EoM} for the tensor fields $\bar{h}_{ij}^{\rm lt}$ and $\bar{h}_{ij}^{\rm tt}$ can be obtained from the solution of the equation of motion resulting from the action \eqref{eq:DGP_eff_act_covar_form}
\begin{equation}
\label{eq:DGP_grav_psi_EoM}
\left( {_4}\square + M_c \sqrt{-{_4}\square} \right) \psi_{\mu\nu} = - \frac{2}{M_4^2} {T_{\rm m}}_{\mu\nu}, \quad \psi_{\mu\nu} = \bar{h}_{\mu\nu} - \frac{1}{2} \eta_{\mu\nu} \bar{h},
\end{equation}
by acting on it with the corresponding projectors \cite{Weinberg:1972kfs,maggiore2008}
\begin{align}
& \bar{h}_{ij}^{\rm lt} = \Lambda_{ij,kl}^{\rm lt} \psi_{kl}, \quad \bar{h}_{ij}^{\rm tt} = \Lambda_{ij,kl}^{\rm tt} \psi_{kl} \\
& \partial_i \Lambda_{ij,kl}^{\rm lt} \neq 0, \quad \partial_i \Lambda_{ij,kl}^{\rm tt} = 0, \quad \Lambda_{ii,kl}^{\rm lt} = \Lambda_{ij,kk}^{\rm lt} = \Lambda_{ii,kl}^{\rm tt} = \Lambda_{ij,kk}^{\rm tt} = 0.
\end{align}
In particular, far away from the source these projectors are written as \cite{Weinberg:1972kfs,maggiore2008}
\begin{align}
& \Lambda_{ij,kl}^{\rm lt} = n_i n_k P_{jl} + n_j n_k P_{il}, \quad n_i \Lambda_{ij,kl}^{\rm lt} = n_k P_{jl}, \quad \Lambda_{ii,kl}^{\rm lt} = \Lambda_{ij,kk}^{\rm lt} = 0, \\
& \Lambda_{ij,kl}^{\rm tt} = P_{ik} P_{jl} - \frac{1}{2} P_{ij} P_{kl}, \quad n_i \Lambda_{ij,kl}^{\rm tt} = 0, \quad \Lambda_{ii,kl}^{\rm tt} = \Lambda_{ij,kk}^{\rm tt} = 0, \\
& n_i = x_i/r, \quad P_{ij} = \delta_{ij} - n_i n_j.
\end{align}
Thus, we compute the emitted parts of the fields $\bar{h}_{ij}^{\rm lt}$ and $\bar{h}_{ij}^{\rm tt}$ in a unified manner as the corresponding projections of the emitted part of the field $\psi_{ij}$. We calculate the latter by analogy with the derivation presented in Ref. \cite{Khlopunov:2022ubp}, which calculates the contribution of the EMT of the scalar field binding the binary system on the brane into the source of gravitational radiation.

The retarded solution of the Eq. \eqref{eq:DGP_grav_psi_EoM} is given by the integral \cite{Dvali:2001gm,Dvali:2001gx,Khlopunov:2022jaw}
\begin{align}
& \psi_{ij} = \frac{2}{\pi M_5^3} \int_{0}^{\infty} d\mu \, \rho(\mu) \int d^4x' \, G_4(x-x'|\mu) \, {T_{\rm m}}_{ij}(x'), \\
& G_4(x|\mu) = \frac{\theta(t)}{2\pi} \left \lbrack \delta(x^2) J_0(\mu \sqrt{x^2}) - \frac{1}{2} \frac{\theta(x^2)}{\sqrt{x^2}} \mu J_1(\mu \sqrt{x^2}) \right \rbrack, \quad \rho(\mu) = \frac{M_c^2}{\mu^2 + M_c^2},
\end{align}
where $G_4(x|\mu)$ is the retarded Green function of the four-dimensional massive field, and $\rho(\mu)$ is the spectral function of Kaluza-Klein decomposition of effective DGP graviton on the brane. Defining the retarded coordinate time as
\begin{equation}
t_{\rm ret}(\mathbf{x}') = t - \sqrt{(\mathbf{x} - \mathbf{x'})^2},
\end{equation}
we rewrite the retarded field $\psi_{ij}$ in the following form
\begin{align}
\psi_{ij} & = \frac{1}{2 \pi^2 M_5^3} \int_{0}^{\infty} d\mu \, \rho(\mu) \int d^3x' \Bigg( \int dt' \frac{\delta(t' - t_{\rm ret})}{\sqrt{(\mathbf{x} - \mathbf{x'})^2}} J_0\!\left( \mu \sqrt{(t - t')^2 - (\mathbf{x} - \mathbf{x}')^2} \right) \nonumber \\ & - \int_{-\infty}^{t_{\rm ret}} dt' \, \frac{\mu J_1\!\left( \mu \sqrt{(t - t')^2 - (\mathbf{x} - \mathbf{x}')^2} \right)}{\sqrt{(t - t')^2 - (\mathbf{x} - \mathbf{x}')^2}} \Bigg) {T_{\rm m}}_{ij}(x').
\end{align}
To eliminate the local term with the $\delta$-function, we integrate by parts the second term of the integrand using the relation
\begin{equation}
\label{eq:ret_mass_field_by_parts_aux_rel}
\frac{\mu J_1\!\left( \mu \sqrt{(t - t')^2 - (\mathbf{x} - \mathbf{x}')^2} \right)}{\sqrt{(t - t')^2 - (\mathbf{x} - \mathbf{x}')^2}} = \frac{1}{(t-t')} \frac{d}{dt'} J_0\!\left( \mu \sqrt{(t - t')^2 - (\mathbf{x} - \mathbf{x}')^2} \right).
\end{equation}
As a result, after the integration by parts we obtain
\begin{align}
\psi_{ij} & = \frac{1}{2 \pi^2 M_5^3} \int_{0}^{\infty} d\mu \, \rho(\mu) \int d^3x' \int_{-\infty}^{t_{\rm ret}} dt' \frac{1}{t-t'} J_0\!\left( \mu \sqrt{(t - t')^2 - (\mathbf{x} - \mathbf{x}')^2} \right) \nonumber \\ & \times \left( \frac{d}{dt'} {T_{\rm m}}_{ij}(x') + \frac{1}{t-t'} {T_{\rm m}}_{ij}(x') \right).
\end{align}

We assume that the observation point is in the radiation zone $r \gg {\cal R}$, where ${\cal R} \sim \lambdabar_{\rm GW}$ denotes the radius of the near zone \cite{Pati:2000vt} (for the definitions of near and radiation zone see, also, \cite{Khlopunov:2022ubp}). In the non-relativistic limit ${\cal R} \gg {\cal S}$ (where $\cal S$ is the characteristic size of the source) the retarded coordinate time is written up to the leading contribution as $t_{\rm ret} \simeq t - r = t_{\rm r}$. Accordingly, the field in the radiation zone (wave zone) is given by the integral
\begin{align}
\psi_{ij} & = \frac{1}{2 \pi^2 M_5^3} \int_{0}^{\infty} d\mu \, \rho(\mu) \int d^3x' \int_{-\infty}^{t_{\rm r}} dt' \frac{1}{t_{\rm r}-t'+r} J_0\!\left( \mu \sqrt{(t_{\rm r} - t')^2 + 2r(t_{\rm r} - t')} \right) \nonumber \\ & \times \left( \frac{d}{dt'} {T_{\rm m}}_{ij}(x') + \frac{1}{t_{\rm r}-t'+r} {T_{\rm m}}_{ij}(x') \right).
\end{align}
Here, the leading contribution into the time integral comes from the region $t_{\rm r} - t' \ll r$, due to the presence of the Bessel function $J_0$ in the integrand. Thus, up to the leading contribution the field in the wave zone is written as
\begin{equation}
\psi_{ij} = \frac{1}{2 \pi^2 M_5^3 r} \int_{0}^{\infty} d\mu \, \rho(\mu) \int_{-\infty}^{t_{\rm r}} dt' J_0\!\left( \mu \sqrt{2r(t_{\rm r} - t')} \right) \frac{d}{dt'} \int d^3x' \, {T_{\rm m}}_{ij}(x').
\end{equation}
Note that here the integrals over time and spatial coordinates have decoupled.

It is convenient to rewrite the field source in a different form. Due to the conservation of the brane localised matter's energy-momentum tensor $\partial^\alpha {T_{\rm m}}_{\alpha\beta} = 0$, we rewrite it as \cite{maggiore2008}
\begin{equation}
\int d^3x' \, {T_{\rm m}}_{ij}(x') = \frac{1}{2} \frac{d^2}{dt'^2} \int d^3x' \, {T_{\rm m}}_{00}(x') x'_i x'_j = \frac{1}{2} \ddot{M}_{ij}(t').
\end{equation}
As a result, the field in the wave zone takes the form
\begin{equation}
\psi_{ij} = \frac{1}{4 \pi^2 M_5^3 r} \int_{0}^{\infty} d\mu \, \rho(\mu) \int_{-\infty}^{t_{\rm r}} dt' \, \dddot{M}_{ij}(t') J_0\!\left( \mu \sqrt{2r(t_{\rm r} - t')} \right).
\end{equation}
Finally, as the projectors extracting from the field $\psi_{ij}$ the dDOF of tensor field are traceless in both pairs of indices, we can equivalently replace in the expressions for $\bar{h}_{ij}^{\rm lt}$ and $\bar{h}_{ij}^{\rm tt}$ the second mass moment of the source $M_{ij}$ with its quadrupole moment
\begin{align}
\label{eq:DGP_grav_tens_in_rad_zone}
& \bar{h}_{ij}^{\rm lt/tt} = \frac{1}{4 \pi^2 M_5^3 r} \int_{0}^{\infty} d\mu \, \rho(\mu) \int_{-\infty}^{t_{\rm r}} dt' \, \dddot{Q}_{ij}^{\rm lt/tt}(t') J_0\!\left( \mu \sqrt{2r(t_{\rm r} - t')} \right), \\
& Q_{ij} = M_{ij} - \frac{1}{3} \delta_{ij} M_{kk}, \quad Q_{ij}^{\rm lt/tt} = \Lambda_{ij,kl}^{\rm lt/tt} Q_{kl}.
\end{align}

Extracting the emitted parts of the fields derivatives, we take into account that diffe\-ren\-tia\-tion of the projectors increases the asymptotics of the fields in inverse powers of the spatial distance $r$
\begin{equation}
\partial_i n_j \propto r^{-1}.
\end{equation}
Therefore, the corresponding terms do not contribute into the long-range part of the fields derivatives. As a result, eliminating the local terms in the fields derivatives, resulting from the differentiation of the upper integration limit in the time integral, by use of the integration by parts with the Eq. \eqref{eq:ret_mass_field_by_parts_aux_rel}, we arrive at the emitted parts of the derivatives of the fields $\bar{h}_{ij}^{\rm lt}$ and $\bar{h}_{ij}^{\rm tt}$ in the form
\begin{align}
\label{eq:DGP_grav_dyn_tens_rad}
& \partial_\mu \bar{h}_{ij}^{\rm lt/tt} = - \frac{c_\mu}{4 \pi^2 M_5^3 r} \mathbb{Q}_{ij}^{\rm lt/tt}(t_{\rm r}, r), \quad c^\mu = \lbrace 1, \mathbf{n} \rbrace, \\
& \mathbb{Q}_{ij}^{\rm lt/tt} = \int_{0}^{\infty} d\mu \, \rho(\mu) \int_{-\infty}^{t_{\rm r}} dt' \, \ddddot{Q}_{ij}^{\rm lt/tt}(t') J_0\!\left( \mu \sqrt{2r(t_{\rm r} - t')} \right).
\end{align}
Analogous expression for the effective field on the brane in the wave zone has been obtained in Ref. \cite{Khlopunov:2022jaw} within the scalar field analog of the DGP model by using the Rohrlich-Teitelboim approach to radiation \cite{Rohrlich1961,Teitelboim1970} (see, also, \cite{Kosyakov:1992qx,Kosyakov1999,Galtsov:2020hhn}). Note that, in accordance with the Huygens principle violation in the five-dimensional bulk of DGP model \cite{hadamard2014lectures,courant2008methods,Ivanenko_book}, the radiation field depends on the entire history of source motion preceding the retarded time.

\subsection{Scalar field in the wave zone}

For the scalar field $\varphi$ all the calculations are similar to those presented above, so here we only briefly describe the main steps.

From the equation of motion \eqref{eq:DGP_phi_EoM}, given the Eq. \eqref{eq:ret_mass_field_by_parts_aux_rel}, we find that the scalar field in the wave zone in the non-relativistic approximation is given by the integral
\begin{equation}
\varphi = - \frac{1}{12 \pi^2 M_5^3 r} \int_{0}^{\infty} d\mu \, \rho(\mu) \int_{-\infty}^{t_{\rm r}} dt' \, J_0\!\left( \mu \sqrt{2r(t_{\rm r}-t')} \right) \frac{d}{dt'} \int d^3x' \, T_{\rm m}(x').
\end{equation}
Given the conservation of the energy-momentum tensor of matter on the brane, we rewrite the spatial integral of its trace as \cite{maggiore2008}
\begin{equation}
\int d^3x' \, T_{\rm m}(x') = - M(t') + \frac{1}{2} \ddot{M}_{kk}(t'), \quad M(t') = \int d^3x' \, {T_{\rm m}}_{00}(x').
\end{equation}
As a result, given the conservation of the matter EMT, the scalar field in the wave zone is given by \cite{maggiore2008}
\begin{equation}
\varphi = - \frac{1}{24 \pi^2 M_5^3 r} \int_{0}^{\infty} d\mu \, \rho(\mu) \int_{-\infty}^{t_{\rm r}} dt' \, \dddot{M}_{kk}(t') J_0\!\left( \mu \sqrt{2r(t_{\rm r}-t')} \right).
\end{equation}
Finally, eliminating the local terms by integration by parts using the Eq. \eqref{eq:ret_mass_field_by_parts_aux_rel}, we extract the long-range part of the scalar field derivative in the form analogous to the emitted part of the tensor field \eqref{eq:DGP_grav_dyn_tens_rad}
\begin{align}
\label{eq:DGP_grav_dyn_sc_rad}
& \partial_\mu \varphi = \frac{c_\mu}{24 \pi^2 M_5^3 r} \mathbb{M}(t_{\rm r}, r), \quad c^\mu = \lbrace 1, \mathbf{n} \rbrace, \\
& \mathbb{M} = \int_{0}^{\infty} d\mu \, \rho(\mu) \int_{-\infty}^{t_{\rm r}} dt' \, \ddddot{M}_{kk}(t') J_0\!\left(\mu \sqrt{2r(t_{\rm r}-t')} \right).
\end{align}

\subsection{Quadrupole formula in DGP gravity}

Having extracted the emitted parts of dDOF of DGP gravity, we now calculate the effective gravitational radiation power of an arbitrary non-relativistic source on the brane.

Substituting the Eqs. \eqref{eq:DGP_grav_dyn_tens_rad} and \eqref{eq:DGP_grav_dyn_sc_rad} into the effective EMT of dDOF of DGP gravity \eqref{eq:DGP_grav_eff_EMT_dyn}, we find the effective gravitational radiation energy flux density on the brane as
\begin{equation}
{T_{\rm d}}^{0k} = \frac{M_4^2}{64 \pi^4 M_5^6 r^2} n^k \left( \frac{1}{6} \mathbb{M}^2 + \mathbb{Q}_{ij}^{\rm lt} \mathbb{Q}_{ij}^{\rm lt} + \mathbb{Q}_{ij}^{\rm tt} \mathbb{Q}_{ij}^{\rm tt} \right).
\end{equation}
Thus, integrating the radiation energy flux over the distant 2-sphere on the brane enclosing the source of the gravitational field, we obtain the analog of the quadrupole formula for the angular distribution of the effective gravitational radiation power of an arbitrary non-relativistic source on the brane
\begin{equation}
\label{eq:DGP_quadrupole_form}
\frac{dW_{\rm DGP}}{d\Omega_2} = \frac{M_4^2}{64 \pi^4 M_5^6} \left( \frac{1}{6} \mathbb{M}^2 + \mathbb{Q}_{ij}^{\rm lt} \mathbb{Q}_{ij}^{\rm lt} + \mathbb{Q}_{ij}^{\rm tt} \mathbb{Q}_{ij}^{\rm tt} \right).
\end{equation}
Here, the contributions of three additional dDOF of DGP graviton $\varphi$ and $\bar{h}_{ij}^{\rm lt}$, corresponding to three additional polarizations of gravitational waves in DGP model, are written out explicitly. In turn, the term $\propto \mathbb{Q}_{ij}^{\rm tt} \mathbb{Q}_{ij}^{\rm tt}$ is the analog of the standard quadrupole formula for the gra\-vi\-ta\-tio\-nal radiation power in GR \cite{maggiore2008}.

Note that the integrals over the history of source motion in the long-range parts of the gravitational field \eqref{eq:DGP_grav_dyn_tens_rad} and \eqref{eq:DGP_grav_dyn_sc_rad} contain the same damping factor $J_0 (\mu \sqrt{2r(t_{\rm r}-t')})$ as the long-range part of the scalar field analog of DGP model \cite{Khlopunov:2022jaw}. Therefore, the estimates for the intensity of leakage of gravitational waves into the extra dimension obtained in Ref. \cite{Khlopunov:2022jaw} within the scalar-field analog of DGP model are still valid in the case of DGP model of gravity.

\subsection{Estimating the parameters of Deffayet-Menou formula}

Let us estimate the parameter $n$ from the Deffayet-Menou formula \eqref{eq:DGP_Deff_men_form} \cite{Deffayet:2007kf}, characterizing the intensity of leakage of gravitational waves into the extra dimension, by use of the Eq. \eqref{eq:DGP_grav_tens_in_rad_zone} for the gravitational field in the wave zone.

As we consider the flat brane embedded into the five-dimensional Minkowski space as a background solution, we replace the luminosity distance $d_L$ in the Deffayet-Menou formula \eqref{eq:DGP_Deff_men_form} by the spatial distance $r$
\begin{equation}
\label{eq:DGP_Deff_men_form_in_Mink}
\bar{h} \propto \frac{1}{r ( 1 + \left( r/R_c \right)^{n/2} )^{1/n}}.
\end{equation}
To estimate the parameter $n$, characterizing the steepness of transition of the gravitational field asymptotics from the four-dimensional to the five-dimensional regime, we expand the Eq. \eqref{eq:DGP_Deff_men_form_in_Mink} in the region $r \ll R_c$ up to the leading order contribution as
\begin{equation}
\label{eq:Deffayet_Menou_form_decomp}
\bar{h} = \frac{1}{r} \left( 1 - \frac{1}{n} (r/R_c)^{n/2} + {\cal O}\left( (r/R_c)^n \right) \right).
\end{equation}
We obtain a similar expansion for the Eq. \eqref{eq:DGP_grav_tens_in_rad_zone} and determine from it the parameter $n$.

We assume that we are in the intermediate region ${\cal R} \ll r \ll R_c$ and neglect the tensor structure of the gravitational field \eqref{eq:DGP_grav_tens_in_rad_zone}
\begin{equation}
\label{eq:DGP_grav_field_in_wave_zone_aux}
\bar{h} \propto \frac{1}{r} \int_{0}^{\infty} d\mu \, \rho(\mu) \int_{-\infty}^{t_{\rm r}} dt' \, \dddot{Q}(t') J_0\!\left( \mu \sqrt{2r(t_{\rm r} - t')} \right).
\end{equation}
For simplicity, we consider the periodic source $Q(t') \propto \sin \omega t'$. Thus, after the change of integration variable $s = t_{\rm r} - t'$, the integral over the history of motion in the Eq. \eqref{eq:DGP_grav_field_in_wave_zone_aux} is calculated as \cite{zwillinger2014table}
\begin{equation}
\int_{-\infty}^{t_{\rm r}} dt' \, \dddot{Q}(t') J_0\!\left( \mu \sqrt{2r(t_{\rm r} - t')} \right) \propto \int_{0}^{\infty} ds \, \sin(\omega s) J_0\!\left(\mu \sqrt{2rs}\right) \propto \cos \frac{r \mu^2}{2\omega}.
\end{equation}
Here we have omitted the factors depending on the frequency of the source motion $\omega$ and its other characteristics, being interested only in the dependence of the gravitational field on the distance from the source. Calculating the integral over the Kaluza-Klein masses in the Eq. \eqref{eq:DGP_grav_field_in_wave_zone_aux} \cite{zwillinger2014table}, we arrive at
\begin{equation}
\label{eq:DGP_grav_field_in_wave_zone_aux_1}
\bar{h} \propto \frac{1}{r} \left \lbrack \cos x^2 - \sqrt{2} \cos \left( x^2 + \frac{\pi}{4} \right) C(x) - \sqrt{2} \sin \left( x^2 + \frac{\pi}{4} \right) S(x) \right \rbrack, \quad x = \sqrt{\frac{rM_c^2}{2\omega}},
\end{equation}
where $C(x)$ and $S(x)$ are the Fresnel cosine and sine integrals, correspondingly \cite{zwillinger2014table}. Introducing the dimensionless distance $\bar{r} = r/R_c$ and frequency $\bar{\omega} = \omega / M_c$, we expand the Eq. \eqref{eq:DGP_grav_field_in_wave_zone_aux_1} in power of $\bar{r} \ll 1$ up to the leading order \cite{zwillinger2014table}
\begin{equation}
\label{eq:DGP_grav_field_in_wave_zone_aux_2}
\bar{h} \propto \frac{1}{r} \left( 1 - \sqrt{\frac{\bar{r}}{2\bar{\omega}}} + {\cal O}(\bar{r}) \right).
\end{equation}

Comparing the Eqs. \eqref{eq:Deffayet_Menou_form_decomp} and \eqref{eq:DGP_grav_field_in_wave_zone_aux_2}, we find that the transition steepness parameter $n$ from the Deffayet-Menou formula \eqref{eq:DGP_Deff_men_form} should have the value $n=1$. Moreover, the crossover radius of DGP model $R_c$ should be replaced by the effective crossover radius
\begin{equation}
R_c \quad \Longrightarrow \quad R_{\rm eff} = 2 \bar{\omega} R_c,
\end{equation}
In accordance with the results of Ref. \cite{Khlopunov:2022jaw} and the infrared transparency of the bulk in DGP model \cite{Dvali:2000xg,Brown:2016gwv}, the obtained effective crossover radius $R_{\rm eff}$, beyond which the leakage of gravitational waves into the extra dimension manifests, is determined not only by the parameters of DGP model, but also by the frequency of gravitational-wave signal. In particular, in the sensitivity range of modern and future gravitational-wave observatories \cite{Moore:2014lga} the effective crossover radius $R_{\rm eff}$ is many orders of magnitude larger than the crossover radius $R_c$
\begin{equation}
\omega \sim 10^2 \ \text{Hz}, \quad M_c \sim 10^{-42} \ \text{GeV} \quad \Longrightarrow \quad R_{\rm eff} \sim 10^{20} R_c,
\end{equation}
significantly obscuring the potential experimental detection of the effect of leakage of gra\-vi\-ta\-tio\-nal waves into the extra dimension \cite{Pardo:2018ipy,Corman:2020pyr,Corman:2021avn,Khlopunov:2022jaw}.

Thus, based on the expansion of gravitational field in the wave zone \eqref{eq:DGP_grav_field_in_wave_zone_aux_2}, we conjecture that the Deffayet-Menou formula \eqref{eq:DGP_Deff_men_form} should be modified as
\begin{equation}
h \propto \frac{1}{r(1+\sqrt{r/R_{\rm eff}})}, \quad R_{\rm eff} = 2 \bar{\omega} R_c, \quad \bar{\omega} = \omega/M_c
\end{equation}
for the case of gravitational waves propagation along the flat brane embedded into the five-dimensional Minkowski bulk.

\section{Conclusions}\label{VI}

The aim of this paper was to study the process of gravitational radiation in the DGP gravity model within the framework of classical field theory. Considering small perturbations of the gravitational field on the background of a flat tensionless brane embedded in the five-dimensional Minkowski bulk, we obtained an analogue of the quadrupole formula for the effective power of gravitational radiation of an arbitrary non-relativistic source on the brane \eqref{eq:DGP_quadrupole_form}. Also, based on the obtained expression for the gravitational field on the brane in the wave zone \eqref{eq:DGP_grav_dyn_tens_rad} and \eqref{eq:DGP_grav_dyn_sc_rad}, we estimated the parameters of the Deffayet-Menou formula \eqref{eq:DGP_Deff_men_form} \cite{Deffayet:2007kf}. It is shown that the parameter $n$, which characterizes the steepness of the transition of the asymptotics of the gravitational field from the four-dimensional to the five-dimensional regime, has the value $n=1$, while the crossover radius of the DGP model $R_c$ should be replaced in this formula by the effective crossover radius $R_{\rm eff} = 2 \bar{\omega} R_c$ (where $\bar{\omega} = \omega / M_c$), depending on the frequency of the gravitational-wave signal $\omega$.

Since the effective quadratic action of DGP gravity on the brane \eqref{eq:DGP_grav_eff_act} is a non-local theory, we generalized Noether's method for constructing the stress-energy tensor to the case of such non-local theories. As a result, it was shown that the effective stress-energy tensor of DGP gravity on the brane \eqref{eq:DGP_grav_EMT_eff} can be calculated from the effective non-local action \eqref{eq:DGP_grav_eff_act} using the standard formula for the canonical stress-energy tensor \cite{Landau:1975pou,Jackson:1998nia,Stepanyanz_book}, neglecting non-local mass terms in the action. The proposed method for constructing the effective energy-momentum tensor was additionally verified within the scalar-field analogue of the DGP model by calculating the brane-localized part of its full five-dimensional Hamiltonian \eqref{eq:DGP_sc_Hamilt_local}. Based on the proposed procedure, the effective energy-momentum tensor of the dynamical degrees of freedom of DGP gravity \eqref{eq:DGP_grav_eff_EMT_dyn} was constructed, determining the effective energy flux of gravitational radiation on the brane.

The problem of gravitational radiation in the DGP model was recently discussed in \cite{Poddar:2021yjd} using the effective field theory approach to radiation problems \cite{Cardoso:2008gn,Porto:2016pyg}. However, this analysis, based on momentum space calculations, does not provide information about the structure of the gravitational field in the wave zone and the role of the metastable nature of the effective graviton on the brane in the formation and leakage of gravitational radiation into the extra dimension. Our approach is aimed precisely at these features of gravitational radiation in the DGP model and obtaining explicit predictions for future GW experiments. In particular, we tested the constraints on the intensity of gravitational wave leakage into the extra dimension and the possibility of its detection by modern and future gravitational wave observatories obtained in \cite{Khlopunov:2022jaw} within the framework of the scalar-field analogue of the DGP model.

\section*{Acknowledgements}

The work of M. Kh. is supported by the Foundation for the Advancement of Theoretical Physics and Mathematics “BASIS” grant No. 20-2-10-8-1.

\appendix

\section{Gravitational radiation of oscillating mass}\label{A}

In GR, the linear equation of motion of the gravitational field in the Lorentz gauge on the Minkowski space background has the form \cite{maggiore2008}
\begin{equation}
{_4}\square \bar{h}_{\mu\nu} = - 16 \pi G {T_{\rm m}}_{\mu\nu}, \quad \bar{h}_{\mu\nu} = h_{\mu\nu} - \frac{1}{2} \eta_{\mu\nu} h, \quad \partial^\mu \bar{h}_{\mu\nu} = 0,
\end{equation}
while the Isaacson EMT \eqref{eq:4D_GR_Isaac_EMT} is written in terms of the reduced metric perturbations as
\begin{equation}
\label{eq:4D_GR_Isaac_EMT_in_reduce_pert}
\bar{h} = - h \quad \Longrightarrow \quad t_{\mu\nu} = \frac{1}{32 \pi G} \left \langle \partial_\mu \bar{h}^{\alpha\beta} \partial_\nu \bar{h}_{\alpha\beta} - \frac{1}{2} \partial_\mu \bar{h} \partial_\nu \bar{h} \right \rangle.
\end{equation}

The gravitational field of the particle of mass $m$ moving along a fixed worldline $z^\mu(\tau)$ parameterized by its proper time is found as (see, e.g., \cite{Kosyakov:1992qx})
\begin{equation}
\label{eq:4D_GR_pp_field}
{T_{\rm m}}_{\mu\nu} = m \int d\tau \, v_\mu v_\nu \, \delta^{(4)}(x-z), \quad v^\mu = \frac{d z^\mu}{d \tau} \quad \Longrightarrow \quad \bar{h}_{\mu\nu} = \frac{4 G m}{\hat{\rho}} \hat{v}_\mu \hat{v}_\nu, \quad \hat{\rho} = \hat{v}^\alpha (x_\alpha - \hat{z}_\alpha),
\end{equation}
where the retarded proper time $\hat{\tau}$ is defined as the solution of the equation
\begin{equation}
(x_\alpha - \hat{z}_\alpha )^2 = 0, \quad x^0 \geq \hat{z}^0,
\end{equation}
and all the hatted quantities correspond to this moment of proper time. Note also that the Lorentz-invariant distance $\hat{\rho}$ coincides with the spatial distance from the observation point to the particle in the Lorentz frame comoving with the particle at the time $\hat{\tau}$ and is equivalent to it far away from the particle \cite{Kosyakov:1992qx}. Thus, the long-range part of the gravitational field derivative, which determines the gravitational radiation energy flux through the distant 2-surface enclosing the particle is found as \cite{Kosyakov:1992qx}
\begin{equation}
\partial_\mu \bar{h}_{\alpha\beta} = - \frac{4 G m}{\hat{\rho}} \hat{c}_\mu \left \lbrack \hat{a}_\alpha \hat{v}_\beta + \hat{v}_\alpha \hat{a}_\beta + (\hat{a} \hat{c}) \hat{v}_\alpha \hat{v}_\beta \right \rbrack, \quad a^\mu = \frac{d^2 z^\mu}{d\tau^2}, \quad x^\mu - \hat{z}^\mu = \hat{\rho} \hat{c}^\mu, \quad \hat{c}_\mu^2 = 0.
\end{equation}
In the non-relativistic limit $|\mathbf{v}| \ll 1$, at the distance from the particle's region of motion $|\mathbf{x}| \gg |\mathbf{z}|$ the components of the emitted part of the gravitational field are written as
\begin{align}
\label{eq:decomp_aux_1}
& \partial_\mu \bar{h}_{00} = - \frac{4 G m}{r} c_\mu n_k \! \left. a_k \right \vert_{t_{\rm r}}, \quad c^\mu = \left \lbrace 1, n^i \right \rbrace, \quad n^i = x^i/r, \quad t_{\rm r} = t - r, \\
\label{eq:decomp_aux_2}
& \partial_\mu \bar{h}_{0i} = \frac{4 G m}{r} c_\mu \! \left. a_i \right \vert_{t_{\rm r}}, \quad \partial_\mu \bar{h}_{ij} = - \frac{4 G m}{r} c_\mu \! \left. \left( \bar{a}_i \bar{v}_j + \bar{v}_i \bar{a}_j \right) \right \vert_{t_{\rm r}},
\end{align}
Note that in the non-relativistic limit the different components of the gravitational field have different orders of smallness.

Substituting the expansions (\ref{eq:decomp_aux_1}--\ref{eq:decomp_aux_2}) into the Eq. \eqref{eq:4D_GR_Isaac_EMT_in_reduce_pert}, in the non-relativistic limit we find the following expression for the gravitational radiation energy flux of the particle
\begin{equation}
\label{eq:4D_GR_wrong_grav_rad}
\frac{dW}{d\Omega_2} = t^{0i} n^i r^2 = \frac{G m^2}{2 \pi} \! \left. \left \langle \frac{1}{2} (\mathbf{n} \bar{\mathbf{a}})^2 - 2 \bar{\mathbf{a}}^2 \right \rangle \right \vert_{t_{\rm r}}.
\end{equation}
Note that we have obtained the contribution of a lower order of smallness compared to the quadrupole formula \cite{maggiore2008}. This is due to the Lorentz gauge preserving the ndDOF of gravitational field, which provide this contribution into the gravitational radiation energy flux. In particular, the gravitational radiation power of an oscillating point mass $\mathbf{z}(t) = \left \lbrace 0, 0, R_0 \cos \omega_0 t \right \rbrace$, calculated by use of the obtained \textit{incorrect} formula \eqref{eq:4D_GR_wrong_grav_rad}, turns out to be negative
\begin{equation}
W = - \frac{11}{6} G m^2 R_0^2 \omega_0^4 < 0,
\end{equation}
due to the contribution of the ndDOF of gravitational field. Thus, to correctly calculate the gravitational radiation without fixing the transverse-traceless (unitary) gauge, one has to separately construct the EMT of the dDOF of gravitational field.

\bibliographystyle{JHEP}
\bibliography{main}

\providecommand{\href}[2]{#2}\begingroup\raggedright\begin{thebibliography}{10}

\bibitem{Clifton:2011jh}
T.~Clifton, P.G.~Ferreira, A.~Padilla and C.~Skordis, \emph{{Modified Gravity
  and Cosmology}},
  \href{https://doi.org/10.1016/j.physrep.2012.01.001}{\emph{Phys. Rept.}
  {\bfseries 513} (2012) 1} [\href{https://arxiv.org/abs/1106.2476}{{\ttfamily
  1106.2476}}].

\bibitem{Li:2020uaz}
B.~Li and K.~Koyama, \emph{{Modified Gravity}}, WSP (2020),
  \href{https://doi.org/10.1142/11090}{10.1142/11090}.

\bibitem{Green:1987sp}
M.B.~Green, J.H.~Schwarz and E.~Witten, \emph{{Superstring Theory. Vol. 1:
  Introduction}}, Cambridge Monographs on Mathematical Physics (7, 1988).

\bibitem{Rubakov:2001kp}
V.A.~Rubakov, \emph{{Large and infinite extra dimensions: An Introduction}},
  \href{https://doi.org/10.1070/PU2001v044n09ABEH001000}{\emph{Phys. Usp.}
  {\bfseries 44} (2001) 871}
  [\href{https://arxiv.org/abs/hep-ph/0104152}{{\ttfamily hep-ph/0104152}}].

\bibitem{Barvinsky:2005ak}
A.O.~Barvinsky, \emph{{Cosmological branes and macroscopic extra dimensions}},
  \href{https://doi.org/10.1070/PU2005v048n06ABEH002618}{\emph{Phys. Usp.}
  {\bfseries 48} (2005) 545}.

\bibitem{Cheng:2010pt}
H.-C.~Cheng, \emph{{Introduction to Extra Dimensions}},  in \emph{{Theoretical
  Advanced Study Institute in Elementary Particle Physics}: {Physics of the
  Large and the Small}}, pp.~125--162, 2011,
  \href{https://doi.org/10.1142/9789814327183_0003}{DOI}
  [\href{https://arxiv.org/abs/1003.1162}{{\ttfamily 1003.1162}}].

\bibitem{Maartens:2010ar}
R.~Maartens and K.~Koyama, \emph{{Brane-World Gravity}},
  \href{https://doi.org/10.12942/lrr-2010-5}{\emph{Living Rev. Rel.} {\bfseries
  13} (2010) 5} [\href{https://arxiv.org/abs/1004.3962}{{\ttfamily
  1004.3962}}].

\bibitem{Arkani-Hamed:1998jmv}
N.~Arkani-Hamed, S.~Dimopoulos and G.R.~Dvali, \emph{{The Hierarchy problem and
  new dimensions at a millimeter}},
  \href{https://doi.org/10.1016/S0370-2693(98)00466-3}{\emph{Phys. Lett. B}
  {\bfseries 429} (1998) 263}
  [\href{https://arxiv.org/abs/hep-ph/9803315}{{\ttfamily hep-ph/9803315}}].

\bibitem{Randall:1999ee}
L.~Randall and R.~Sundrum, \emph{{A Large mass hierarchy from a small extra
  dimension}}, \href{https://doi.org/10.1103/PhysRevLett.83.3370}{\emph{Phys.
  Rev. Lett.} {\bfseries 83} (1999) 3370}
  [\href{https://arxiv.org/abs/hep-ph/9905221}{{\ttfamily hep-ph/9905221}}].

\bibitem{Randall:1999vf}
L.~Randall and R.~Sundrum, \emph{{An Alternative to compactification}},
  \href{https://doi.org/10.1103/PhysRevLett.83.4690}{\emph{Phys. Rev. Lett.}
  {\bfseries 83} (1999) 4690}
  [\href{https://arxiv.org/abs/hep-th/9906064}{{\ttfamily hep-th/9906064}}].

\bibitem{Dvali:2000hr}
G.R.~Dvali, G.~Gabadadze and M.~Porrati, \emph{{4-D gravity on a brane in 5-D
  Minkowski space}},
  \href{https://doi.org/10.1016/S0370-2693(00)00669-9}{\emph{Phys. Lett. B}
  {\bfseries 485} (2000) 208}
  [\href{https://arxiv.org/abs/hep-th/0005016}{{\ttfamily hep-th/0005016}}].

\bibitem{Dvali:2000xg}
G.R.~Dvali and G.~Gabadadze, \emph{{Gravity on a brane in infinite volume extra
  space}}, \href{https://doi.org/10.1103/PhysRevD.63.065007}{\emph{Phys. Rev.
  D} {\bfseries 63} (2001) 065007}
  [\href{https://arxiv.org/abs/hep-th/0008054}{{\ttfamily hep-th/0008054}}].

\bibitem{Deffayet:2000uy}
C.~Deffayet, \emph{{Cosmology on a brane in Minkowski bulk}},
  \href{https://doi.org/10.1016/S0370-2693(01)00160-5}{\emph{Phys. Lett. B}
  {\bfseries 502} (2001) 199}
  [\href{https://arxiv.org/abs/hep-th/0010186}{{\ttfamily hep-th/0010186}}].

\bibitem{Deffayet:2001pu}
C.~Deffayet, G.R.~Dvali and G.~Gabadadze, \emph{{Accelerated universe from
  gravity leaking to extra dimensions}},
  \href{https://doi.org/10.1103/PhysRevD.65.044023}{\emph{Phys. Rev. D}
  {\bfseries 65} (2002) 044023}
  [\href{https://arxiv.org/abs/astro-ph/0105068}{{\ttfamily
  astro-ph/0105068}}].

\bibitem{Deffayet:2002sp}
C.~Deffayet, S.J.~Landau, J.~Raux, M.~Zaldarriaga and P.~Astier,
  \emph{{Supernovae, CMB, and gravitational leakage into extra dimensions}},
  \href{https://doi.org/10.1103/PhysRevD.66.024019}{\emph{Phys. Rev. D}
  {\bfseries 66} (2002) 024019}
  [\href{https://arxiv.org/abs/astro-ph/0201164}{{\ttfamily
  astro-ph/0201164}}].

\bibitem{Duff:1986hr}
M.J.~Duff, B.E.W.~Nilsson and C.N.~Pope, \emph{{Kaluza-Klein Supergravity}},
  \href{https://doi.org/10.1016/0370-1573(86)90163-8}{\emph{Phys. Rept.}
  {\bfseries 130} (1986) 1}.

\bibitem{Arkani-Hamed:1998sfv}
N.~Arkani-Hamed, S.~Dimopoulos and G.R.~Dvali, \emph{{Phenomenology,
  astrophysics and cosmology of theories with submillimeter dimensions and TeV
  scale quantum gravity}},
  \href{https://doi.org/10.1103/PhysRevD.59.086004}{\emph{Phys. Rev. D}
  {\bfseries 59} (1999) 086004}
  [\href{https://arxiv.org/abs/hep-ph/9807344}{{\ttfamily hep-ph/9807344}}].

\bibitem{Ezquiaga:2018btd}
J.M.~Ezquiaga and M.~Zumalac\'arregui, \emph{{Dark Energy in light of
  Multi-Messenger Gravitational-Wave astronomy}},
  \href{https://doi.org/10.3389/fspas.2018.00044}{\emph{Front. Astron. Space
  Sci.} {\bfseries 5} (2018) 44}
  [\href{https://arxiv.org/abs/1807.09241}{{\ttfamily 1807.09241}}].

\bibitem{Yu:2019jlb}
H.~Yu, Z.-C.~Lin and Y.-X.~Liu, \emph{{Gravitational waves and extra
  dimensions: a short review}},
  \href{https://doi.org/10.1088/0253-6102/71/8/991}{\emph{Commun. Theor. Phys.}
  {\bfseries 71} (2019) 991}
  [\href{https://arxiv.org/abs/1905.10614}{{\ttfamily 1905.10614}}].

\bibitem{Andriot2017}
D.~Andriot and G.~Lucena~G{\'o}mez, \emph{{Signatures of extra dimensions in
  gravitational waves}}, \href{https://doi.org/10.1088/1475-7516/2019/05/E01,
  10.1088/1475-7516/2017/06/048}{\emph{JCAP} {\bfseries 1706} (2017) 048}
  [\href{https://arxiv.org/abs/1704.07392}{{\ttfamily 1704.07392}}].

\bibitem{Chu:2021uea}
Y.-Z.~Chu, \emph{{Electromagnetic and gravitational radiation in all
  dimensions: A classical field theory treatment}},
  \href{https://doi.org/10.1103/PhysRevD.104.084074}{\emph{Phys. Rev. D}
  {\bfseries 104} (2021) 084074}
  [\href{https://arxiv.org/abs/2107.14744}{{\ttfamily 2107.14744}}].

\bibitem{Khlopunov:2022ubp}
M.~Khlopunov and D.V.~Gal'tsov, \emph{{Gravitational radiation from a binary
  system in odd-dimensional spacetime}},
  \href{https://doi.org/10.1088/1475-7516/2022/04/014}{\emph{JCAP} {\bfseries
  04} (2022) 014} [\href{https://arxiv.org/abs/2201.11804}{{\ttfamily
  2201.11804}}].

\bibitem{Yu:2016tar}
H.~Yu, B.-M.~Gu, F.P.~Huang, Y.-Q.~Wang, X.-H.~Meng and Y.-X.~Liu,
  \emph{{Probing extra dimension through gravitational wave observations of
  compact binaries and their electromagnetic counterparts}},
  \href{https://doi.org/10.1088/1475-7516/2017/02/039}{\emph{JCAP} {\bfseries
  02} (2017) 039} [\href{https://arxiv.org/abs/1607.03388}{{\ttfamily
  1607.03388}}].

\bibitem{Visinelli:2017bny}
L.~Visinelli, N.~Bolis and S.~Vagnozzi, \emph{{Brane-world extra dimensions in
  light of GW170817}},
  \href{https://doi.org/10.1103/PhysRevD.97.064039}{\emph{Phys. Rev.}
  {\bfseries D97} (2018) 064039}
  [\href{https://arxiv.org/abs/1711.06628}{{\ttfamily 1711.06628}}].

\bibitem{Lin:2020wnp}
Z.-C.~Lin, H.~Yu and Y.-X.~Liu, \emph{{Constraint on the radius of
  five-dimensional dS spacetime with GW170817 and GRB 170817A}},
  \href{https://doi.org/10.1103/PhysRevD.101.104058}{\emph{Phys. Rev. D}
  {\bfseries 101} (2020) 104058}
  [\href{https://arxiv.org/abs/2001.06581}{{\ttfamily 2001.06581}}].

\bibitem{Lin:2022hus}
Z.-C.~Lin, H.~Yu and Y.-X.~Liu, \emph{{Shortcut in codimension-2 brane
  cosmology in light of GW170817}},
  \href{https://doi.org/10.1140/epjc/s10052-023-11328-x}{\emph{Eur. Phys. J. C}
  {\bfseries 83} (2023) 190}
  [\href{https://arxiv.org/abs/2202.04866}{{\ttfamily 2202.04866}}].

\bibitem{Shiromizu:1999wj}
T.~Shiromizu, K.-i.~Maeda and M.~Sasaki, \emph{{The Einstein equation on the
  3-brane world}},
  \href{https://doi.org/10.1103/PhysRevD.62.024012}{\emph{Phys. Rev. D}
  {\bfseries 62} (2000) 024012}
  [\href{https://arxiv.org/abs/gr-qc/9910076}{{\ttfamily gr-qc/9910076}}].

\bibitem{Maeda:2003ar}
K.-i.~Maeda, S.~Mizuno and T.~Torii, \emph{{Effective gravitational equations
  on brane world with induced gravity}},
  \href{https://doi.org/10.1103/PhysRevD.68.024033}{\emph{Phys. Rev. D}
  {\bfseries 68} (2003) 024033}
  [\href{https://arxiv.org/abs/gr-qc/0303039}{{\ttfamily gr-qc/0303039}}].

\bibitem{Barvinsky:2003jf}
A.O.~Barvinsky and S.N.~Solodukhin, \emph{{Echoing the extra dimension}},
  \href{https://doi.org/10.1016/j.nuclphysb.2003.10.011}{\emph{Nucl. Phys.}
  {\bfseries B675} (2003) 159}
  [\href{https://arxiv.org/abs/hep-th/0307011}{{\ttfamily hep-th/0307011}}].

\bibitem{Khlopunov:2023hnl}
M.~Khlopunov, \emph{{Non-local tails in radiation in odd dimensions}},
  \href{https://doi.org/10.1088/1475-7516/2023/10/019}{\emph{JCAP} {\bfseries
  10} (2023) 019} [\href{https://arxiv.org/abs/2308.08997}{{\ttfamily
  2308.08997}}].

\bibitem{Chakraborty:2017qve}
S.~Chakraborty, K.~Chakravarti, S.~Bose and S.~SenGupta, \emph{{Signatures of
  extra dimensions in gravitational waves from black hole quasinormal modes}},
  \href{https://doi.org/10.1103/PhysRevD.97.104053}{\emph{Phys. Rev. D}
  {\bfseries 97} (2018) 104053}
  [\href{https://arxiv.org/abs/1710.05188}{{\ttfamily 1710.05188}}].

\bibitem{Mishra:2021waw}
A.K.~Mishra, A.~Ghosh and S.~Chakraborty, \emph{{Constraining extra dimensions
  using observations of black hole quasi-normal modes}},
  \href{https://arxiv.org/abs/2106.05558}{{\ttfamily 2106.05558}}.

\bibitem{Chakravarti:2018vlt}
K.~Chakravarti, S.~Chakraborty, S.~Bose and S.~SenGupta, \emph{{Tidal Love
  numbers of black holes and neutron stars in the presence of higher
  dimensions: Implications of GW170817}},
  \href{https://doi.org/10.1103/PhysRevD.99.024036}{\emph{Phys. Rev. D}
  {\bfseries 99} (2019) 024036}
  [\href{https://arxiv.org/abs/1811.11364}{{\ttfamily 1811.11364}}].

\bibitem{Cardoso:2019vof}
V.~Cardoso, L.~Gualtieri and C.J.~Moore, \emph{{Gravitational waves and higher
  dimensions: Love numbers and Kaluza-Klein excitations}},
  \href{https://doi.org/10.1103/PhysRevD.100.124037}{\emph{Phys. Rev. D}
  {\bfseries 100} (2019) 124037}
  [\href{https://arxiv.org/abs/1910.09557}{{\ttfamily 1910.09557}}].

\bibitem{Chakravarti:2019aup}
K.~Chakravarti, S.~Chakraborty, K.S.~Phukon, S.~Bose and S.~SenGupta,
  \emph{{Constraining extra-spatial dimensions with observations of GW170817}},
  \href{https://doi.org/10.1088/1361-6382/ab8355}{\emph{Class. Quant. Grav.}
  {\bfseries 37} (2020) 105004}
  [\href{https://arxiv.org/abs/1903.10159}{{\ttfamily 1903.10159}}].

\bibitem{Gabadadze:2004dq}
G.~Gabadadze, \emph{{Looking at the cosmological constant from infinite-volume
  bulk}},  in \emph{{From Fields to Strings: Circumnavigating Theoretical
  Physics: A Conference in Tribute to Ian Kogan}}, pp.~1061--1130, 8, 2004,
  \href{https://doi.org/10.1142/9789812775344_0025}{DOI}
  [\href{https://arxiv.org/abs/hep-th/0408118}{{\ttfamily hep-th/0408118}}].

\bibitem{Charmousis:2006pn}
C.~Charmousis, R.~Gregory, N.~Kaloper and A.~Padilla, \emph{{DGP
  Specteroscopy}},
  \href{https://doi.org/10.1088/1126-6708/2006/10/066}{\emph{JHEP} {\bfseries
  10} (2006) 066} [\href{https://arxiv.org/abs/hep-th/0604086}{{\ttfamily
  hep-th/0604086}}].

\bibitem{Brown:2016gwv}
K.~Brown, H.~Mathur and M.~Verostek, \emph{{Exploring extra dimensions with
  scalar fields}}, \href{https://doi.org/10.1119/1.5024221}{\emph{Am. J. Phys.}
  {\bfseries 86} (2018) 327}
  [\href{https://arxiv.org/abs/1608.06547}{{\ttfamily 1608.06547}}].

\bibitem{Khlopunov:2022jaw}
M.~Khlopunov and D.V.~Gal'tsov, \emph{{Leakage of gravitational waves into an
  extra dimension in the DGP model}},
  \href{https://doi.org/10.1088/1475-7516/2022/10/062}{\emph{JCAP} {\bfseries
  10} (2022) 062} [\href{https://arxiv.org/abs/2209.04262}{{\ttfamily
  2209.04262}}].

\bibitem{Nicolis:2004qq}
A.~Nicolis and R.~Rattazzi, \emph{{Classical and quantum consistency of the DGP
  model}}, \href{https://doi.org/10.1088/1126-6708/2004/06/059}{\emph{JHEP}
  {\bfseries 06} (2004) 059}
  [\href{https://arxiv.org/abs/hep-th/0404159}{{\ttfamily hep-th/0404159}}].

\bibitem{Gorbunov:2005zk}
D.~Gorbunov, K.~Koyama and S.~Sibiryakov, \emph{{More on ghosts in DGP model}},
  \href{https://doi.org/10.1103/PhysRevD.73.044016}{\emph{Phys. Rev. D}
  {\bfseries 73} (2006) 044016}
  [\href{https://arxiv.org/abs/hep-th/0512097}{{\ttfamily hep-th/0512097}}].

\bibitem{Koyama:2007za}
K.~Koyama, \emph{{Ghosts in the self-accelerating universe}},
  \href{https://doi.org/10.1088/0264-9381/24/24/R01}{\emph{Class. Quant. Grav.}
  {\bfseries 24} (2007) R231}
  [\href{https://arxiv.org/abs/0709.2399}{{\ttfamily 0709.2399}}].

\bibitem{Gregory:2008bf}
R.~Gregory, \emph{{The Three burials of Melquiades DGP}},
  \href{https://doi.org/10.1143/PTPS.172.71}{\emph{Prog. Theor. Phys. Suppl.}
  {\bfseries 172} (2008) 71} [\href{https://arxiv.org/abs/0801.1603}{{\ttfamily
  0801.1603}}].

\bibitem{Gabadadze:2003ck}
G.~Gabadadze and M.~Shifman, \emph{{Softly massive gravity}},
  \href{https://doi.org/10.1103/PhysRevD.69.124032}{\emph{Phys. Rev. D}
  {\bfseries 69} (2004) 124032}
  [\href{https://arxiv.org/abs/hep-th/0312289}{{\ttfamily hep-th/0312289}}].

\bibitem{Luty:2003vm}
M.A.~Luty, M.~Porrati and R.~Rattazzi, \emph{{Strong interactions and stability
  in the DGP model}},
  \href{https://doi.org/10.1088/1126-6708/2003/09/029}{\emph{JHEP} {\bfseries
  09} (2003) 029} [\href{https://arxiv.org/abs/hep-th/0303116}{{\ttfamily
  hep-th/0303116}}].

\bibitem{Koyama:2005tx}
K.~Koyama, \emph{{Are there ghosts in the self-accelerating brane universe?}},
  \href{https://doi.org/10.1103/PhysRevD.72.123511}{\emph{Phys. Rev. D}
  {\bfseries 72} (2005) 123511}
  [\href{https://arxiv.org/abs/hep-th/0503191}{{\ttfamily hep-th/0503191}}].

\bibitem{Deffayet:2007kf}
C.~Deffayet and K.~Menou, \emph{{Probing Gravity with Spacetime Sirens}},
  \href{https://doi.org/10.1086/522931}{\emph{Astrophys. J.} {\bfseries 668}
  (2007) L143} [\href{https://arxiv.org/abs/0709.0003}{{\ttfamily 0709.0003}}].

\bibitem{Pardo:2018ipy}
K.~Pardo, M.~Fishbach, D.E.~Holz and D.N.~Spergel, \emph{{Limits on the number
  of spacetime dimensions from GW170817}},
  \href{https://doi.org/10.1088/1475-7516/2018/07/048}{\emph{JCAP} {\bfseries
  07} (2018) 048} [\href{https://arxiv.org/abs/1801.08160}{{\ttfamily
  1801.08160}}].

\bibitem{Corman:2021avn}
M.~Corman, A.~Ghosh, C.~Escamilla-Rivera, M.A.~Hendry, S.~Marsat and
  N.~Tamanini, \emph{{Constraining cosmological extra dimensions with
  gravitational wave standard sirens: From theory to current and future
  multimessenger observations}},
  \href{https://doi.org/10.1103/PhysRevD.105.064061}{\emph{Phys. Rev. D}
  {\bfseries 105} (2022) 064061}
  [\href{https://arxiv.org/abs/2109.08748}{{\ttfamily 2109.08748}}].

\bibitem{Corman:2020pyr}
M.~Corman, C.~Escamilla-Rivera and M.A.~Hendry, \emph{{Constraining extra
  dimensions on cosmological scales with LISA future gravitational wave siren
  data}}, \href{https://doi.org/10.1088/1475-7516/2021/02/005}{\emph{JCAP}
  {\bfseries 02} (2021) 005}
  [\href{https://arxiv.org/abs/2004.04009}{{\ttfamily 2004.04009}}].

\bibitem{Poddar:2021yjd}
T.K.~Poddar, S.~Mohanty and S.~Jana, \emph{{Gravitational radiation from binary
  systems in massive graviton theories}},
  \href{https://doi.org/10.1088/1475-7516/2022/03/019}{\emph{JCAP} {\bfseries
  03} (2022) 019} [\href{https://arxiv.org/abs/2105.13335}{{\ttfamily
  2105.13335}}].

\bibitem{Cardoso:2008gn}
V.~Cardoso, O.J.C.~Dias and P.~Figueras, \emph{{Gravitational radiation in
  d\ensuremath{>}4 from effective field theory}},
  \href{https://doi.org/10.1103/PhysRevD.78.105010}{\emph{Phys. Rev. D}
  {\bfseries 78} (2008) 105010}
  [\href{https://arxiv.org/abs/0807.2261}{{\ttfamily 0807.2261}}].

\bibitem{Porto:2016pyg}
R.A.~Porto, \emph{{The effective field theorist\textquoteright{}s approach to
  gravitational dynamics}},
  \href{https://doi.org/10.1016/j.physrep.2016.04.003}{\emph{Phys. Rept.}
  {\bfseries 633} (2016) 1} [\href{https://arxiv.org/abs/1601.04914}{{\ttfamily
  1601.04914}}].

\bibitem{Birnholtz:2013ffa}
O.~Birnholtz and S.~Hadar, \emph{{Action for reaction in general dimension}},
  \href{https://doi.org/10.1103/PhysRevD.89.045003}{\emph{Phys. Rev. D}
  {\bfseries 89} (2014) 045003}
  [\href{https://arxiv.org/abs/1311.3196}{{\ttfamily 1311.3196}}].

\bibitem{hadamard2014lectures}
J.~Hadamard, \emph{Lectures on Cauchy's Problem in Linear Partial Differential
  Equations}, Dover Publications (2014).

\bibitem{courant2008methods}
R.~Courant and D.~Hilbert, \emph{Methods of Mathematical Physics: Partial
  Differential Equations}, Wiley Classics Library, Wiley (2008).

\bibitem{Ivanenko_book}
D.D.~Ivanenko and A.A.~Sokolov, \emph{{Classical Fields Theory}} (1953).

\bibitem{Cardoso:2002pa}
V.~Cardoso, O.J.C.~Dias and J.P.S.~Lemos, \emph{{Gravitational radiation in
  D-dimensional space-times}},
  \href{https://doi.org/10.1103/PhysRevD.67.064026}{\emph{Phys. Rev.}
  {\bfseries D67} (2003) 064026}
  [\href{https://arxiv.org/abs/hep-th/0212168}{{\ttfamily hep-th/0212168}}].

\bibitem{Galtsov:2001iv}
D.V.~Galtsov, \emph{{Radiation reaction in various dimensions}},
  \href{https://doi.org/10.1103/PhysRevD.66.025016}{\emph{Phys. Rev.}
  {\bfseries D66} (2002) 025016}
  [\href{https://arxiv.org/abs/hep-th/0112110}{{\ttfamily hep-th/0112110}}].

\bibitem{Galtsov:2020hhn}
D.V.~Gal'tsov and M.~Khlopunov, \emph{{Synchrotron radiation in odd
  dimensions}}, \href{https://doi.org/10.1103/PhysRevD.101.084054}{\emph{Phys.
  Rev. D} {\bfseries 101} (2020) 084054}
  [\href{https://arxiv.org/abs/2003.00261}{{\ttfamily 2003.00261}}].

\bibitem{Rohrlich1961}
F.~Rohrlich, \emph{The definition of electromagnetic radiation},
  \href{https://doi.org/10.1007/BF02785607}{\emph{Il Nuovo Cimento (1955-1965)}
  {\bfseries 21} (1961) 811}.

\bibitem{Teitelboim1970}
C.~Teitelboim, \emph{Splitting of the maxwell tensor: Radiation reaction
  without advanced fields},
  \href{https://doi.org/10.1103/PhysRevD.1.1572}{\emph{Phys. Rev. D} {\bfseries
  1} (1970) 1572}.

\bibitem{Kosyakov:1992qx}
B.P.~Kosyakov, \emph{{Radiation in electrodynamics and in Yang-Mills theory}},
  \href{https://doi.org/10.1070/PU1992v035n02ABEH002218}{\emph{Sov. Phys. Usp.}
  {\bfseries 35} (1992) 135}.

\bibitem{Hinterbichler:2011tt}
K.~Hinterbichler, \emph{{Theoretical Aspects of Massive Gravity}},
  \href{https://doi.org/10.1103/RevModPhys.84.671}{\emph{Rev. Mod. Phys.}
  {\bfseries 84} (2012) 671} [\href{https://arxiv.org/abs/1105.3735}{{\ttfamily
  1105.3735}}].

\bibitem{deRham:2014zqa}
C.~de~Rham, \emph{{Massive Gravity}},
  \href{https://doi.org/10.12942/lrr-2014-7}{\emph{Living Rev. Rel.} {\bfseries
  17} (2014) 7} [\href{https://arxiv.org/abs/1401.4173}{{\ttfamily
  1401.4173}}].

\bibitem{Moore:2014lga}
C.J.~Moore, R.H.~Cole and C.P.L.~Berry, \emph{{Gravitational-wave sensitivity
  curves}}, \href{https://doi.org/10.1088/0264-9381/32/1/015014}{\emph{Class.
  Quant. Grav.} {\bfseries 32} (2015) 015014}
  [\href{https://arxiv.org/abs/1408.0740}{{\ttfamily 1408.0740}}].

\bibitem{Poisson:2009pwt}
E.~Poisson, \emph{{A Relativist's Toolkit: The Mathematics of Black-Hole
  Mechanics}}, Cambridge University Press (12, 2009),
  \href{https://doi.org/10.1017/CBO9780511606601}{10.1017/CBO9780511606601}.

\bibitem{Gibbons:1976ue}
G.W.~Gibbons and S.W.~Hawking, \emph{{Action Integrals and Partition Functions
  in Quantum Gravity}},
  \href{https://doi.org/10.1103/PhysRevD.15.2752}{\emph{Phys. Rev. D}
  {\bfseries 15} (1977) 2752}.

\bibitem{Hawking:1995fd}
S.W.~Hawking and G.T.~Horowitz, \emph{{The Gravitational Hamiltonian, action,
  entropy and surface terms}},
  \href{https://doi.org/10.1088/0264-9381/13/6/017}{\emph{Class. Quant. Grav.}
  {\bfseries 13} (1996) 1487}
  [\href{https://arxiv.org/abs/gr-qc/9501014}{{\ttfamily gr-qc/9501014}}].

\bibitem{Arnowitt:1962hi}
R.L.~Arnowitt, S.~Deser and C.W.~Misner, \emph{{The Dynamics of general
  relativity}}, \href{https://doi.org/10.1007/s10714-008-0661-1}{\emph{Gen.
  Rel. Grav.} {\bfseries 40} (2008) 1997}
  [\href{https://arxiv.org/abs/gr-qc/0405109}{{\ttfamily gr-qc/0405109}}].

\bibitem{Wald:1984rg}
R.M.~Wald, \emph{{General Relativity}}, Chicago Univ. Pr., Chicago, USA (1984),
  \href{https://doi.org/10.7208/chicago/9780226870373.001.0001}{10.7208/chicago/9780226870373.001.0001}.

\bibitem{Israel:1966rt}
W.~Israel, \emph{{Singular hypersurfaces and thin shells in general
  relativity}}, \href{https://doi.org/10.1007/BF02710419}{\emph{Nuovo Cim. B}
  {\bfseries 44S10} (1966) 1}.

\bibitem{Rubakov:2003zb}
V.A.~Rubakov, \emph{{Strong coupling in brane induced gravity in
  five-dimensions}},  \href{https://arxiv.org/abs/hep-th/0303125}{{\ttfamily
  hep-th/0303125}}.

\bibitem{Maggiore:2005qv}
M.~Maggiore, \emph{{A Modern introduction to quantum field theory}}, Oxford
  Master Series in Physics (2005).

\bibitem{Stepanyanz_book}
K.V.~Stepanyantz, \emph{Classical Field Theory (in Russian)}, Fizmatlit (2009).

\bibitem{Fierz:1939ix}
M.~Fierz and W.~Pauli, \emph{{On relativistic wave equations for particles of
  arbitrary spin in an electromagnetic field}},
  \href{https://doi.org/10.1098/rspa.1939.0140}{\emph{Proc. Roy. Soc. Lond. A}
  {\bfseries 173} (1939) 211}.

\bibitem{Isaacson:1968hbi}
R.A.~Isaacson, \emph{{Gravitational Radiation in the Limit of High Frequency.
  I. The Linear Approximation and Geometrical Optics}},
  \href{https://doi.org/10.1103/PhysRev.166.1263}{\emph{Phys. Rev.} {\bfseries
  166} (1968) 1263}.

\bibitem{Isaacson:1968zza}
R.A.~Isaacson, \emph{{Gravitational Radiation in the Limit of High Frequency.
  II. Nonlinear Terms and the Ef fective Stress Tensor}},
  \href{https://doi.org/10.1103/PhysRev.166.1272}{\emph{Phys. Rev.} {\bfseries
  166} (1968) 1272}.

\bibitem{maggiore2008}
M.~Maggiore, \emph{Gravitational Waves: Volume 1: Theory and Experiments},
  Gravitational Waves, OUP Oxford (2008).

\bibitem{Flanagan:2005yc}
E.E.~Flanagan and S.A.~Hughes, \emph{{The Basics of gravitational wave
  theory}}, \href{https://doi.org/10.1088/1367-2630/7/1/204}{\emph{New J.
  Phys.} {\bfseries 7} (2005) 204}
  [\href{https://arxiv.org/abs/gr-qc/0501041}{{\ttfamily gr-qc/0501041}}].

\bibitem{Landau:1975pou}
L.D.~Landau and E.M.~Lifschits, \emph{{The Classical Theory of Fields}},
  vol.~Volume 2 of \emph{Course of Theoretical Physics}, Pergamon Press, Oxford
  (1975).

\bibitem{Jackson:1998nia}
J.D.~Jackson, \emph{{Classical Electrodynamics}}, Wiley (1998).

\bibitem{Lifshitz:1945du}
E.~Lifshitz, \emph{{Republication of: On the gravitational stability of the
  expanding universe}},
  \href{https://doi.org/10.1007/s10714-016-2165-8}{\emph{J. Phys. (USSR)}
  {\bfseries 10} (1946) 116}.

\bibitem{Bardeen:1980kt}
J.M.~Bardeen, \emph{{Gauge Invariant Cosmological Perturbations}},
  \href{https://doi.org/10.1103/PhysRevD.22.1882}{\emph{Phys. Rev. D}
  {\bfseries 22} (1980) 1882}.

\bibitem{Stewart:1990fm}
J.M.~Stewart, \emph{{Perturbations of Friedmann-Robertson-Walker cosmological
  models}}, \href{https://doi.org/10.1088/0264-9381/7/7/013}{\emph{Class.
  Quant. Grav.} {\bfseries 7} (1990) 1169}.

\bibitem{vanDam:1970vg}
H.~van Dam and M.J.G.~Veltman, \emph{{Massive and massless Yang-Mills and
  gravitational fields}},
  \href{https://doi.org/10.1016/0550-3213(70)90416-5}{\emph{Nucl. Phys. B}
  {\bfseries 22} (1970) 397}.

\bibitem{Zakharov:1970cc}
V.I.~Zakharov, \emph{{Linearized gravitation theory and the graviton mass}},
  {\emph{JETP Lett.} {\bfseries 12} (1970) 312}.

\bibitem{Weinberg:1972kfs}
S.~Weinberg, \emph{{Gravitation and Cosmology}: {Principles and Applications of
  the General Theory of Relativity}}, John Wiley and Sons, New York (1972).

\bibitem{Dvali:2001gm}
G.R.~Dvali, G.~Gabadadze, M.~Kolanovic and F.~Nitti, \emph{{The Power of brane
  induced gravity}},
  \href{https://doi.org/10.1103/PhysRevD.64.084004}{\emph{Phys. Rev. D}
  {\bfseries 64} (2001) 084004}
  [\href{https://arxiv.org/abs/hep-ph/0102216}{{\ttfamily hep-ph/0102216}}].

\bibitem{Dvali:2001gx}
G.R.~Dvali, G.~Gabadadze, M.~Kolanovic and F.~Nitti, \emph{{Scales of
  gravity}}, \href{https://doi.org/10.1103/PhysRevD.65.024031}{\emph{Phys. Rev.
  D} {\bfseries 65} (2002) 024031}
  [\href{https://arxiv.org/abs/hep-th/0106058}{{\ttfamily hep-th/0106058}}].

\bibitem{Pati:2000vt}
M.E.~Pati and C.M.~Will, \emph{{PostNewtonian gravitational radiation and
  equations of motion via direct integration of the relaxed Einstein equations.
  1. Foundations}},
  \href{https://doi.org/10.1103/PhysRevD.62.124015}{\emph{Phys. Rev. D}
  {\bfseries 62} (2000) 124015}
  [\href{https://arxiv.org/abs/gr-qc/0007087}{{\ttfamily gr-qc/0007087}}].

\bibitem{Kosyakov1999}
B.P.~Kosyakov, \emph{Exact solutions of classical electrodynamics and the
  yang-mills-wong theory in even-dimensional space-time},
  \href{https://doi.org/10.1007/BF02557347}{\emph{Theoretical and Mathematical
  Physics} {\bfseries 119} (1999) 493}.

\bibitem{zwillinger2014table}
D.~Zwillinger, \emph{Table of Integrals, Series, and Products}, Elsevier
  Science (2014).

\end{thebibliography}\endgroup

\end{document}